\documentclass[acmtog, authorversion]{acmart}
\def\BibTeX{{\rm B\kern-.05em{\sc i\kern-.025em b}\kern-.08em
    T\kern-.1667em\lower.7ex\hbox{E}\kern-.125emX}}

\copyrightyear{2017}

\acmJournal{TOG}
\acmVolume{36}
\acmNumber{4}
\acmArticle{39}
\acmYear{2017}
\acmMonth{3}

\title{Interactive Diversity Optimization of Environments}

\author{Glen Berseth*}
\orcid{0}
\affiliation{%
  \institution{University of British Columbia}
  }
\author{Mahyar Khayatkhoei*}
\affiliation{%
  \institution{Rutgers University}
}
\author{Brandon Haworth*}
\author{Muhammad Usman*}
\affiliation{%
  \institution{York University}
}

\author{Mubbasir Kapadia}
\affiliation{%
  \institution{Rutgers University}
}

\author{Petros Faloutsos}
\affiliation{%
  \institution{York University, Toronto Rehab Institute}
}

\keywords{diversity optimization, space analysis, architecture}

\setcopyright{rightsretained}

\usepackage{amssymb}
\usepackage{xspace}
\usepackage{subfig}

\usepackage{amsmath}
\usepackage{booktabs}
\usepackage{float}
\usepackage{wrapfig}

\usepackage{graphicx}
\usepackage{multirow}
\usepackage{verbatim}

\usepackage{tikz}
\usetikzlibrary{shapes,snakes}
\usetikzlibrary{positioning}
\usetikzlibrary{calc}
\usepackage{caption}
\usepackage{siunitx}

\usepackage{rotating}
\usepackage{algorithm}
\usepackage[noend]{algorithmic}

\newcommand{\func}[1]{\operatorname{\mathit{#1}}}  

\newcommand{\refsec}[1]{Section~\ref{#1}}
\newcommand{\refeq}[1]{Eq.~(\ref{#1})}
\newcommand{\reffig}[1]{Fig.~\ref{#1}}
\newcommand{\reftab}[1]{Table~\ref{#1}}
\newcommand{\appendixx}[0]{Appendix\xspace}

\newcommand{\DOME}{$\mu$DOME\xspace}

\renewcommand\vec[1]{\ensuremath\boldsymbol{#1}}
\newcommand{\function}[1]{\ensuremath{\textit{#1}}}

\newcommand{\todo}[1]{\textcolor{red}{Todo:#1}}

\newcommand{\glen}[1]{\textcolor{blue}{Glen:#1}}
\newcommand{\petros}[1]{\textcolor{orange}{Petros:#1}}
\newcommand{\mubbasir}[1]{\textcolor{purple}{Mubbasir:#1}}

\newcommand{\mahyar}[1]{\textcolor{brown}{Mahyar:#1}}

\newcommand{\valueWithUnits}[2]{#1~\normalsize $#2$\normalsize}
\definecolor{myCol}{RGB}{200,0,0}
\newcommand{\newChanges}[2]{#2}

\newcommand{\ignore}[1]{}

\newcommand{\nodeText}{\textnormal{node}\xspace}
\newcommand{\edgeText}{\textnormal{edge}\xspace}
\newcommand{\nodes}{\textnormal{nodes}\xspace}
\newcommand{\edges}{\textnormal{edges}\xspace}

\newcommand{\architecturalGraphText}{\textnormal{architectural graph}\xspace}
\newcommand{\spaceSyntaxText}{\textnormal{Space-Syntax}\xspace}

\newcommand{\architecturalGraph}{\ensuremath{G_{A}}\xspace}
\newcommand{\visibilityGraph}{\ensuremath{G_{V}}\xspace}
\newcommand{\edge}{\ensuremath{e}\xspace}
\newcommand{\node}{\ensuremath{n}\xspace}
\newcommand{\edgesE}{\ensuremath{E}\xspace}
\newcommand{\nodesN}{\ensuremath{N}\xspace}

\newcommand{\roq}{\emph{Region of Query}\xspace}
\newcommand{\ror}{\emph{Region of Reference}\xspace}

\newcommand{\Vg}{V}
\newcommand{\Eg}{E}

\newcommand{\gridNodes}{\ensuremath{V}\xspace}

\newcommand{\adjmat}{M_{adj}}
\newcommand{\adjvec}{V_{adj}}
\newcommand{\front}{F}
\newcommand{\child}{C}
\newcommand{\parent}{P}
\newcommand{\dc}{D}

\newcommand{\constrantParameter}{\vectorParameters}

\newcommand{\diversityMembers}{\ensuremath{\mathcal{D}}\xspace}
\newcommand{\diversityMembersCovar}{\ensuremath{\vec{\Sigma}}\xspace}
\newcommand{\diversityMember}{\ensuremath{\textbf{p}_{m}}\xspace}
\newcommand{\diversityMemberWeight}{\ensuremath{k}\xspace}
\newcommand{\diversityMemberMinWeight}{\ensuremath{k_{m}}\xspace}
\newcommand{\thresholdFunctions}{\ensuremath{\mathcal{T}}\xspace}

\newcommand{\constraint}{\ensuremath{c}\xspace}

\newcommand{\minDistanceThreshold}{\ensuremath{d_{min}}\xspace}
\newcommand{\threshold}{\ensuremath{z}\xspace}
\newcommand{\sampleSize}{\ensuremath{\lambda}\xspace}

\newcommand{\parameterSpace}{\ensuremath{\mathcal{P}}\xspace}

\newcommand{\clearance}{\ensuremath{clr}\xspace}
\newcommand{\alignment}{\ensuremath{align}\xspace}
\newcommand{\sumWallLengths}{\ensuremath{\mathcal{S}}\xspace}
\newcommand{\walls}{\ensuremath{wall}\xspace}

\newcommand{\vectorParameters}{\ensuremath{\textbf{p}}\xspace}
\newcommand{\vectorObjectiveWeights}{\ensuremath{\textbf{w}}\xspace}

\newcommand{\memberDistance}[2]{\ensuremath{\func{dn}(#1,#2)}\xspace}
\newcommand{\memberMinDistance}[2]{\ensuremath{\func{d}(#1,#2)}\xspace}

\newcommand{\penaltyFun}[1]{\ensuremath{\func{g}(#1)}\xspace}
\newcommand{\hierachicalPenaltyFun}[1]{\ensuremath{\func{g}_{h}(#1)}\xspace}
\newcommand{\functionArea}[1]{\ensuremath{\func{A}(#1)}\xspace}

\newcommand{\functionDiversity}[1]{\ensuremath{\func{div}(#1)}\xspace}

\newcommand{\functionConstraints}[1]{\ensuremath{\func{C}(#1)}\xspace}

\newcommand{\functionThreshold}[3]{\ensuremath{\func{t}(#1,#2,#3)}\xspace}

\newcommand{\wall}{\ensuremath{e}\xspace}

\newcommand{\objectiveDegree}[1]{\ensuremath{K(#1)}\xspace}
\newcommand{\objectiveDepth}[1]{\ensuremath{D(#1)}\xspace}
\newcommand{\objectiveEntropy}[1]{\ensuremath{H(#1)}\xspace}
\newcommand{\objectiveMulti}[1]{\ensuremath{m(#1)}\xspace}

\newcommand{\objective}[1]{\ensuremath{F(#1)}\xspace}
\newcommand{\objectiveVector}[0]{\ensuremath{{\bf f}\xspace}}

  \DeclareMathOperator*{\argmax}{arg\,max}

  \newcolumntype{M}{>{\centering\arraybackslash}m{\dimexpr.31\linewidth-1\tabcolsep}}

\begin{document}

\begin{abstract}

The design of a building requires an architect to balance a wide range of constraints:  aesthetic, geometric, usability, lighting, safety, etc. At the same time, there are often a multiplicity of diverse designs that can meet these constraints equally well. Architects must use their skills and artistic vision to explore these rich but highly constrained design spaces. A number of computer-aided design tools use automation to provide useful analytical data and optimal designs with respect to certain fitness criteria. However, this automation can come at the expense of a designer's creative control. 

We propose \DOME, a user-in-the-loop system for computer-aided design exploration that balances automation and control by efficiently exploring, analyzing, and filtering the space of environment layouts to better inform an architect's decision-making. At each design iteration, \DOME provides a set of diverse designs which satisfy user-defined constraints and optimality criteria within a user defined parameterization of the design space. The user then selects a design and performs a similar optimization  with the same or different parameters and objectives. This exploration process can be repeated as many times as the designer wishes. Our user studies indicates that \DOME, with its diversity-based approach, improves the efficiency and effectiveness of even novice users with minimal training, without compromising the quality of their designs.

\end{abstract}


%
%

\ccsdesc[500]{Computing methodologies~Animation}
\ccsdesc[300]{Computing methodologies~Physics-based Simulation}






\maketitle

\renewcommand{\thefootnote}{\fnsymbol{footnote}}
\footnote{These authors contributed equally to this work.}
\renewcommand{\thefootnote}{\arabic{footnote}}

\section{Introduction}
\label{sec:Intro}


Building design is both an art and a science. An architect must balance a wide variety of potentially competing objectives, such as space utilization, accessibility,  visibility of certain areas, and safety regulations, while at the same time exercising artistic and creative control. The space of possible designs  is extremely high-dimensional and continuous even for a small building, such as a single family home, let alone a museum the size of the Louvre. Searching this space for good solutions that meet different optimization criteria while balancing constraints is a challenging combinatorial problem.



Traditional \newChanges{word requested by reviews}{manual} design approaches rely on an architect's intuition and expertise to find suitable design solutions by ignoring or simplifying constraints, making heuristic, rather than optimal decisions, and accepting potentially sub-optimal results. Computer-aided design tools help address these challenges by leveraging automation to predictively analyze and evaluate building layouts.
Earlier methods are limited to simply computing quantitative measures  for a design, typically in the form of charts, tables, or heat maps.   
 Recent computer-aided design approaches  not only provide analysis information, but can also produce optimal designs using the recent advances in optimization techniques and brute force computing power.  However, these methods do not account for how people act and interact in these environments, because it is hard to quantify and incorporate into the optimization process. Furthermore, these approaches present a trade-off between automation and human control, where designers are limited to automatically synthesized designs which meet optimality considerations but may disregard designer constraints. 
 
 


 , we aim to combine combinatorial optimization  and human insight into  a framework for  exploring creative designs. We propose \DOME,  a user-in-the-loop computer-aided design tool that  employs diversity optimization to help architects and designers explore, analyze, and improve their work.   
 A key aspect of our approach is that the optimization process  itself is tuned for exploring alternatives (diversity) rather  than simply producing one optimal design at each invocation.

Within \DOME, a user first selects a set of environment elements and specifies which  of the associated parameters may be explored by the system.
Then, the user selects one or more metrics to serve as the optimization objectives in addition to the regions in the environment where the metrics should be computed on a  regions of interest. 
For example, a user may wish to increase the visibility of a painting in a room with respect to the entrance(s) while maintaining an ordered room layout with sufficient clearance between walls for people to pass through.
The user's selections define  a constrained multi-objective optimization problem.
A key novelty of our approach is that instead of simply solving for a single optimal configuration, we solve for a set of diverse candidate solutions.
Our formulation introduces a diversity term in the objective formulation. This requires the solver to \emph{focus} the search to meet optimality criteria, while simultaneously \emph{broadening} its exploration to maximize diversity of its candidate solutions. The process of balancing multiple objectives during optimization is a well known challenge, which is rendered even more difficult by the presence of a diversity term.  To address this issue, we propose a novel hierarchical multi-objective optimization algorithm which balances optimality and diversity while remaining efficient for interactive use. 


An important issue, for any application that computationally evaluates an environment, is the choice of evaluation quantities. 
There are different metrics that quantify useful aspects of an environment. In this work, we use metrics that focus on the utility of an environment with respect to its inhabitants
use three measures defined by Space Syntax~\cite{bafna2003}. These metrics, in general terms, capture the way people interact with an environment by quantifying the visibility, accessibility, and organization of the space. These metrics are expensive to compute for large environments, especially as part of multiple optimization iterations. To mitigate their computational cost, we develop GPU accelerated versions of the metrics.  We perform a sensitivity analysis to identify the minimum critical resolution of the environment discretization beyond which the measures converge, thus allowing us to compute these measures at a coarse granularity without noticeable loss in accuracy. These  improvements offer significant performance gains, enabling the system to operate interactively for mid-scale environments.


Our  framework can serve in a range of assisting roles, from an efficient way to evaluate alternate configurations which accomplish the similar objectives, all the way to a design brainstorming assistant. We have integrated \DOME within an industry standard architectural design system, Autodesk Revit\textregistered. Our results demonstrate the value of our approach in iteratively optimizing and refining existing floor plans for a wide range of environments including an office, an art gallery, a subway station,  a museum, and  a maze. We have also performed user studies with experts and novice users to evaluate the usability and efficacy of \DOME. The SUS score of \DOME is $71.73$, which suggests that even novices were able  to use our system with minimal training. The results indicate that subjects using \DOME were able to produce more optimal designs, in comparison to subject who didn't use \DOME and experts preferred designs from \DOME. 
 Our contributions can be summarized as follows:
\begin{itemize}
\setlength\itemsep{0em}
\item We propose a user-in-loop system for computer-assisted exploration of building designs. 
\item We introduce an efficient hierarchical multi-objective optimization method to balance optimality and diversity of alternative designs.
\item We develop GPU-accelerated measures for spatial analysis of scenes.
\item We integrate \DOME within the Autodesk Revit {\textregistered} pipeline for demonstration and evaluation. 
\item We performed user studies to show the effectiveness of \DOME. 
\end{itemize}





\begin{figure*}[t]
	\centering
	\includegraphics[clip,width=\textwidth]{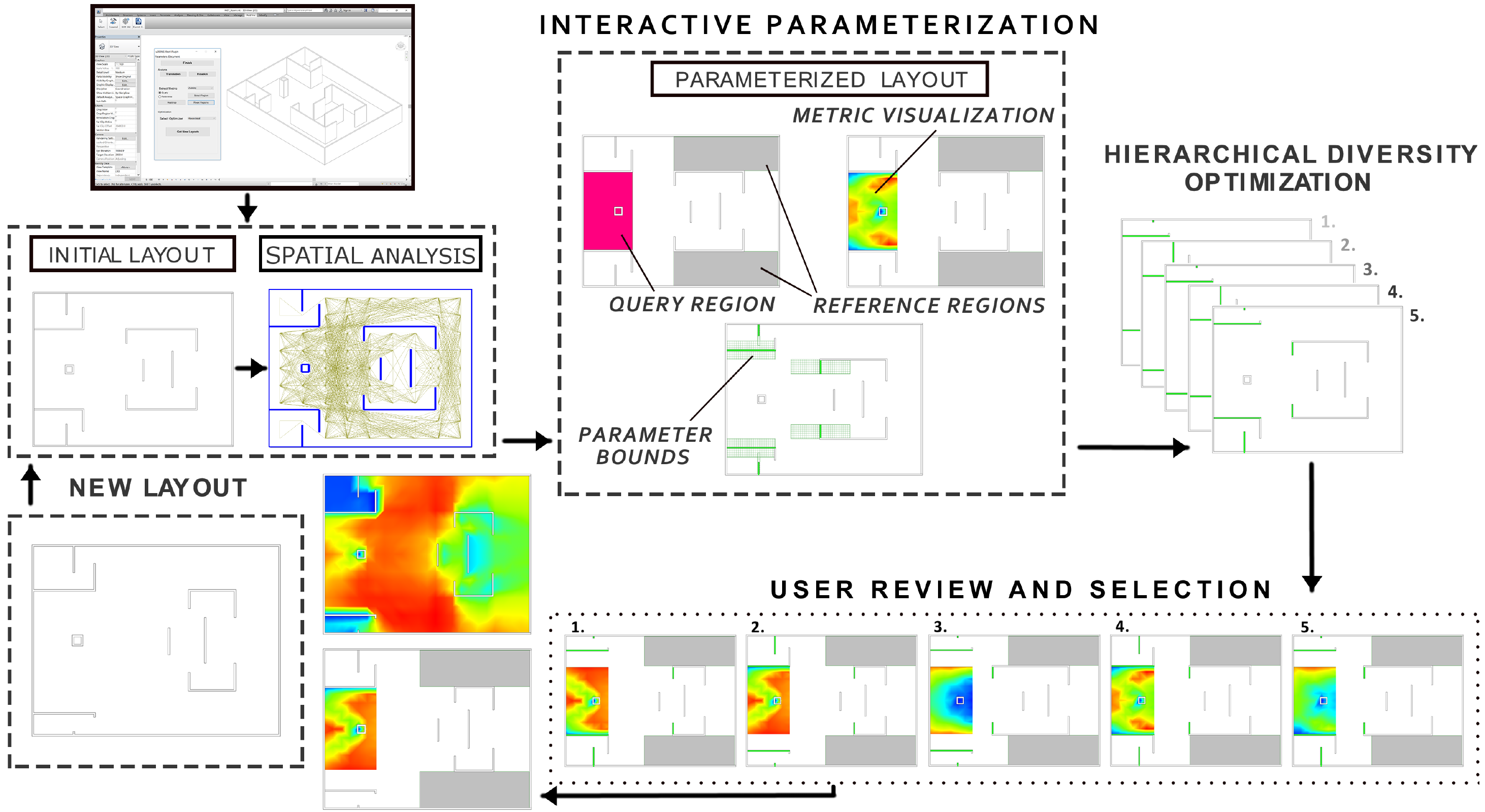}

\caption{\label{figure:DOME-FrameWork} \DOME Framework Overview. Starting with an initial environment design, the user specifies permissible alterations to the layout as bounds on the degree to which different environment elements may be transformed. The user then specifies one or more focal regions in the environment for which different spatial measures are computed, to quantify visibility, accessibility, and organization of the space. A multi-objective hierarchical diversity optimization produces a set of diverse near optimal solutions with respect to user-defined optimality criteria, from which the user may select one and repeat the process as desired.}

\end{figure*}

\section{Related Work} 
\label{section-related-work}
Computer-aided design (CAD) methods have garnered increasing attention from both researchers and practitioners in recent years, as they allow designers and casual users to leverage automation at all stages of the design process. This has led to an evolution of CAD tools for architectural design that increasingly use computational resources to analyze, evaluate, and optimize the layout of buildings subject to various criteria. \\ 


\noindent \textbf{Automated Architectural Design. } There is a growing interest in using optimization techniques to explore design spaces for near-optimal solutions given certain problem criteria~\cite{block2014advances,pottmann2014architectural,Peng:2016:CND:2897824.2925935}.  Galle~\shortcite{Galle:1981:AEG:358800.358804} focused on exhaustively searching possible layout configurations for small-scale environments. Since then, evolutionary approaches~\cite{michalek2002interactive,yi2014performance} have been used to curb the infeasibility of brute-force methods for larger design spaces. Liu \emph{et al}.~\shortcite{DBLP:journals/vc/LiuYAM13} introduced functional, design, and fabrication constraints as objective measures to guide the optimization process. Data-driven approaches~\cite{Merrell:2010:CRB:1882261.1866203} learn layout configurations from existing databases, which are used to automatically generate new layouts for computer graphics applications. Design objectives can be modelled as forces applied to physical features to generate layout designs automatically~\cite{arvin2002modeling}. A sophisticated optimization scheme takes into account the visibility, accessibility, and other hierarchical spatial relationships between interior objects to produce realistic interior design configurations~\cite{craigyu2011furniture}. Optimization methods can also successfully account for different physical aspects considered important to architecture such as  sunlight~\cite{yi2014performance}, materials, energy savings~\cite{caldas2002design} or even acoustics~\cite{bassuet2014computational}.  \\

\noindent \textbf{Interactive Design Solutions.} While automated approaches can take into account  objective criteria, architectural design inherently involves subjective decisions about aesthetics, domain expertise, and hard-to-quantify criteria such as human activity and its relationship to the environment. These challenges are mitigated by proposing computer-assisted, interactive tools that keep the user in the design loop, while using automation to inform the designers decision-making~\cite{shi2013performance,felkner2013interactive,turrin2011design,CGF:CGF12314}. Harada \emph{et al.}~\shortcite{Harada:1995:IPM:218380.218443} uses shape grammars to support the interactive manipulation of architectural layouts. Recent works have proposed optimisation-based interactive design tools to facilitate furniture arrangement using interior design principles~\cite{craigyu2011furniture,Merrell:2011:IFL:2010324.1964982}. Akase \emph{et al}.~\shortcite{Akase:2014:WMR:2636240.2636849} proposed an online room design framework where the objective function entirely relies upon the user's evaluation. \\ 


\noindent \textbf{Automatic Exploration of Diverse Designs.} To better balance automation and the user's creative control, researchers have proposed approaches for exploring multi-dimensional search spaces to find multiple, diverse, yet optimal solutions which can be provided as suggestions to the designer. This provides the designer with more control, allowing them to harness the power of computation to efficiently explore large design spaces, in domains including multi-body dynamics~\cite{Twigg:2007:MBC:1275808.1276395,6781622}, light selection and image rendering~\cite{Marks:1997:DGG:258734.258887}. Introducing diversity as part of the optimization formulation makes the problem significantly more challenging, with many proposed solutions including constraint programming~\cite{hebrard2005finding}, evolutionary methods~\cite{Ursem2002}, and domain-independent methods~\cite{srivastava2007domain,coman2011generating}. In this work, we use a round robin approach that introduces a minimal number of optimization parameters. \\


\noindent \textbf{Architectural Metrics}
\spaceSyntaxText is an established framework for spatial analysis \cite{hillier,peponis,turner1999making,bafna2003}. It includes a wide range of spatial measures, which have been shown to correlate with human behaviour \cite{dara2006architecture,davies2006isovists,meilinger2012isovists,emo2012wayfinding}. In this work, we  use a set of static measures grounded in \spaceSyntaxText, however, our framework is independent of this particular choice and can easily incorporate other spatial measures.

\noindent \textbf{Human-Factored Architectural Layout Analysis and Optimization.} A key challenge in the analysis of environment designs is to account for factors related to its human occupants, which are difficult to quantify. Fruin~\shortcite{fruin1971pedestrian} uses crowd density as a proxy to estimate the level of service (LOS) of environments. Fisher \emph{et al}.~\shortcite{Fisher:2015:ASS:2816795.2818057} synthesized functional 3D scenes by deducing possible human activities. AlHalawani and Mitra~\shortcite{DBLP:conf/isvc/AlHalawaniM15} proposed an approach for optimizing object placement in a warehouse by analyzing traffic congestion. Recent work has collect network related city features and classified different cities based on these features~\cite{AlHalawani:2014:MLW:2771589.2771605}. 
Crowd simulation methods are perhaps the most accurate proxy of measuring real human movement, but are  computationally too expensive for  interactive optimization applications~\cite{doi:10.2200/S00673ED1V01Y201509CGR020}. 
Berseth \emph{et al}.~\shortcite{berseth2015environment} uses crowd simulation to optimally place a small number of environmental elements in small-scale evacuation scenarios. Feng \emph{et al}.~\shortcite{midlayout} learns the relationship between crowd flow and various layout alternatives. The estimated model  is then used to automatically reconfigure the layout in order to optimize for human factors such as flow. \\ 
 
 \noindent \textbf{Our Work. } Our work strives to keep the user central to the design process, while leveraging computation to inform the user of factors  which are difficult to interpret (e.g., human occupancy), and efficiently explore the design spaces. 

\section{Overview} 
\label{section-overview}
An overview of the major components of \DOME is illustrated in Figure~\ref{figure:DOME-FrameWork}. 
In subsequent sections, each part of \DOME is delineated in detail with examples. \\ 

\noindent \textbf{Environment Parameterization.} 
Given an initial environment  layout, a user first selects elements (e.g., disjoint structures such as pillars, junctions, or walls), and specifies limits on different degrees of freedom of these elements. These attributes  represent a user defined parametrization of the environment layout, which together with the associated limits, model the space of admissible configurations of the environment. 
This affords both subjectivity as well as strict adherence to constraints such as structural integrity of the building. See Section~\ref{section-environment-parameterization} for details. \\


\noindent \textbf{Spatial Analysis.}
\DOME constructs a discrete graph representation of free space in the environment, \DOME computes different spatial metrics to quantify visibility, accessibility, and organization of the space.
While any metrics may be computed over the environment, these measures are predictive of spatial utilization and human movement, and serve as the basis for quantitatively analysing the environment. The user may optionally restrict the computation of these measures to specific regions of interest. For example, the user may wish to maximize the visibility of a key location with respect to the exits in a room. See Section~\ref{section-environment-analysis} for details. \\


\noindent \textbf{Multi-Objective Diversity Optimization.} The environment parameters, designer constraints, and spatial measures are used to formulate an optimization problem over the space of environment configurations. We desire to keep the user central to the design process while using automation to provide multiple diverse suggestions for improving the current design. To facilitate this, we formulate our objective formulation to generate structurally diverse layouts, while preserving the aforementioned optimality criteria. \DOME efficiently searches through the space of permissible environment configurations to identify diverse, yet optimal candidates using a novel hierarchical multi-objective optimization algorithm. See Section~\ref{section-optimization} for details. \\

\noindent \textbf{User-in-the Loop Iterative Design.} The designer reviews each of the candidate designs which are then used  as the basis for subsequent alterations through a tightly coupled design and optimization process. Using \DOME, designers can  leverage computation to account for difficult to interpret features such as accessibility and visibility of an environment with respect to its human occupants. The diverse layouts are provided to the user as suggestions, together with visualisations of the spatial measures. The designer may browse these and make an informed decision on which candidate best suits their vision. See Section~\ref{section-multi-objective} for details.

\section{Environment Parameterization} 
\label{section-environment-parameterization}

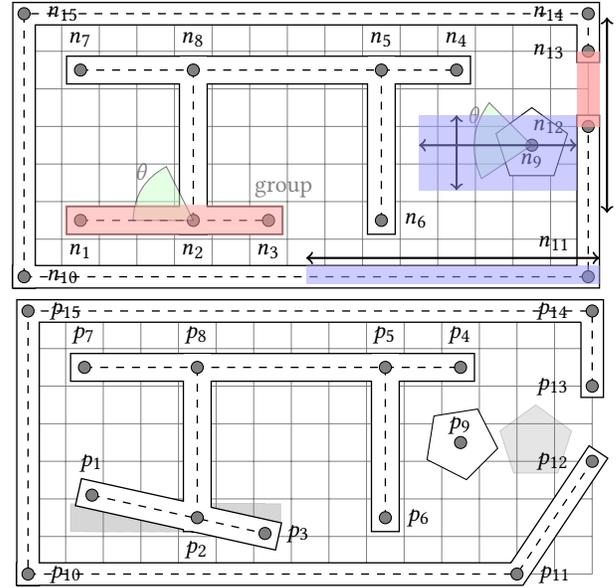
\begin{figure}
\centering
\begin{tikzpicture}
\draw[step=5mm, gray, very thin, shift={(-0.25,-0.25)}] (-0.5,-0.5) grid (7,3); 
 \draw[line width = 0.5, double distance = 10,line cap=rect](0,0)  --(2.5,0)  (1.5,0)--(1.5,2.0) (0,2.0)--(5.0,2.0) (4.0,2.0)--(4.0,0)-- (4.0,0) ;
\draw[line width = 0.5, dashed](0,0)--(2.5,0)  (1.5,0)--(1.5,2.0) (0,2.0)--(5.0,2.0) (4.0,2.0)--(4.0,0)-- (4.0,0);
\draw[solid, fill=gray] (0,0) circle (0.08)   node[below = 0.2cm ] {$\node_1$}; 
\draw[solid, fill=gray] (2.5,0) circle (0.08) node[below = 0.2cm ] {$\node_3$};
\draw[solid, fill=gray] (0,2.0) circle(0.08) node[above=0.2cm] {$\node_7$};
\draw[solid,fill=gray] (5.0,2.0) circle(0.08) node[above=0.2cm] {$\node_4$};
\draw[solid,fill=gray] (4.0,2.0) circle(0.08) node[above=0.2cm] {$\node_5$};
\draw[solid,fill=gray] (4.0,0) circle(0.08) node[right=0.2cm] {$\node_6$};
\draw[solid,fill=gray] (1.5,0) circle(0.08) node[below=0.2cm] {$\node_2$};
\draw[solid,fill=gray] (1.5,2.0) circle(0.08)  node[above=0.2cm] {$\node_{8}$};
\node [draw, fill=white, thin, scale=1.0, minimum size=1.0cm, regular polygon, regular polygon sides=5] at (6 ,1) {};
\draw[solid,fill=gray] (6,1.0) circle(0.08) node[below] {$\node_9$} ;
\draw[line width = 0.5, shift={(-0.25,-0.25)}, double distance = 8,line cap=rect] (-0.5,-0.5) -- (7,-0.5) -- (7,1.5) (7,2.5)--(7,3)-- (-0.5,3)--(-0.5,-0.5) ;
\draw[line width = 0.5, dashed, shift={(-0.25,-0.25)} ] (-0.5,-0.5) -- (7,-0.5) -- (7,1.5) (7,2.5)--(7,3)-- (-0.5,3)--(-0.5,-0.5) ;
\draw[solid, fill=gray, shift={(-0.25,-0.25)}] (-0.5,-0.5)  circle (0.08)   node[right = 0.2cm ] {$\node_{10}$}; 
\draw[solid, fill=gray, shift={(-0.25,-0.25)}] (7,-0.5)  circle (0.08)  ; 
\node at (6.3,-0.3) { $\node_{11}$};
\draw[solid, fill=gray, shift={(-0.25,-0.25)}]  (7,1.5)  circle (0.08)   node[left = 0.2cm ] {$\node_{12}$}; 
\draw[solid, fill=gray, shift={(-0.25,-0.25)}] (7,2.5) circle (0.08)   node[left = 0.2cm ] {$\node_{13}$}; 
\draw[solid, fill=gray, shift={(-0.25,-0.25)}](7,3) circle (0.08)   node[left = 0.2cm ] {$\node_{14}$}; 
\draw[solid, fill=gray,	shift={(-0.25,-0.25)}] (-0.5,3) circle (0.08)   node[right = 0.2cm ] {$\node_{15}$}; 
\coordinate (b) at (3,-1);
\coordinate (c) at (6,1);
\coordinate (a) at (5,2);
\draw[fill=green!20, opacity=0.5] (c) -- ($(c)!8mm!(b)$) to[bend left] node[above=0.1]{$\theta$} ($(c)!8mm!(a)$)  -- cycle ;

\coordinate (b) at (-1.5,0);
\coordinate (c) at (1.5,0);
\coordinate (a) at (0.5,2);
\draw[fill=green!20, opacity=0.5] (c) -- ($(c)!8mm!(b)$) to[bend left] node[above]{$\theta$} ($(c)!8mm!(a)$)  -- cycle ;

\draw[<->, thick] (4.5,1) -- (6.6,1);
\draw[<->, thick] (5.0,0.4) -- (5.0,1.4);
\draw[fill=blue!40,opacity=0.5,draw=none] (4.5,0.4) rectangle (6.6,1.4);

\draw[fill=red!40,opacity=0.5,draw=red!40] (-0.2,-0.2) rectangle (2.7,0.2) node[above] { group};
\draw[fill=red!40,opacity=0.7,draw=none] (6.6,1.25) rectangle (6.9,2.25);
\draw[<->, thick] (7,0.1) -- (7, 2.7);

\draw[<->, thick] (3,-0.5) -- (6.9,-0.5);
\draw[fill=blue!40,opacity=0.5,draw=none] (3,-0.85) rectangle (6.9, -0.6);

\end{tikzpicture}
\begin{tikzpicture}
 \draw[line width = 0.5, thin,dashed,opacity=0.2, double distance = 10,line cap=rect](0,0)  --(2.5,0) ;

\draw[step=5mm, gray, very thin, shift={(-0.25,-0.25)}] (-0.5,-0.5) grid (7,3); 
 \draw[line width = 0.5, double distance = 10,line cap=rect](0.10,0.3)  --(2.4,-0.21)  (1.5,0.0)--(1.5,2.0) (0,2.0)--(5.0,2.0) (4.0,2.0)--(4.0,0)-- (4.0,0) ;
\draw[line width = 0.5, dashed](0.10,0.3)--(2.4,-0.21)  (1.5,0.2)--(1.5,2.0) (0,2.0)--(5.0,2.0) (4.0,2.0)--(4.0,0)-- (4.0,0);

\draw[solid, fill=gray] (0.10,0.3) circle (0.08)   node[above = 0.2cm ] {$p_1$}; 
\draw[solid, fill=gray] (2.4,-0.21) circle (0.08) node[right = 0.2cm ] {$p_3$};
\draw[solid, fill=gray] (0,2.0) circle(0.08) node[above=0.2cm] {$p_7$};
\draw[solid,fill=gray] (5.0,2.0) circle(0.08) node[above=0.2cm] {$p_4$};
\draw[solid,fill=gray] (4.0,2.0) circle(0.08) node[above=0.2cm] {$p_5$};
\draw[solid,fill=gray] (4.0,0) circle(0.08) node[right=0.2cm] {$p_6$};
\draw[solid,fill=gray] (1.5,0) circle(0.08) node[below=0.2cm] {$p_2$};
\draw[solid,fill=gray] (1.5,2.0) circle(0.08)  node[above=0.2cm] {$p_{8}$};
\node [draw, rotate =45, fill=white, thin, scale=1.0, minimum size=1.0cm, regular polygon, regular polygon sides=5] at (5 ,1) {};
\node [draw, fill=gray,opacity=0.2, thin, scale=1.0, minimum size=1.0cm, regular polygon, regular polygon sides=5] at (6 ,1) {};
\draw[solid, fill=gray] (5,1.0) circle(0.08) node[above] {$p_9$} ;
\draw[line width = 0.5, shift={(-0.25,-0.25)}, double distance = 8,line cap=rect] (-0.5,-0.5) -- (6,-0.5) -- (7,1) (7,2.0)--(7,3)-- (-0.5,3)--(-0.5,-0.5) ;
\draw[line width = 0.5, dashed, shift={(-0.25,-0.25)} ] (-0.5,-0.5) -- (6,-0.5) -- (7,1) (7,2.0)--(7,3)-- (-0.5,3)--(-0.5,-0.5) ;
\draw[solid, fill=gray, shift={(-0.25,-0.25)}] (-0.5,-0.5)  circle (0.08)   node[right = 0.2cm ] {$p_{10}$}; 
\draw[solid, fill=gray, shift={(-0.25,-0.25)}] (6,-0.5)  circle (0.08)   node[right= 0.2cm ] {$p_{11}$}; 
\draw[solid, fill=gray, shift={(-0.25,-0.25)}]  (7,1)  circle (0.08)   node[left = 0.2cm ] {$p_{12}$}; 
\draw[solid, fill=gray, shift={(-0.25,-0.25)}] (7,2) circle (0.08)   node[left = 0.2cm ] {$p_{13}$}; 
\draw[solid, fill=gray, shift={(-0.25,-0.25)}](7,3) circle (0.08)   node[left = 0.2cm ] {$p_{14}$}; 
\draw[solid, fill=gray,	shift={(-0.25,-0.25)}] (-0.5,3) circle (0.08)   node[right = 0.2cm ] {$p_{15}$}; 
\end{tikzpicture} 
\caption{The layout of a floor plan and the corresponding graph parametrization of the walls, doors and other rigid elements. User selected nodes of the graph,  $\node_i $, can be grouped, translated, scaled, and rotated within user defined  bounds shown in colour and with arcs and arrows. }
\label{figure:building-graph-example}
\end{figure}

The architectural elements of a building and their connections can be represented by an undirected 
 \architecturalGraphText $\architecturalGraph = \langle N, E \rangle$, comprising of a set of \nodes \nodesN $= \{ \node_i \}$ and \edges \edgesE = $\{  \edge_i\}$. Each \nodeText $\node \in R^{2}$ specifies a location in  2D-space. Each \edgeText is a pair of \nodes $\edge = \langle \node_{i}, \node_{j} \rangle$. An example of a building layout and the associated graph abstraction is shown in \reffig{figure:building-graph-example}. In this representation, the  walls are the \edges (\edge) in the graph, while the  \nodes (\node) represent end points and junctions between walls.  
If a connected component in an \architecturalGraphText contains a  single \nodeText and no edges, such as $\node_9$ in \reffig{figure:building-graph-example},  then the node itself  represents an  element with fixed structure. The geometry of each element (wall, kiosk, etc.)  is stored in a database and associated with the corresponding node or edge.

Given an \architecturalGraphText \architecturalGraph, the user can define the design space by parametrizing and constraining the attributes  of selected \nodes or groups of \nodes. For demonstration, we focus primarily on rigid body transformations of position and orientation.

Each element of the parametrization, $q_{i} = \left\langle N_{i}, p_{i}, g_{i}, \constraint_{i} \right\rangle$, contains a set of \nodes $ N_i = \{\node_{j}\}$, a transformation  $g_{i}$ that will be applied to the \nodes, the magnitude $p_{i}$  of the transformation, and the limits or constraints $\constraint_{i}$ on the magnitude $p_{i}$. Grouping the free parameters in a vector $\vectorParameters = \{p_{1}, \ldots, p_{k}\} $ the parametrization of the design space can be compactly represented as  $\architecturalGraph(\vectorParameters)$.

\reffig{figure:building-graph-example} shows an example of a floor plan with sixteen \nodes.  The arrows, arc, and painted regions around 
\nodeText $\node_{9}$ show the user specified range that the \nodeText can translate and rotate within. The group of \nodes $\{\node_1, \node_2, \node_3\}$ (in red) can rotate around $\node_2$ within the specified range.  The \nodes  $\node_{12}$ and $\node_{13}$ can move in the $y-$axis but are constrained to maintain their initial distance, forming another group.

\section{Spatial Analysis} 
\label{section-environment-analysis}

\begin{figure*}[t]
	\centering
	\captionsetup[subfloat]{farskip=1pt,captionskip=4pt}
	\subfloat[Visibility]{\includegraphics[trim={0cm 0cm 2.6cm 0cm},clip, width=0.22\textwidth]{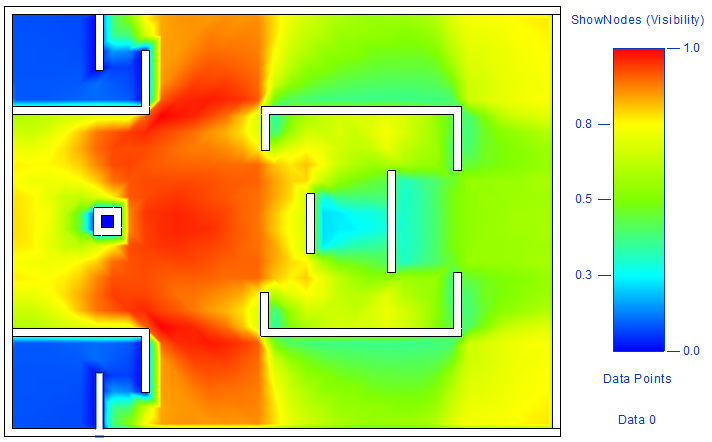}}\hfill
	\subfloat[Tree depth]{\includegraphics[trim={0cm 0cm 2.7cm 0cm},clip, width=0.22\textwidth]{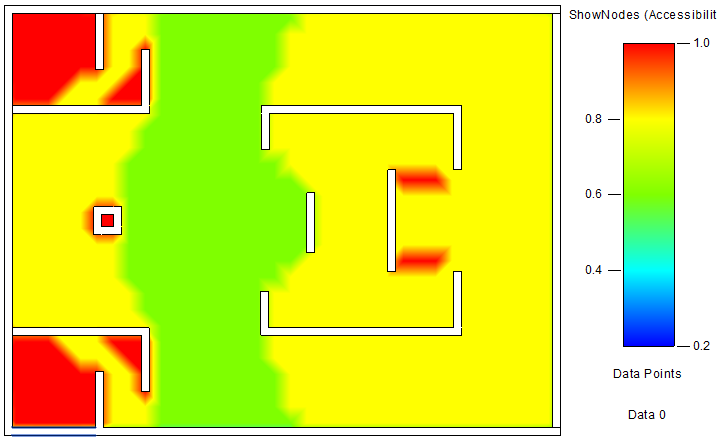}}\hfill
	\subfloat[Entropy]{\includegraphics[trim={0cm 0cm 3cm 0cm},clip, width=0.22\textwidth]{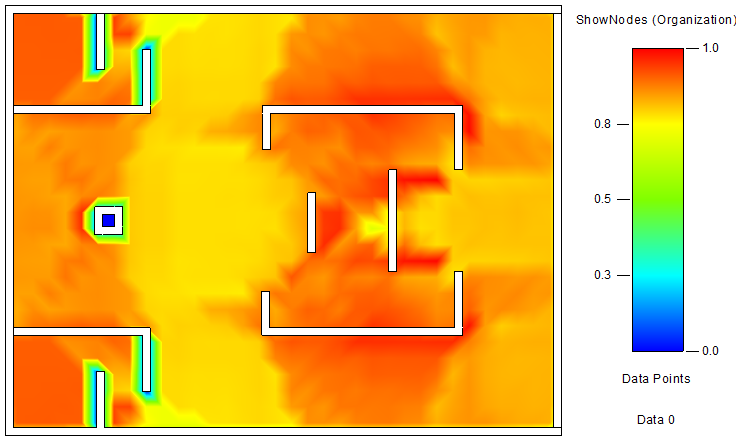}}\hfill
	\subfloat[Visibility for \roq]{\includegraphics[trim={0cm 0cm 0cm 0cm},clip, width=0.275\textwidth]{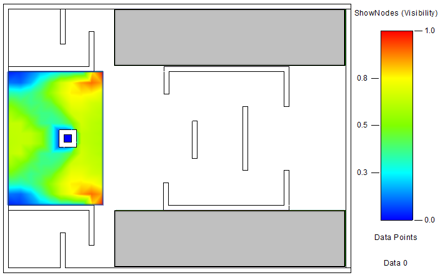}}\hfill
	\caption{Metrics values for a room in the Metropolitan Museum of Art. Heatmap colours indicate value from blue (low) to red (high). (a) Degree of visibility, where redder areas show more integrated regions, and are good candidates for placement of fire exits, signs, main event, etc. (b) Tree depth, where bluer areas have lower depth and are easier to access. (c) Entropy, where redder areas have high entropy (order), resulting in better human environmental cognition and easier planning at those points. (d) Degree of visibility in {\roq} with respect to {\ror} which is shown in grey. Notice that the degree values (which are a function of {\ror}) are different in comparison to (a).
	}
	\label{figure:metrics}
\end{figure*}

\begin{figure}
	\centering
	\resizebox{.95\linewidth}{!}{\input{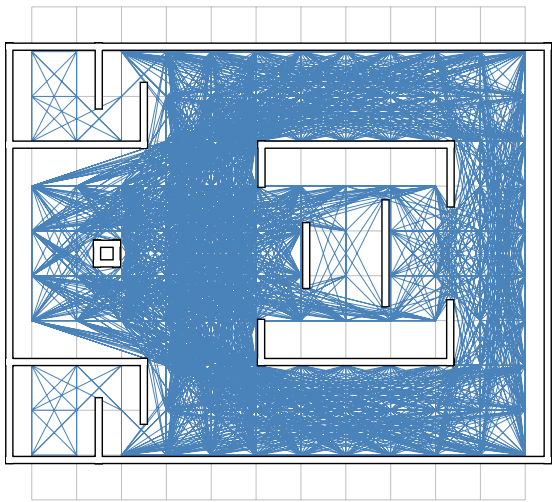}}	
	\caption{Visibility graph (\visibilityGraph) of the environment space which is discretized as finite grid ($\gridNodes$). Note that the sampling frequency is reduced in this figure for visualization - real examples use denser sampling.}
	\label{figure:visibility-graph}
\end{figure}

Spatial analysis aims to quantify attributes of an environment that directly affect  how people use the environment.  The ideal measures should be general and intuitive, and should cover all of the most important aspects of an environment. Spatial analysis focuses primarily on static measures that are computed geometrically. 

There are different approaches to represent space for the purposes of defining and computing spatial measures~\cite{desyllas}.  We chose visibility graphs, as they are easier to compute and tend to be more informative than  alternative representations, such as axial maps \cite{turner1999making,desyllas}. 
Our method is not restricted to specific metrics, however for this work we compute \spaceSyntaxText metrics.

\subsection{Visibility Graph}
\label{sec:visibilityGraph}


To construct a visibility graph, $\visibilityGraph(\Vg,\Eg)$, we first sample the environment with a finite grid $\gridNodes$, and then create an edge, $e \in \Eg$, between every pair of nodes that share an unobstructed line of sight, see \reffig{figure:visibility-graph}. 

In most prior work, every vertex of  grid $\gridNodes$  is a vertex of the visibility graph. In many cases it may be  useful to define the visibility graph, and consequently the associated measures, on  specific regions of interest. For instance, we may be more interested in the accessibility of certain doors, or the visibility of a an exit sign, from specific hallways in the environment.  To support this important feature, we allow the user to define two sets of grid vertices, the  {\roq} with vertices $V_q \subset \gridNodes$,  and the {\ror} with vertices $V_r \subset \gridNodes$, see \reffig{figure:DOME-FrameWork}. We then construct a visibility graph from these two sets of vertices by computing the lines of sight between the vertices in the {\roq} and the vertices in the {\ror}. 
The user defined regions provide greater flexibility to the user, giving them more control over the spatial queries to be performed on the layout. Putting everything together, the visibility graph depends on the architectural graph, its parametrization, and the regions of interest: 
\begin{equation}
\visibilityGraph = \phi (\gridNodes_r,\gridNodes_q, \architecturalGraph(\vectorParameters)).
\end{equation}
The spatial measures described in the next section are computed only for the vertices of the region of query. 

\subsection{Metrics}
\label{sec:finalMetrics}

Given a visibility graph $\visibilityGraph(\Vg,\Eg)$, metrics are computed that characterize meaningful
relationships between floor plans and human behaviour. While \DOME can incorporate many metrics that have been proposed~\cite{bafna2003,turner,hillier1987syntactic,jiang}, we find the following measures sufficient.

\textbf{Degree of Visibility.} The degree of visibility, $k_i$, of a vertex $v_i \in \Vg$ is the number of edges incident to the vertex, in other words the number of its immediate  (1-hop) neighbours  $N_i$.
Regions with high degree of visibility can be considered to be more connected,  safe, or important~\cite{turner,bafna2003}. If one wants to install a public safety sign, then positioning it in a high visibility region might be appropriate~\cite{holscher}. In \reffig{figure:metrics}(a), red areas have the highest degree of visibility while blues indicate the lowest. 

\textbf{Tree Depth.}  Let $\visibilityGraph^{i} \subseteq \visibilityGraph$ be the largest connected component that contains  vertex $v_i \in \Vg$. The minimum height Tr\'{e}meaux tree rooted at $v_i$ is the tree depth, $d_i$.  Tree depth has a few intuitive interpretations. First, it measures how far  $\visibilityGraph^{i}$ is from being a star~\cite{Neetil:2012:SGS:2230458}. Second,  a vertex with large tree depth is connected to other regions of the environment through a long sequence of  vertices. 
 Thus, tree depth often relates to the notion of \textit{accessibility} in an environment~\cite{turner}. Tree depth values, together with context dependent information, allow a user to  make flow and congestion estimations on specific areas of a layout.
\reffig{figure:metrics}(b) shows the computed depth values in heatmap form, where a lower value (blue) is better.

\textbf{Entropy.} Let  $\visibilityGraph^{i} \subseteq \visibilityGraph$ be the largest connected component  that contains  vertex $v_i \in \Vg$. Given a Tr\'{e}meaux tree $T_i$  rooted at vertex $i$ with $n_i^j$ vertices at each  level $j$, we define a probability distribution $p(j|i)$ for $T_i$ over the domain $j \in [1, |\Vg_i|]$, where $\Vg_i$ is the set of vertices in $\visibilityGraph^{i}$, and through this distribution we define
the entropy $h_i$ at vertex $i$ as follows:
\begin{eqnarray}
& & p(j|i) = \frac{n_i^j}{\sum\limits_{j' \in L_i} n_i^{j'}}, \nonumber \\
& &  h_i = -\sum_{j \in L_i} p(j|i)  \log_2 p(j|i).
\end{eqnarray}
Technically, $p(j|i)$ is the probability that a vertex in  $\Vg_i$ will be at level $j$  of the tree  $T_i$. In more intuitive terms,
entropy measures the \textit{organization} of an environment. Low entropy at a vertex means that the decision tree rooted at the vertex is unbalanced, or in other words the branching factor varies widely from level to level. This unbalance can materialize both as bottlenecks or areas with too many options which may disorient a person moving through the associated areas. In some sense, while tree depth relates to path lengths, entropy relates to the uniformity of the paths: the higher the entropy, the more uniform the branching, and thus better organization. Typically higher uniformity affords easier pedestrian decision making and navigation~\cite{turner,holscher}.
\ignore{
This unbalance make affect 
 that is how easy it is for pedestrians to make planning decisions and navigate~\cite{turner}. Lower entropy at a certain vertex, means that the sequence of decisions required to reach other regions of the environment from that vertex is unbalanced \glen{Why unbalanced? The metric is a sum over each layer, there is no recursive computation such that neighbouring layers affect each other. I guess you need bottlenecks in the environment to produce more height in the tree?}\mahyar{unbalanced because lower entropy means the tree constructed at that vertex has an unbalanced probability distribution over its levels, and thus it has a lot of connections on a certain level and little connection on other levels. Adding bottlenecks increases the entropy, but the location of bottlenecks with respect to each other determines the amount of increase}. 
For example,  the number of visible vertices dramatically and discontinuously changes as one exits a small narrow room  into a large open space.  In essence, while tree depth measures  path lengths, entropy measures their uniformity. An environment with high overall entropy reduces the branching factor for pedestrian planning.
}
\reffig{figure:metrics}(c) shows the  entropy values in a heat map form over a sample environment. Notice how the top and bottom corridor have higher entropy because the decision sequences from those regions are balanced, i.e. the environment appears more organized from those regions point of view.


\reffig{figure:metrics}(d) shows the degree of visibility computed over the {\roq} (shown in heatmap) with respect to {\ror} (shown in grey). 
Notice how changing the reference from the entire environment in part (a) to just the top and bottom hallways has affected the values 
of the metric and therefore our view of the space.


For an entire visibility graph $\visibilityGraph$ with vertices $V$, our metrics are the averages of the corresponding per vertex measures:
\begin{align}
K(\visibilityGraph) &= \bar{k_i}, \;\;\; D(\visibilityGraph) = \bar{d_i}, \;\;\; H(\visibilityGraph) = \bar{h_i}.
\label{eq:expMetrics}
\end{align}

\begin{figure}[t]
	\centering
	\includegraphics[trim={6cm 1cm 6cm 3cm},clip, width=\linewidth]{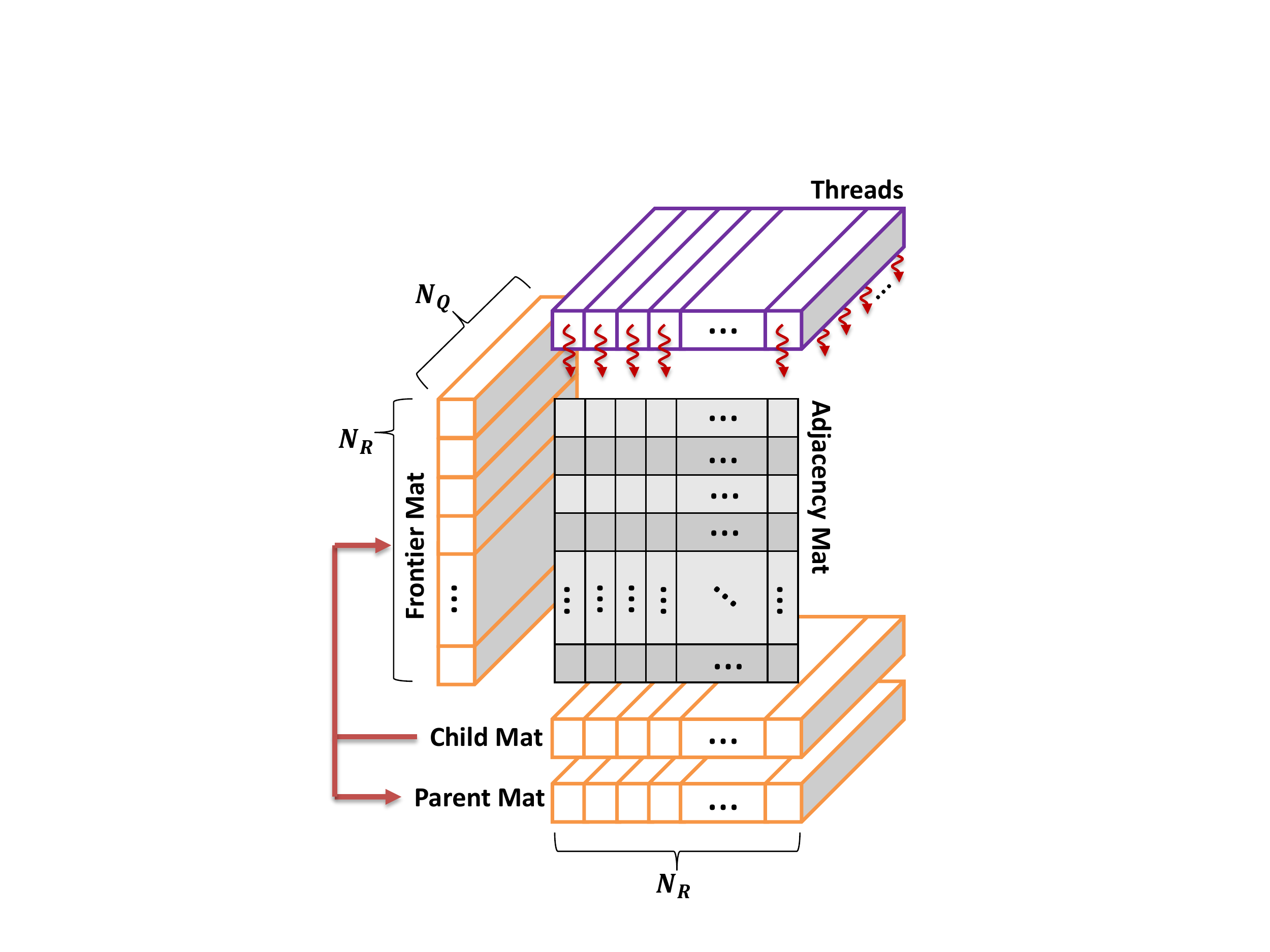}
	\caption{GPU model of the forest construction kernel. $N_R$ and $N_Q$ are the number of reference and query vertices respectively. The red wavy arrows represent individual threads.}
	\label{figure:gpuModel}
\end{figure}

\subsection{Metric Parallelization}

The aforementioned metrics can be computationally expensive. The construction of the visibility graph $\visibilityGraph(\Vg,\Eg)$ is $O(K \cdot N^2)$ where $N = |\Vg|$ and $K$ is the total number of obstacles in an environment. Furthermore, constructing the trees needed to calculate depth and entropy is of order $O(N \cdot b^d)$ where $b$ is the maximum branching factor of the $\visibilityGraph$ and $d$ is the maximum  of all the minimum depths of the Tr\'{e}maux trees constructed at different vertices. This process is $\Omega (N^2)$ which means that, at best, it is as complex as constructing the graph itself, although it is much more complex in practice. In order to mitigate this computational overhead, we off-load the construction of the visibility graph and the forest to the GPU. 

For the purposes of parallelization,  we consider that computing the metrics involves two main tasks: the  construction of  the visibility graph \visibilityGraph for the given environment layout, and the  computation of a set of $N$  trees. Although we discuss each task separately, 
our implementation runs these computations concurrently, and not in isolation.

\subsubsection{Graph Construction}
We represent the  strictly upper triangular part of the $|\Vg| \times |\Vg|$-dimensional symmetric adjacency matrix $\adjmat$ of the graph in row major fashion as a  vector, $\adjvec$ of dimension that is equal to $0.5 \times (|\Vg|-1) \times |\Vg|$. Each pair of $(i,j)$ vertices in $\Vg$ where $i \leq j$ is assigned to a thread which calculates the straight-line between the vertices and checks whether the line intersects any obstacle.  The load assignment is designed to exploit memory alignment and maximize GPU utilization.
\ignore{
We assign each position on $v((i*|\Vg|)+j) \in \adjvec$ where $ i \leq j$ to one thread, which calculates the straight-line between $i$-th and $j$-th vertices in $\Vg$, checks whether the line intersects any obstacle, and updates both $v((i \cdot |\Vg|)+j)$ and $v((j \cdot |\Vg|)+i)$ with 1 if no intersection, and 0 otherwise. Note that the assignment of threads to positions on $\adjvec$ has two properties: first, symmetric positions are not calculated twice, second the adjacent threads are assigned adjacent vertices of $\Vg$. The former property avoids duplicate computations, and the latter makes sure the threads in one warp access aligned positions from the main memory, and therefore can exploit memory alignment. Furthermore, the large pool of threads (with sufficiently random load) results in good load sharing between streaming multiprocessors (SMs) \glen{What are SMs???} and consequently high GPU utilization.
}

\subsubsection{Tree Construction}

Consider the task of performing a Breadth First Search starting at a vertex $s \in \Vg$ and branching until the whole visibility graph is traversed, i.e. all vertices are visited exactly once. We introduce three binary $|\Vg|$-dimensional vectors: (a) frontier $\front^x$ holds the elements of $\Vg$ that must be expanded in the current level $l$, (b) children $\child^l$ holds the elements of $\Vg$ that must be expanded in the next level $l+1$, and (c) parents $\parent$  holds the elements that have been already expanded. We also keep a $|\Vg|$-dimensional integer vector $\dc$  which stores the number of elements visited at each level.

\textbf{A Naive Kernel.} In a CUDA kernel, we assign each vertex, $i$ in $\Vg$ to one thread, that is each thread $i$ is responsible for one row of  the adjacency matrix corresponding to the vertex $i$. The kernel runs, level after level, until a flag is set showing that all vertices have been visited. At each level $l$, each thread $i$ first checks if its vertex $i$ is adjacent to the $j$-th vertex of the graph, second if the $j$-th element is to be expanded ($\front^l(j)$ is set), third if the $j$-th element has not been expanded ($\parent(j)$ is not set), and if so, the thread will set the $i$-th element to be expanded at the next level (set $\child^l(i) = \parent(j) = 1$). After each level, the number of 1s in $\child^l$ is stored in $\dc(l)$, then the child vectors are copied into the frontier ($\front^{l+1} = \child^l$), and the child vector is reset ($\child^{l+1}$ is initialized to zero vector). Note that other information can be stored depending on the required metrics, but in this case the number of visited vertices at each level suffices.

\textbf{Cut-Off Threads.} In the naive kernel, each thread has to check exactly $|\Vg|$ vertices of the frontier at each level, resulting in exactly $L\times|\Vg|$ operations where $L$ is the number of levels. However, each vertex in the graph only needs to be visited once. This fact can be exploited by cutting off threads that have already been visited from the start of each level, that is, stopping thread i whose assigned vertex has already been expanded ($\parent(i)$ is 1) from launching in the first place. This results in each thread having to check at most $|\Vg|$ vertices of the frontier at each level, and in practice greatly reduces the running time.

\textbf{Indexed Frontier.} So far, each thread, if not cut off, has to check all $|\Vg|$ elements of the frontier, even though many may be zero (not to be expanded)
 at many levels. However, each vertex in the graph can only be expanded once, that is, each element of the binary frontier can be 1 exactly once over all levels. Thus, the frontier is changed from a binary vector to an integer vector which stores the indices of the elements to be expanded. This indexed frontier is populated by an intermediate process that first sets all elements of frontier to $0$, then starts filling it from the start with the indices of the positions of 1s in children, instead of just copying the children vector into the frontier at the end of each level. When a zero is encountered in the frontier (i.e. $\front(j) = 0$), the kernel is terminated. This process essentially takes the burden of passing over the whole frontier from every single thread, to one single preprocessing thread. The result is that no thread will pass over the frontier more than once.

\textbf{Forest Construction.} The tree construction process must start at all vertices in the graph. Because one tree construction is completely independent of another, all the tree constructions may run concurrently. Therefore, the same kernel as before  is used but  a new dimension is introduced to all containers (this is essentially concatenating), and then we put each tree process on one row of the device grid. \reffig{figure:gpuModel} shows our final model, where the inward dimension of size $N_Q$ is the result of concatenation, each layer on this dimension belongs to a new tree. 
Thus, when the kernel runs at one level, all of the forest is expanded one level deeper on the device. This allows for having a very large pool of threads, and therefore maximizes the load sharing and consequently the GPU utilization.

\subsection{Penalty Metrics}
\label{subsubsection:penaltyFunctions}

In the context of architectural optimization, a user may wish to impose a number of conditions on certain design  elements, such as  a minimum amount of open space in passages, aesthetic relationships, or building codes. These conditions can be modelled with penalty functions which are treated as soft constraints by the optimizer. A few practical examples of this are described below.


\paragraph*{Clearance} A measure of open space between architectural elements, clearance is computed as the aggregate Minkowski sum of each wall and a disk $D_r$ of radius \valueWithUnits{$0.5$}{m}, which approximates the minimum width of a hallway.
The Minkowski sum between a polygon and a disk dilates the polygon, effectively adding a buffer area around an obstacle or wall for comfortable passage.
\begin{equation}
	\clearance(\architecturalGraph) = \sum\limits_{\wall_{i},\wall_{j} \in \architecturalGraph}^{} \functionArea{(\wall_{i} \oplus D_r) \bigcap (\wall_{j} \oplus D_r)},
\end{equation}
\noindent
where $\functionArea{\cdot}$ computes the area, and $\oplus$ denotes the Minkowski sum between two polygons and $\bigcap$ is the geometric intersection of the two Minkowski sums.
Adjoining walls are excluded from this computation.
 The associated  penalty function is $g_{\clearance} = \clearance(\architecturalGraph)^2$.

\paragraph*{Total Wall Length} This is the sum the of the wall lengths of the \emph{new} environment ($\architecturalGraph^{n}$) with respect to the original environment ($\architecturalGraph^{o}$)
$ \walls(\architecturalGraph^{n}, \architecturalGraph^{o}) = |\sumWallLengths(\architecturalGraph^{n}) - \sumWallLengths(\architecturalGraph^{o})|$, where $\sumWallLengths(\cdot)$ computes a sum over the length of every edge/wall in the graph.
This penalty function is used to constrain the repositioning of elements to not reduce or increase the quantity of wall surface area in an environment. This particular penalty method is appropriate for museums and art galleries where there is a desired amount of wall surface area needed to display an art collection.

\section{Optimization Formulation} 
\label{section-optimization}

The user defined parametrization of the architectural graph, $\architecturalGraph{(\vectorParameters)}$ in \refsec{section-environment-parameterization}, defines the design  domain $\parameterSpace$, with bounds  in \functionConstraints{\vectorParameters}. In this section, we describe the key elements of our objective function.

\subsection{Diversity Objective}
Unlike a typical optimization that produces a single design solution $\vectorParameters^*$,  the \DOME system must produce a set of optimal solutions $\diversityMembers^{*} = \{\vectorParameters_1, ..., \vectorParameters_n \}$ whose members are sufficiently diverse from each other. Therefore, measures of diversity are introduce and maximized. There are a number of techniques to accomplish this, each with their own advantages and disadvantages. For efficiency, instead of augmenting the parameter vector \vectorParameters with additional elements for each member of the diversity set, a round robin technique is used, where one member in $\diversityMembers$ is optimized at a time while keeping the parameters of the other members constant.
 
In practise, enforcing diversity naively can lead to a clustering of solutions~\cite{6781622}. To avoid clustering, we impose a minimum distance between members of $\diversityMembers$. 
Our diversity metric is as follows:
\begin{eqnarray}
		\functionDiversity{\diversityMember, \diversityMembers} = 
		\diversityMemberWeight  (\sum\limits_{j \in \diversityMembers} \memberDistance{ \vectorParameters_j}{\diversityMember}) 
		- \diversityMemberMinWeight \memberMinDistance{\diversityMember}{\diversityMembers},  \label{eq:diversity-metric}
\end{eqnarray}
\begin{eqnarray}
\label{eqn:div-mem-dist}
		\memberMinDistance{\diversityMember}{\diversityMembers} = 
		(\min (0,
		\min_{j \in \diversityMembers, j \neq m} (  \memberDistance{\vectorParameters_j}{\diversityMember}) - \minDistanceThreshold  
		))^{2},
\end{eqnarray}
\noindent
where $\memberDistance{\cdot}{\cdot}$ normalizes its arguments over the parameter constraints and computes the Euclidean distance and 
$\memberMinDistance{\diversityMember}{\diversityMembers}$ is the minimum distance between $\diversityMember$ and all other members in $\diversityMembers$.
Equation~\ref{eqn:div-mem-dist} ensures that diverse members don't cluster by adding a cost when the closest neighbour is less than $d_{min}$ away.
The terms $d_{min}$, \diversityMemberWeight and \diversityMemberMinWeight are experimentally determined hyper-parameters that control the influence of the diversity term.

\subsection{Optimization formulation}
\label{sec:optimizationFormulation}
For a set of optimal solutions, $\diversityMembers$,   the objective vector is aggregated over the entire set.
This results in the following multi-objective optimization problem:
\begin{align}
& \diversityMembers^{*}= \underset{\diversityMembers \subset \parameterSpace }{\argmax}   
\sum\limits_{\vectorParameters \in \diversityMembers}
\langle
		-\penaltyFun{\vectorParameters},
	 	\objectiveDegree{\vectorParameters}, -\objectiveDepth{\vectorParameters} , \objectiveEntropy{\vectorParameters},
			 \functionDiversity{\vectorParameters, \diversityMembers}
	\rangle,
\label{eq:ObjectiveMaximization} \\
&\;\;\;\;\text{s.t. }  \functionConstraints{\vectorParameters}  \nonumber
\end{align}
where a  $\functionConstraints{\vectorParameters}$ are the parameter bounds specified by the user and
$\penaltyFun{\vectorParameters}$ is the penalty function described in \refsec{subsubsection:penaltyFunctions}.
 Solving this problem produces a set of solutions with maximum spatial objectives  in combination with  minimum penalties, and maximum diversity.

The next section discusses our solution to the above optimization.

\section{Multi-Objective Optimization}
\label{section-multi-objective}

There exists several methods that can be used to perform multi-objective optimization~\cite{Marler2004}. 
Scalarized multi-objective optimization combines a vector of objectives with a vector of weights, however, finding a good vector of weights can be challenging, especially when the  objectives are of largely different scales, as they are in our case. Pareto Front-based approaches produce a collection of parameter settings that are optimal trade-offs between the objectives~\cite{wagner2007pareto}. However, they tend to be computationally expensive, and it is unclear how they would handle the diversity term. Hierarchical methods optimize one objective at a time, in order, in a fashion similar to coordinate descent. Each optimized objective becomes part of the objective function in the form of a soft constraint for the optimization of the next objective.
A hierarchical approach appears to be the most practical approach for this problem space.  
It allows for more practical and intuitive control of the trade-off between optimality and  diversity, in the form of a lower bound with respect to the optimal solutions.
See the \appendixx for more details on the multi-objective optimization methods.

Similar optimization problems have been solved in the graphics literature with a combination of Simulated Annealing and the Metropolis-Hastings algorithm~\cite{craigyu2011furniture,Merrell:2011:IFL:2010324.1964982}. The convergence rates of these methods can make them prohibitive for interactive systems. 
This is shown in the engineering literature were Covariance Matrix Adaptation (CMA)~\cite{542381} is more popular for many design reasons~\cite{NGUYEN20141043}, details are described in the \appendixx.
To address these design considerations, a hierarchical optimization solution based on CMA is used, which can manage the same number of parameters with faster convergence rates.

Our optimization approach  aims to produce a set of diverse, near optimal solutions and  is best described with two separate algorithmic steps. 

\subsection{Hierarchical Multi-Objective Optimization}

Instead of optimizing a weighted combination of objectives, in this case the objectives defined in \refeq{eq:ObjectiveMaximization}, the components are optimized as separate objectives. 
For each objective we specify an order (ranking) and a desired minimum improvement threshold \threshold.
The desired \threshold is a ratio between [0, 1] where \threshold dictates a threshold between the default objective value and optimal objective value.
For example if an objective is ranked first with a threshold of $0.7$, then the optimization process will optimize it first, ignoring other metrics. 
After converging to an optimal value, a constraint is added to the second objective that imposes a penalty if the first objective falls below $70\%$ of its optimal value. The process  repeats for all  objectives in the order specified with the near-optimality margins given.

To incorporate  an objective as a constraint during the hierarchical optimization process a \textit{threshold function} is used.
These functions are constant or simply zero when the input is within a given range, and rapidly increase when the input is outside this range.
\begin{equation}
	\functionThreshold{l}{u}{x} = 
	\begin{cases} 
	   (l-x)^{2} & \text{if } x < l \\
	   (x-u)^{2} & \text{if } x > u \\
	   0 & otherwise\\
	 \end{cases}
\end{equation}
For a set of threshold functions \thresholdFunctions the total threshold violation cost is
\begin{equation}
	\hierachicalPenaltyFun{\vectorParameters, \thresholdFunctions} = \sum\limits_{t \in \thresholdFunctions}t(\vectorParameters).
\end{equation}

Algorithm~\ref{alg:multi-hierarchical} describes the hierarchical multi-objective optimization method over the objective vector 
\begin{equation}
\objectiveVector = \langle
- \penaltyFun{\vectorParameters_{0}}, \objectiveDegree{\vectorParameters},  - \objectiveDepth{\vectorParameters}, \objectiveEntropy{\vectorParameters},
\functionDiversity{\vectorParameters_{0}, \diversityMembers}
 \rangle.
 \end{equation}
 
It is important to have the diversity metric be the last objective in this vector.
The diversity metric creates and uses a set of diverse members, the other metrics operate over a single member.
Also, the penalty function should be first, as it is necessary to constrain the optimization of the following metrics.
For each objective, a CMA-based optimization is performed
(lines $9$-$15$). 
At the end of an individual objective optimization, a threshold constraint is created and added to the vector of threshold constraints \thresholdFunctions (lines $16$-$17$). 
At the end of the main loop, the optimal parameter vectors are captured within the thresholding function vector, \thresholdFunctions. 
The last objective, diversity, is optimized using $\function{DivOpt}()$  in Algorithm~\ref{alg:roundRobin}, which searches for a diverse set of near optimal solutions given the set threshold functions constructed. Note that the other objectives are now represented as penalties through the threshold functions.


\begin{algorithm}[tb]
\caption{Hierarchical Multi-Objective Optimization}
\label{alg:multi-hierarchical}
\begin{algorithmic}[1]
\STATE{\textbf{Input:} Number of diversity members, $n$}
\STATE{\textbf{Input:} Vector of objective thresholds, $\mathbf{z} = \langle z_{i} \rangle $}
\STATE{\textbf{Input:} Initial parameter vector, $\vectorParameters_{0}$}
\STATE{\textbf{Input:} Vector of objective functions, $\objectiveVector = \langle f_{i} \rangle $}
\STATE{\textbf{Input:} Variance, $\sigma$, Sample size, $\sampleSize$}
\STATE{ $\thresholdFunctions \leftarrow \{\} $} 
\FOR{ \textbf{each} $f_{i} \in \objectiveVector$}
	\STATE{$\diversityMembersCovar \leftarrow \sigma^{2} I, \vectorParameters_{opt} \leftarrow \vectorParameters_{0}$}
	\WHILE{$\neg$ \function{Terminate}()}
		\FOR{ \textbf{each} $j \in (1, \sampleSize)$}
			\STATE{$\vectorParameters_{j} \leftarrow \ensuremath{\mathcal{N}}(\vectorParameters_{opt}, \diversityMembersCovar)$}
			\STATE{$y_{j} \leftarrow f_{i}(\vectorParameters_{j}) - 	\hierachicalPenaltyFun{\vectorParameters_{j}, \thresholdFunctions}$}
		\ENDFOR
		\STATE{$\langle \vectorParameters_{opt}, \diversityMembersCovar \rangle \leftarrow \function{Update}(\vectorParameters_{opt}, \diversityMembersCovar, \langle y_{j} \rangle , \langle \vectorParameters_{j} \rangle)$ }
	\ENDWHILE
	\STATE{$l \leftarrow f_{i}(\vectorParameters_{0})+ z_{i}\cdot(f_{i}(\vectorParameters_{opt}) -f_{i}(\vectorParameters_{0})) $}
	\STATE{$ \thresholdFunctions \leftarrow \thresholdFunctions \cup \functionThreshold{l}{\infty}{\cdot}$}
\ENDFOR
\STATE{\diversityMembers$ \leftarrow$ \function{DivOpt}$(n,\hierachicalPenaltyFun{\cdot, \thresholdFunctions}, \vectorParameters_{0})$}
\RETURN{\diversityMembers}
\end{algorithmic}
\end{algorithm}

The next section describes the final step of our hierarchical optimization - the diversity objective.

\subsection{Diversity Optimization}


A round-robin method is used to select and optimize each diversity member $m$ one at a time, see Algorithm~\ref{alg:roundRobin}.
Each member is initialized using $\vectorParameters_{0}$ and progressively diverge from each other as the optimization unfolds.
In each round, a single member is selected from \diversityMembers and candidate parameters are sampled using CMA~\cite{542381}(lines $7$-$12$). 
A simple in-order method is used to select the next member in each round. 
More complex, or random, selections may be employed, but we empirically found this strategy to work well.
In lines $10$-$11$, the objective values for those candidates are calculated.
In line $12$, the structures in CMA that influence the optimization evolution are updated.
The termination condition 
is dependent on the optimization progress with respect to the improvements made on $\vectorParameters_{opt}$ and $\hat{y}^{*}$ and the maximum number of function evaluations, which are parameters of CMA~\cite{542381}.



\begin{algorithm}[tb]
\caption{Diversity Optimization:}
\label{alg:roundRobin}
\begin{algorithmic}[1]
\STATE{\textbf{function} \function{DivOpt}($n, f, \vectorParameters_{0}$)}
\STATE{\textbf{Input}: Number of diversity members, $n$}
\STATE{\textbf{Input}: Objective function, $f$}
\STATE{\textbf{Input:} Initial parameters, $\vectorParameters_{0}$}
\STATE{\textbf{Given:} Variance, $\sigma$, Sample size $\sampleSize$}
\FOR{$ i \in (1 \ldots n)$}
	\STATE{$~~ \diversityMembersCovar_{i} \leftarrow \sigma^{2}I, \diversityMembers_{i} \leftarrow \vectorParameters_{0}$}
\ENDFOR
\WHILE{$\neg$  \function{Terminate}() }
	\STATE{Choose $m$ from $P(1 \leq x \leq n)$}
	\FOR{\textbf{each} $i \in (1, \sampleSize)$}
		\STATE{$\vectorParameters_i \leftarrow \ensuremath{\mathcal{N}}(\diversityMembers_{m}, \diversityMembersCovar_{m})$}
		\STATE{$ y_{i} \leftarrow \functionDiversity{\vectorParameters_{i}, \diversityMembers} - f(\vectorParameters_{i})$}
	\ENDFOR
	\STATE{$\langle \diversityMembers_{m}, \diversityMembersCovar_{m} \rangle \leftarrow \function{Update}(\diversityMembers_{m}, \diversityMembersCovar_{m}, \langle y_{i} \rangle, \langle \vectorParameters_{i} \rangle)$ }
\ENDWHILE
\RETURN \diversityMembers
\end{algorithmic}
\end{algorithm}






\subsection{Diversity Set}

\begin{figure*}
	\begin{minipage}[b]{0.496\linewidth} 
		\centering
			\subfloat[\num{0.436603} Default]{\includegraphics[width=0.35\linewidth]{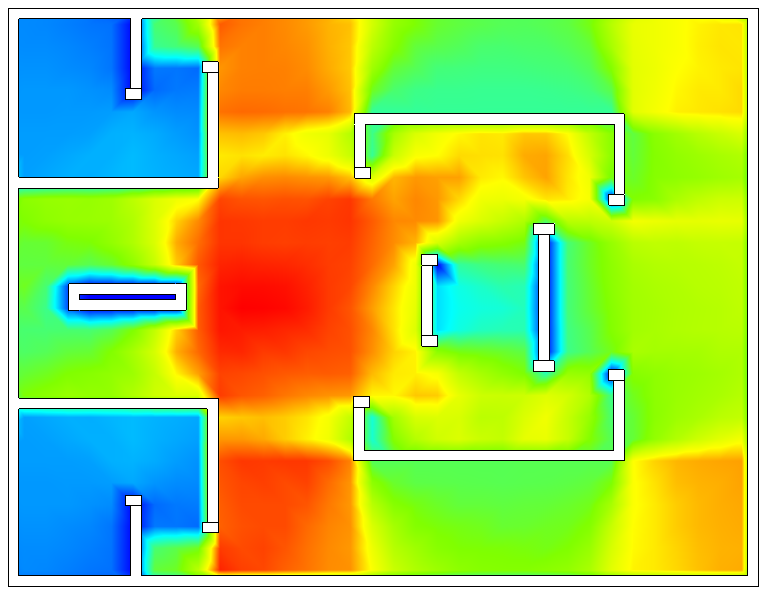}} 
			\subfloat[\num{1.0325}]{\includegraphics[width=0.35\linewidth]{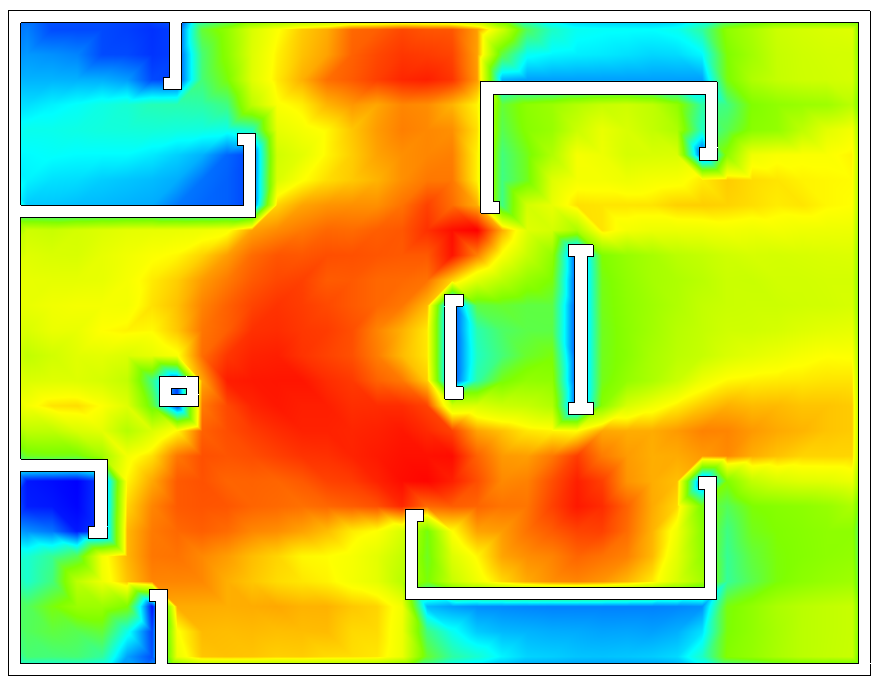}} 
			\subfloat[\num{1.283812}]{\includegraphics[width=0.35\linewidth]{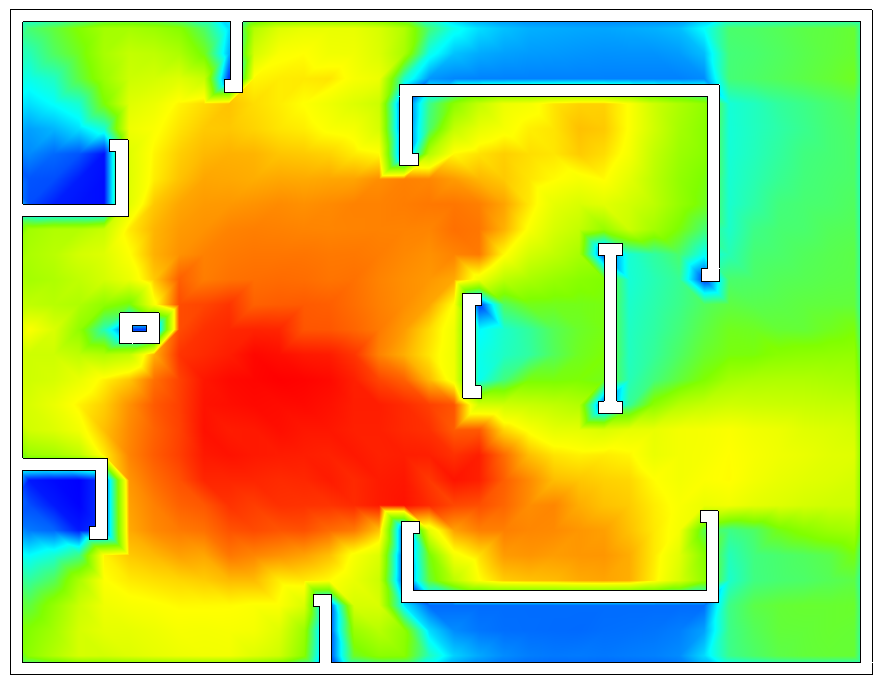}}  \\
			 \subfloat[\num{1.094401}]{\includegraphics[width=0.35\linewidth]{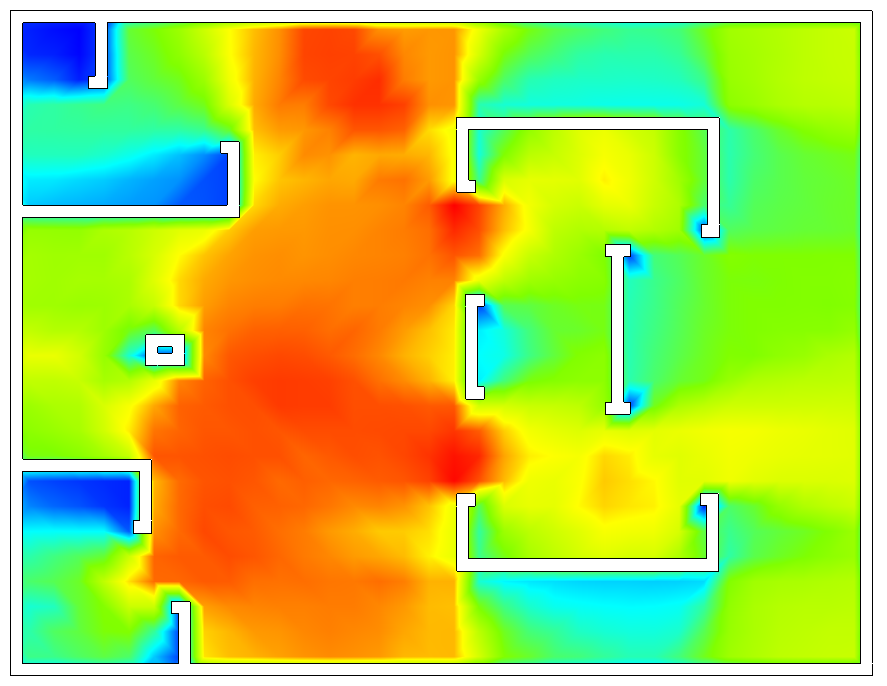}} 
			\subfloat[\num{1.215894}]{\includegraphics[width=0.35\linewidth]{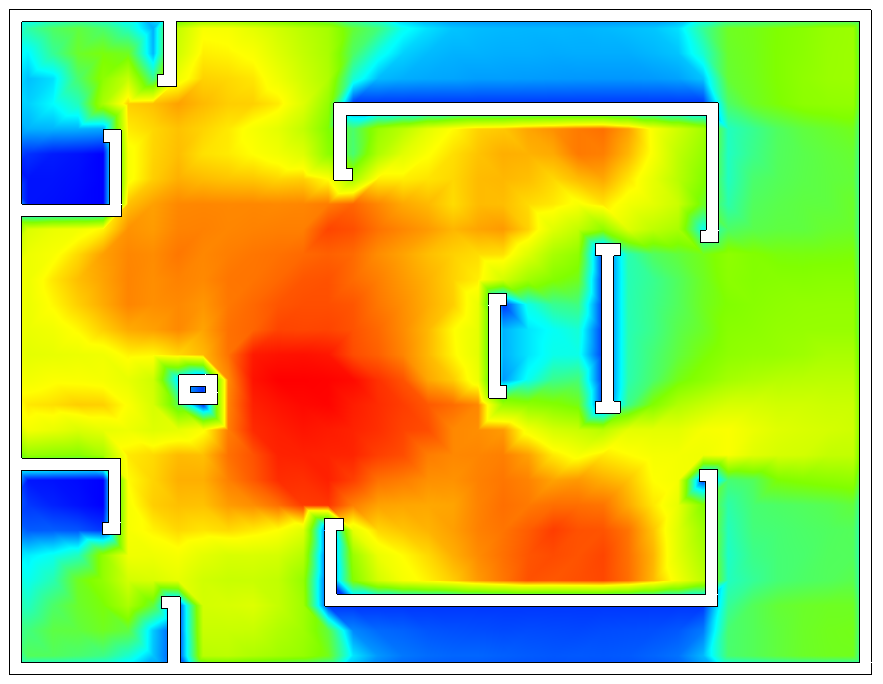}} 
			\subfloat[\num{1.459438}]{\includegraphics[width=0.35\linewidth]{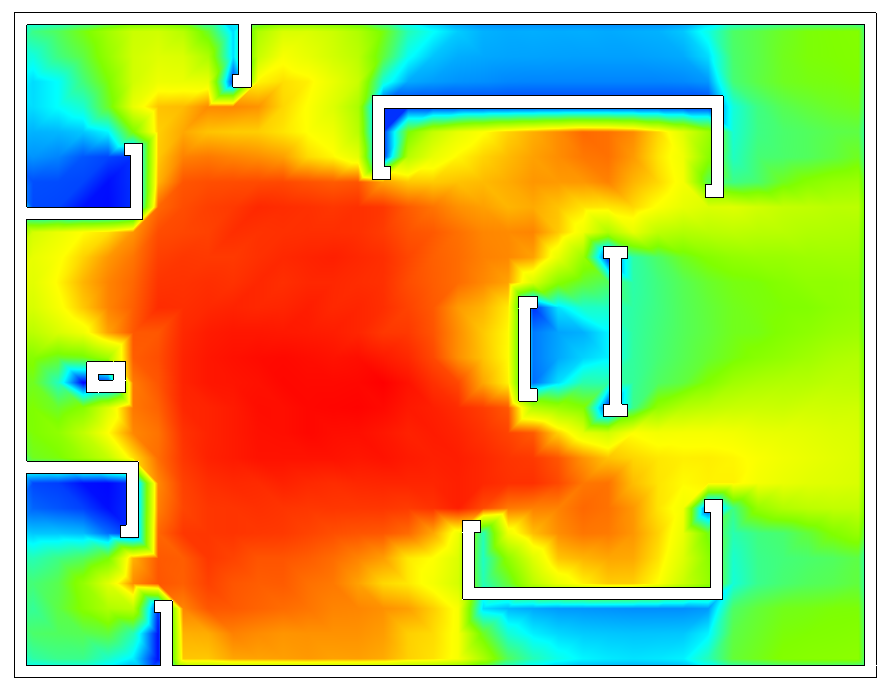}}  
	\end{minipage}
	\begin{minipage}[b]{0.496\linewidth}  
	\centering 
		\subfloat[]{\includegraphics[width=0.8\linewidth]{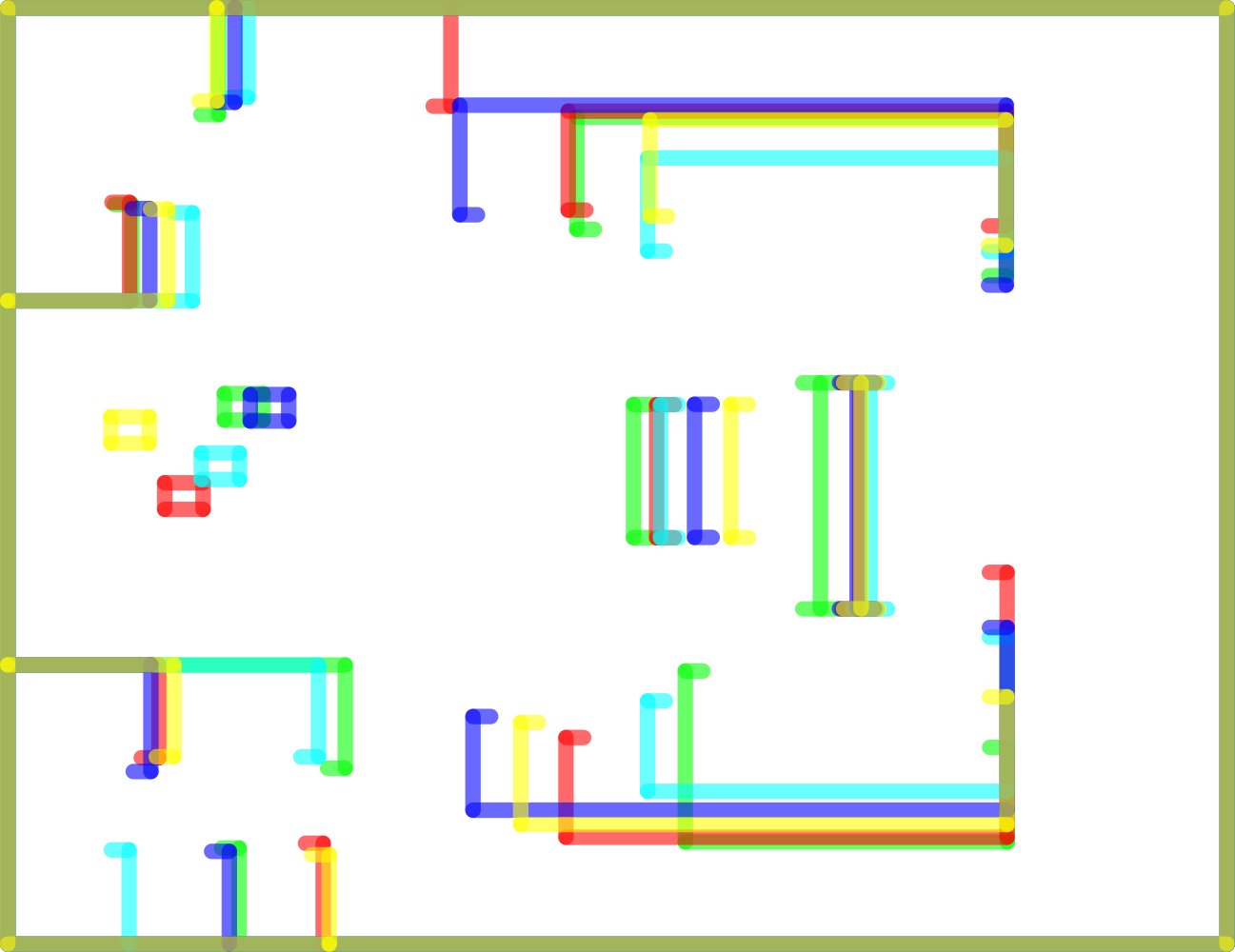}}
	\end{minipage}		
	\caption{\label{figure:diversity-comparison} A diversity set example. The solution with the highest metric value (f), although optimal, can be considered less aesthetically pleasing. It is interesting that all solutions effectively converted the wall, at left centre, into a pillar, which indicates that the layout could potentially be improved by completely removing the wall. (g) For comparison all solutions are shown superimposed.}
\end{figure*}

A key advantage of our approach is the production of a diverse set of solutions. \reffig{figure:diversity-comparison} shows an example set:  (a) is the default layout,  (b-f) are the five members of the diversity set,  and (g) is a superimposition of all solutions to better illustrate their differences.
 

\section{Results}
\label{sec:results}

In this section we discuss the capabilities and limitations of \DOME. First, we explore performance issues that are important for the practical use of the system. Then we present examples that  clarify aspects of the system, and demonstrate its effectiveness. Note that it can be difficult to convey the user-in-the-loop nature of the system with static pictures alone~\cite{Usman:2017:PES:3136457.3136458}. For a more effective demonstration we refer the reader to the accompanying video.  

\subsection{Performance Analysis}
\label{subection-performance}

\begin{figure}[t]
	\centering
	\includegraphics[trim={3.5cm 8.5cm 4.5cm 8.5cm},clip, width=\linewidth]{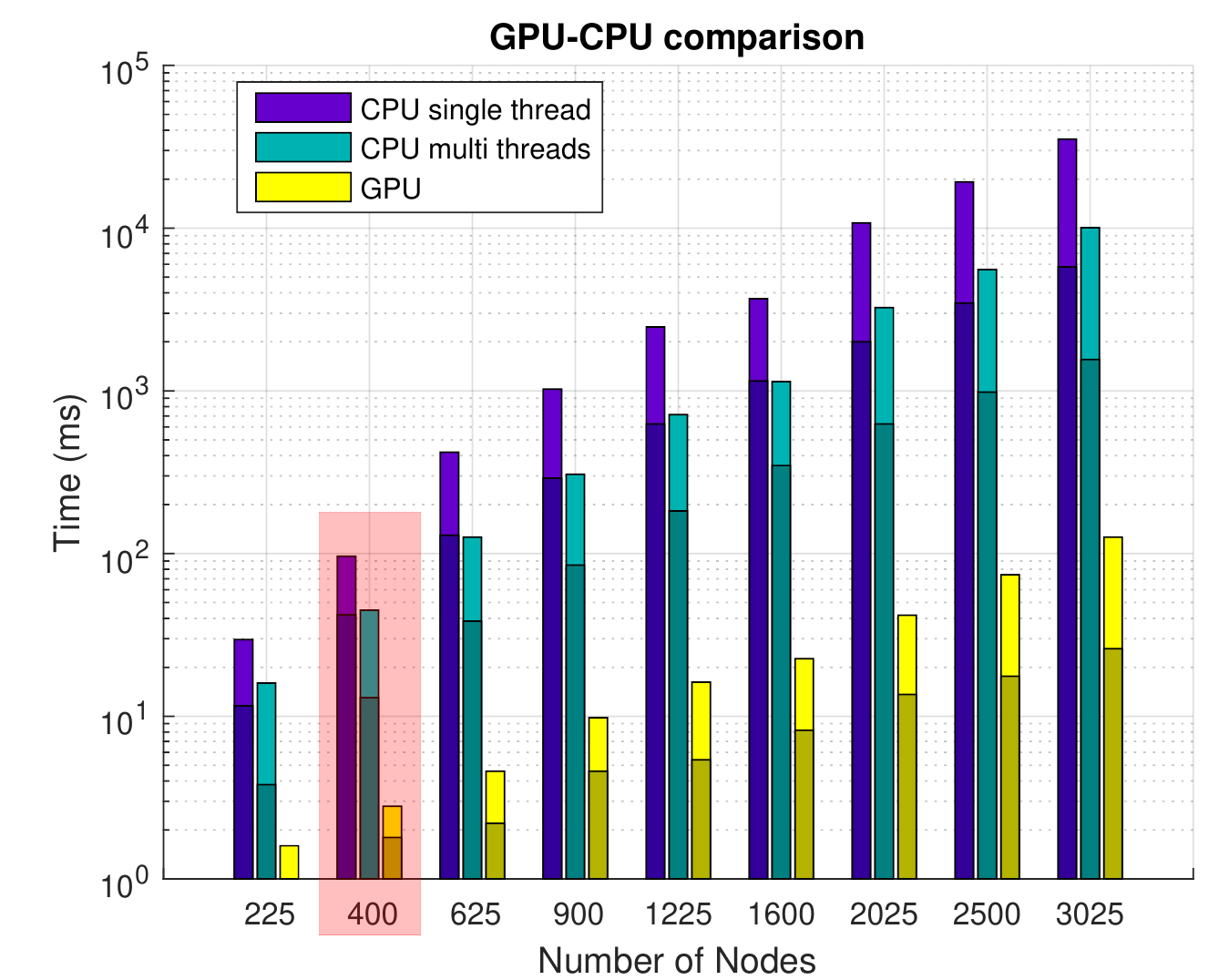}
	\caption{Spatial analysis framework performance analysis of CPU and GPU implementations. The bars show the total time to calculate the three objectives using the corresponding hardware, with the typical use case highlighted in red. Each bar is also divided into darker and lighter shades to depict the time for graph generation and forest construction respectively. Time is in base $10$ logarithmic scale. 
	}
	\label{figure:cpugpu}
\end{figure}

\noindent \textbf{Spatial Metrics.} \reffig{figure:cpugpu} illustrates the comparative performance of our spatial analysis framework (\refsec{section-environment-analysis}) using single-threaded CPU, multi-threaded CPU, and GPU implementations. It is evident that the GPU implementation completes the computation much faster. For example, on  a grid of $900$ vertices, over an effective area of \valueWithUnits{$3600$}{m^2} using a $0.5$ cell per meter granularity, the GPU takes \valueWithUnits{$10$}{ms} compared to the CPU's \valueWithUnits{$300$}{ms} ($4$-threads) on average to generate the visibility graph, construct the corresponding forest, and calculate the objective. This advantage increases as the number of vertices in the graph increases, with an order-of magnitude speedup. This test compares Intel Xenon at 3.5 GHz with GeForce GTX $1070$.  Note that certain operations in our calculations (e.g. entropy calculations) are especially amenable to GPU parallelization. Moreover, the reported times include the initialization process for each granularity which is executed once per optimization; therefore the actual average times over objective calls would be considerably lower. 
In our current implementation, the spatial objective are computed concurrently on the GPU and a weighted sum of the spatial metrics may be used for efficiency.
The performance analyses reported here encompass the entire spatial analysis pipeline averaged over $5$ runs.


\noindent \textbf{GPU Memory Complexity.} The GPU memory required for the objective calculation on the GPU is of $O(N^2)$, more strictly it grows with $8N^2$, where $N$ is the total number of vertices included in the graph.
All example environments in~\reftab{table:optimization-times} take up less than \valueWithUnits{$20$}{MB} memory. 
Note that the provided memory complexity is for one unified run of optimization, a much larger environment can be processed in subsets (chunks) of vertices.

\begin{figure}[t]
	\centering
	\includegraphics[trim={4cm 8.5cm 4.5cm 8.5cm},clip, width=\linewidth]{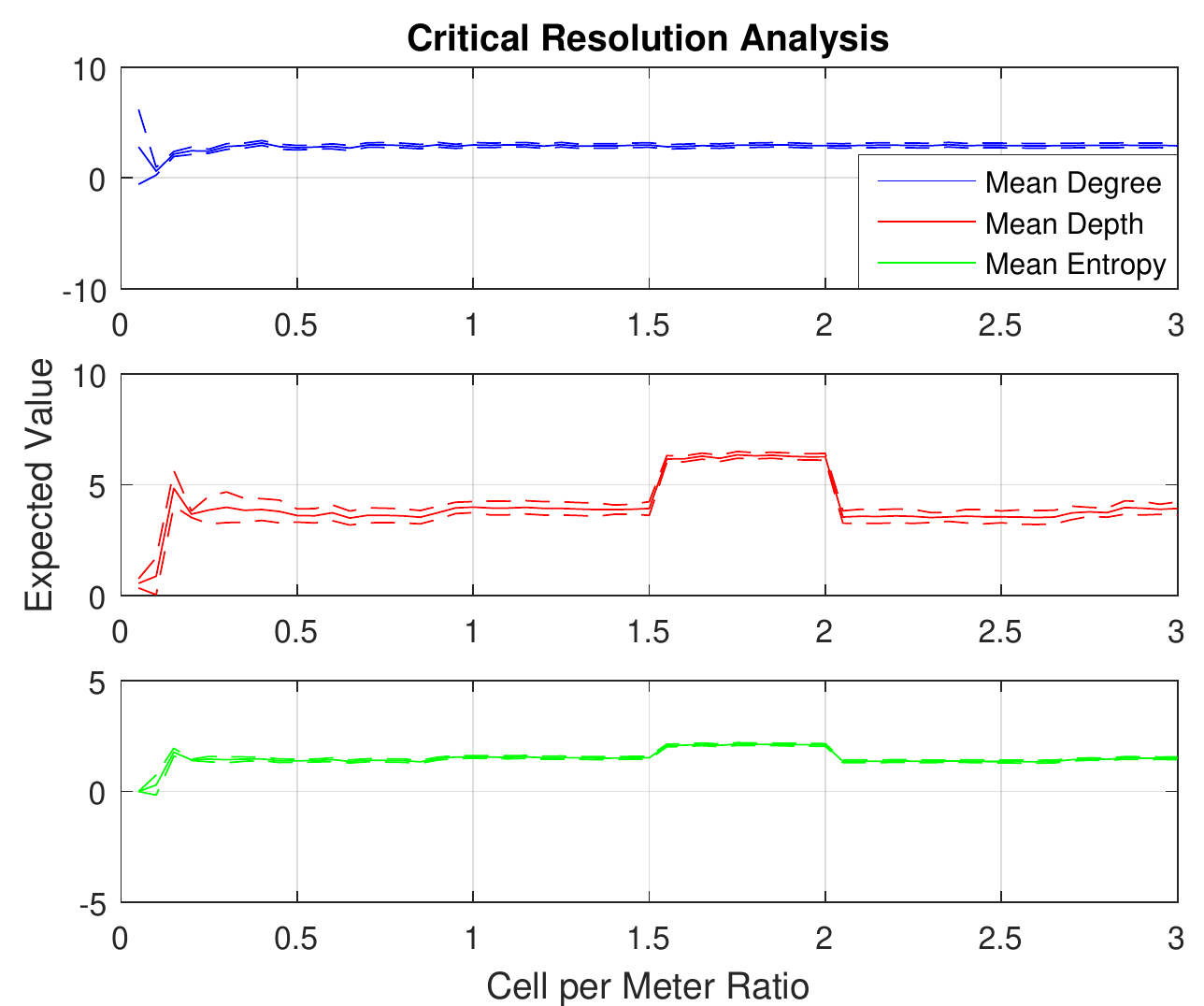}
	\caption{Sensitivity of metric values to visibility graph resolution. The vertical axis is the metric average over $10$ randomly sampled environments. The standard deviation is also provided for each metric.}
	\label{figure:metrics_sens}
\end{figure}

\noindent \textbf{Critical Resolution.} 
The grid resolution determines the number of vertices in the visibility graph,
to identify the minimum resolution needed we perform a sensitivity analysis over the granularity. Each metric is computed over a range of grid resolutions, aggregated over multiple environment layouts. Here, resolution is represented as the number of cells per meter ratio, for example resolution $0.5$ means that in each dimension one cell covers $2$ square meters.  The study results are illustrated in~\reffig{figure:metrics_sens}. These diagrams show that the metrics do not substantially change after a certain sampling frequency, suggesting that a critical value can be identified. The two jumps in the depth and entropy diagrams are caused by discovering new bottlenecks after a certain increase in resolution, which are discarded at higher sampling frequency. For the remaining experiments reported in this paper, we have used a sampling resolution of $0.5$ cells/meter$^2$. 

\begin{figure}[h]
	\centering
	\setlength{\tabcolsep}{1pt}
	\begin{tabular}{c c c}
		Default Floor Plan &  Manual & \DOME-assisted \\
		\includegraphics[ width=0.325\linewidth]{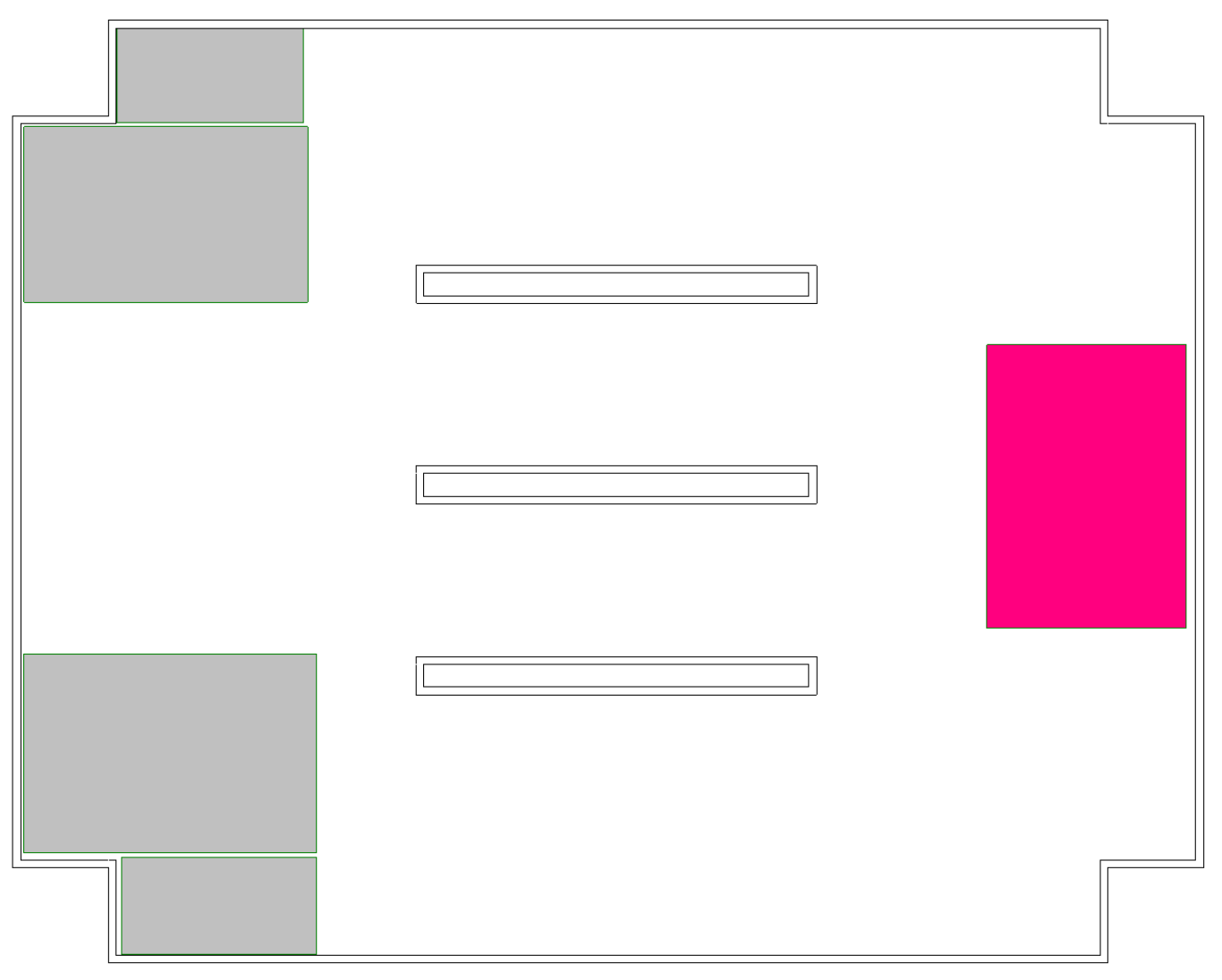} &
		\includegraphics[ width=0.325\linewidth]{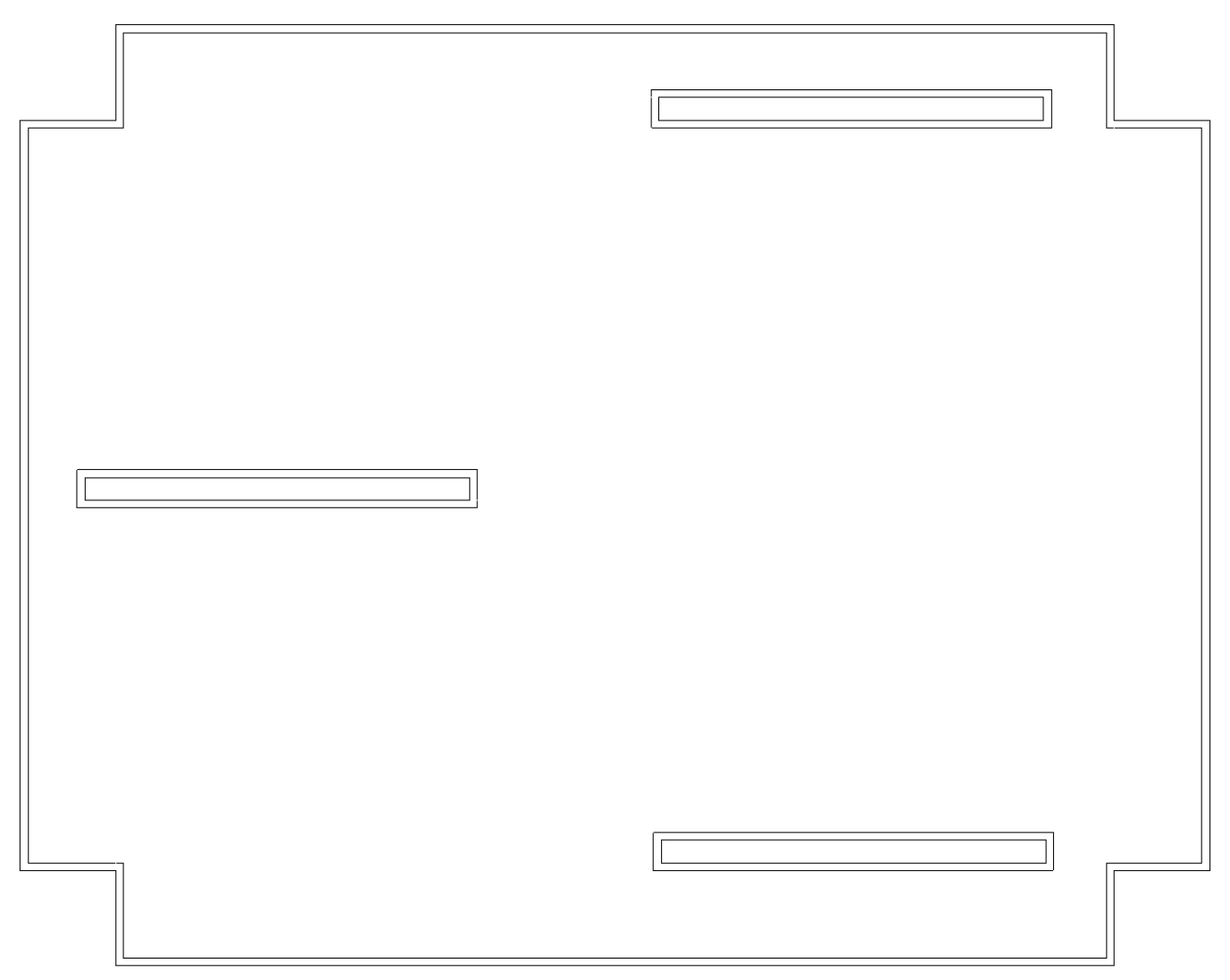} &
		\includegraphics[ width=0.325\linewidth]{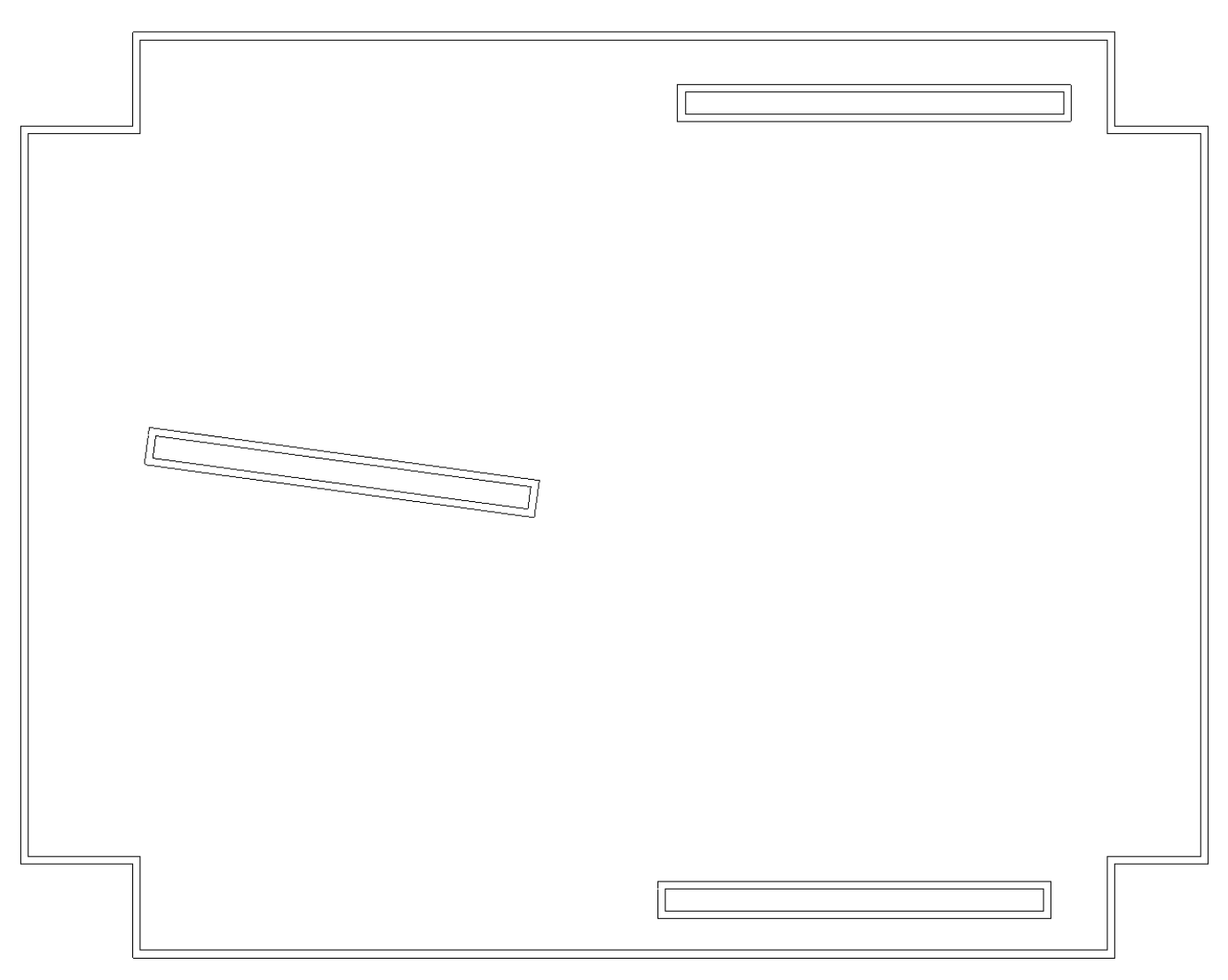} \\
		
		& $f=\num{-52.5488}$ & $f=\num{-51.3088}$ \\
		
		\includegraphics[ width=0.325\linewidth]{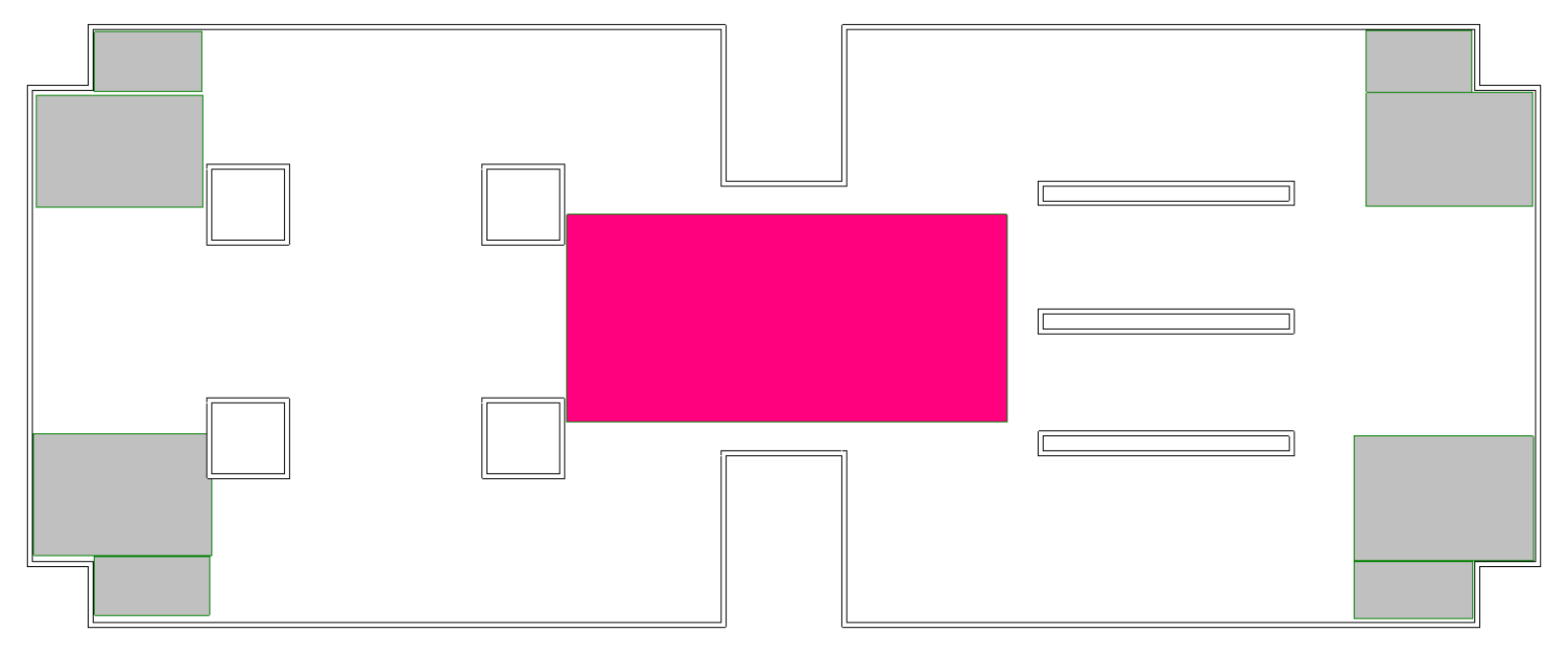} &
		\includegraphics[ width=0.325\linewidth]{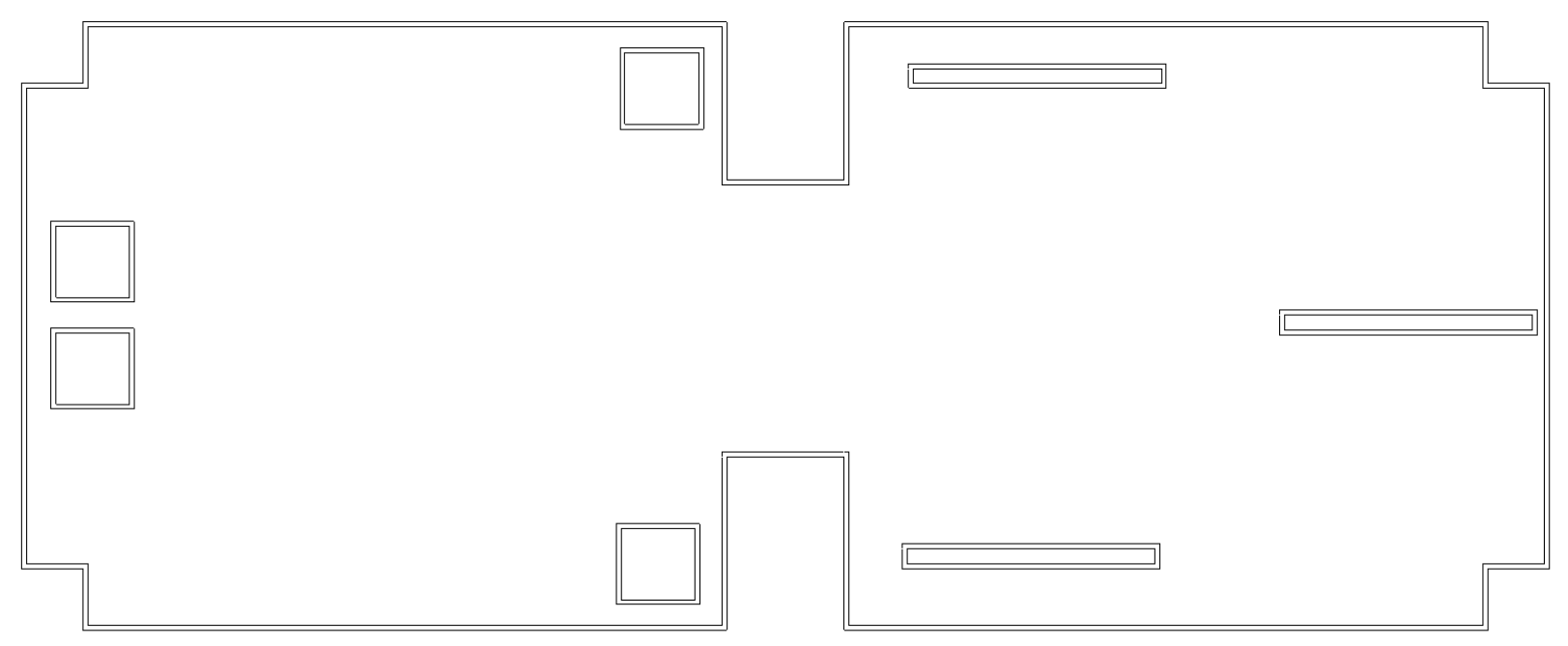} &
		\includegraphics[ width=0.325\linewidth]{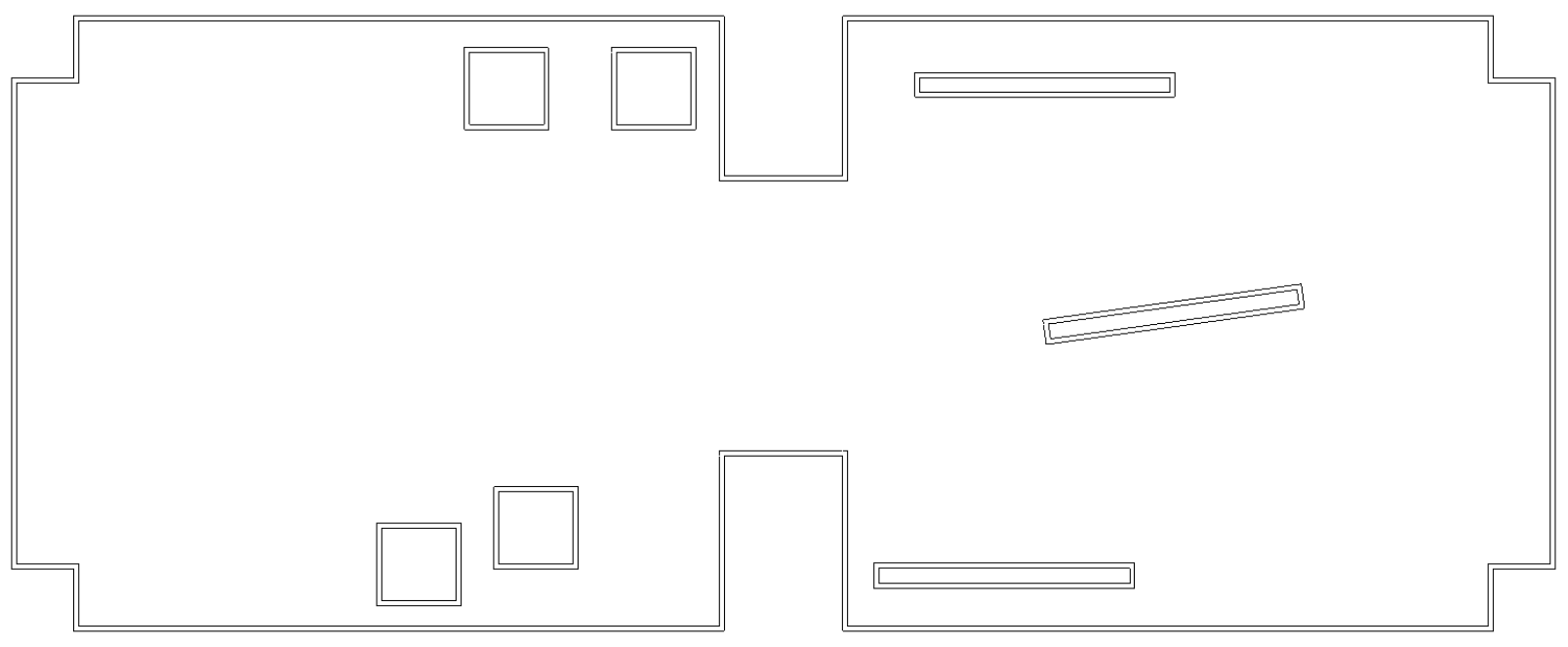} \\

		& $f=\num{-79.9943}$ & $f=\num{-49.7661}$ \\
	\end{tabular}
	\caption{\label{fig:user-designs}Selected user study layouts generated by participants. The top row shows the simple layout, while the  bottom row shows the complex one. Pink regions form the \roq, and grey regions form the \ror. Note the difference between manual design and \DOME assisted results. The corresponding objective values are reported under each layout.}
\end{figure}

\noindent \textbf{Diversity Optimization.} \reftab{table:optimization-times} provides the computation times of diversity optimization for three exemplar environments. These include the environment used in the user study~\reffig{fig:user-designs}, as well as the art gallery and museum illustrated in~\reffig{figure:art-gallery-reduced} and~\reffig{figure:metrics} respectively. For moderately complex designs (hundreds of vertices in the visibility graph), the results show that \DOME maintains interactive running times, taking a few seconds to compute diverse solution candidates. For most practical purposes, we anticipate that users will define optimization problems in focused environment regions, such as a particular room in a larger building, by specifying appropriately sized query and reference regions. For more complex designs with tens of thousands of vertices, such as~\reffig{figure:subway-station-example}, optimizations take close to one hour to complete. While this prevents an interactive design session, the results of our framework can still provide valuable design suggestions and feedback to the designer. 

\begin{table*}
\centering
\begin{tabular}{| c | p{7mm} | p{7mm} | p{7mm} | p{13mm} | p{13mm} | p{7mm} | p{7mm} | p{7mm} | p{9mm} | p{9mm} | p{8mm} | p{8mm} |}
\toprule
Environments & Ref Vertices & Query Vertices & Total Vertices & Effective Size ($m^2$) & Objective Calls (c) & \multicolumn{2}{l|}{Graph (ms/c)} &  \multicolumn{2}{l|}{Forest (ms/c)} & Penalty (ms/c) & \multicolumn{2}{l|}{Total Time (s)}\\ \cline{7-13}
&&&&& & CPU & GPU & CPU & GPU & CPU & CPU & GPU \\ \toprule
Simple Room  & 361 & 25 & 361 & 1444 & 692 & 3.9 & 0.88 & 0.7 & 0.56 & 0.08 & 4.35 & \textbf{2.25} \\ \hline
Large Room  & 1369 & 81 & 1369 & 5476 & 692 & 61.08 & 1.68 & 19.29 & 1.76 & 0.12 & 57.26 & \textbf{3.62} \\ \hline
Museum & 588 & 208 & 588 & 2352 & 772 & 23.37 & 1.04 & 15.98 & 1.91 & 1.06 & 34.04 & \textbf{5.06} \\ \hline
Art Gallery  & 487 & 438 & 915 & 3660 & 732 & 67.59 & 1.35 & 22.73 & 1.93 & 5.36 & 72.69 & \textbf{7.41} \\ \bottomrule 
\end{tabular}
\caption{\label{table:optimization-times} Diversity optimization running times. These results were computed using GeForce GTX $1070$ and Intel Xenon at 3.5 GHz on a range of environments from simple and small to large and complex. Note that while the system is not real-time, it is sufficiently fast for interactive use. 
}
\end{table*}

\noindent \textbf{Optimization Convergence}. 
The convergence or stopping conditions of the optimization algorithm has a dramatic impact on the computational performance as well as the quality of the results. The default termination conditions are overly conservative for this application, leading to long optimization times with negligible effects on quality after the first few iterations~\cite{542381}. 
The termination conditions are adjusted to return results after the optimization has converged to $\sim95\%$ of optimal.
This leads to significant performance gains. 

\subsection{Examples} 


We demonstrate the application of \DOME on a variety of real environments including a portion of the NYC Penn Station subway, the Metropolitan Museum of Art, and a layout based on the Washington Art Museum.
\DOME can also be applied in other interesting ways, for example, to increase or reduce the complexity of a maze-like environments. 

\begin{figure}[htb]
	\centering
	\begin{tabular}{c}
		\includegraphics[width=0.9\linewidth]{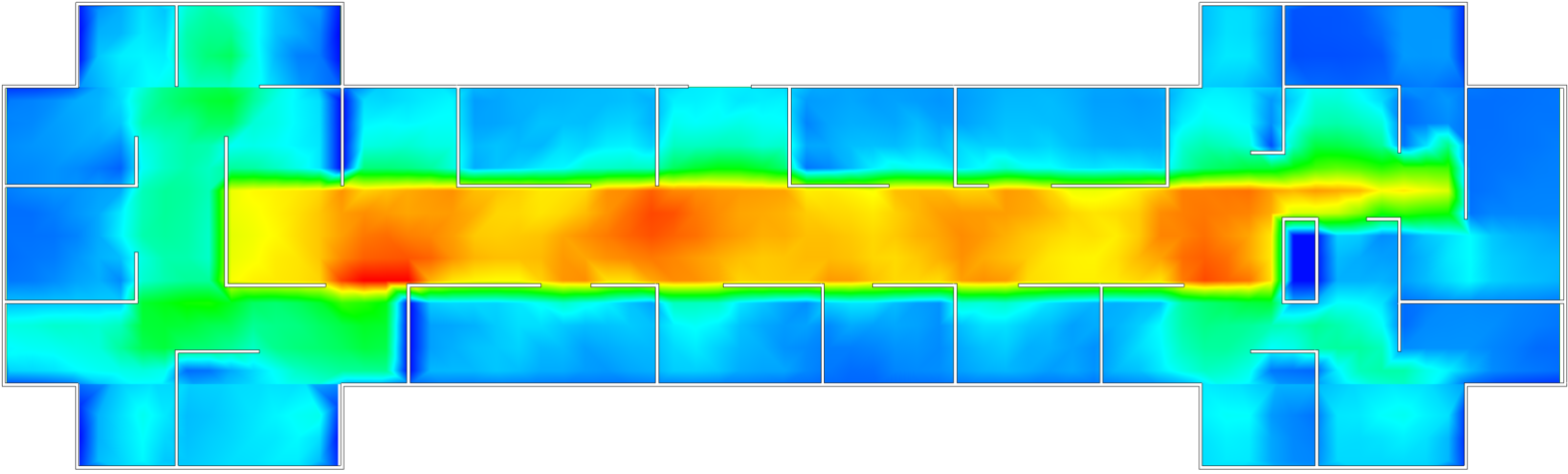} \\
		\objective{\diversityMembers, \vectorObjectiveWeights} = \num{2.4043} \\
		\includegraphics[width=0.9\linewidth]{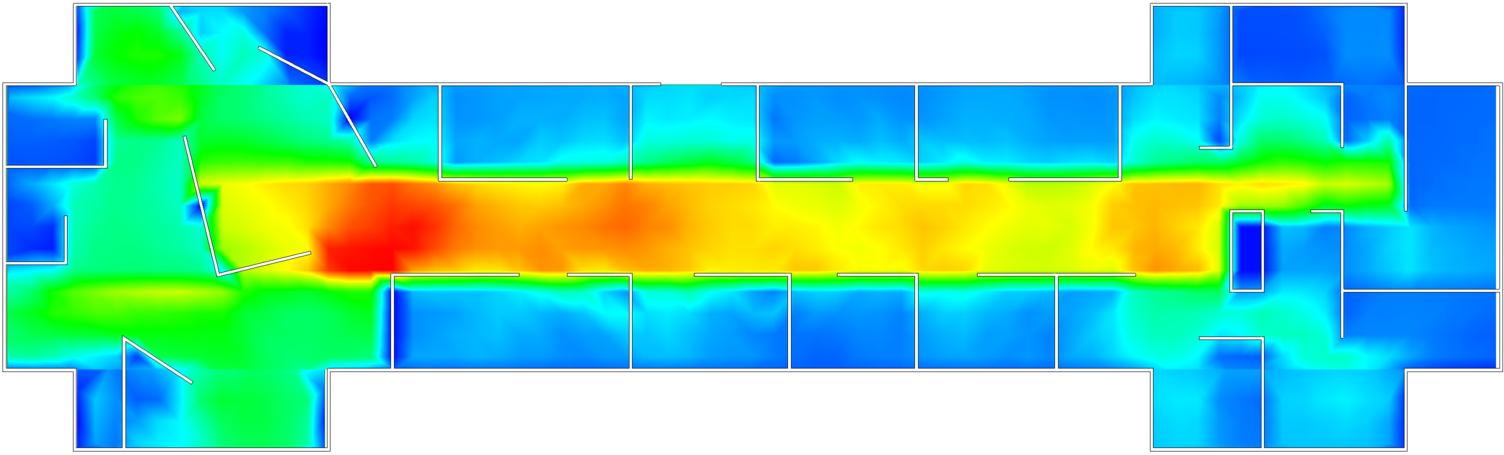} \\
		\objective{\diversityMembers, \vectorObjectiveWeights} = \num{3.0074} \\
		\includegraphics[width=0.9\linewidth]{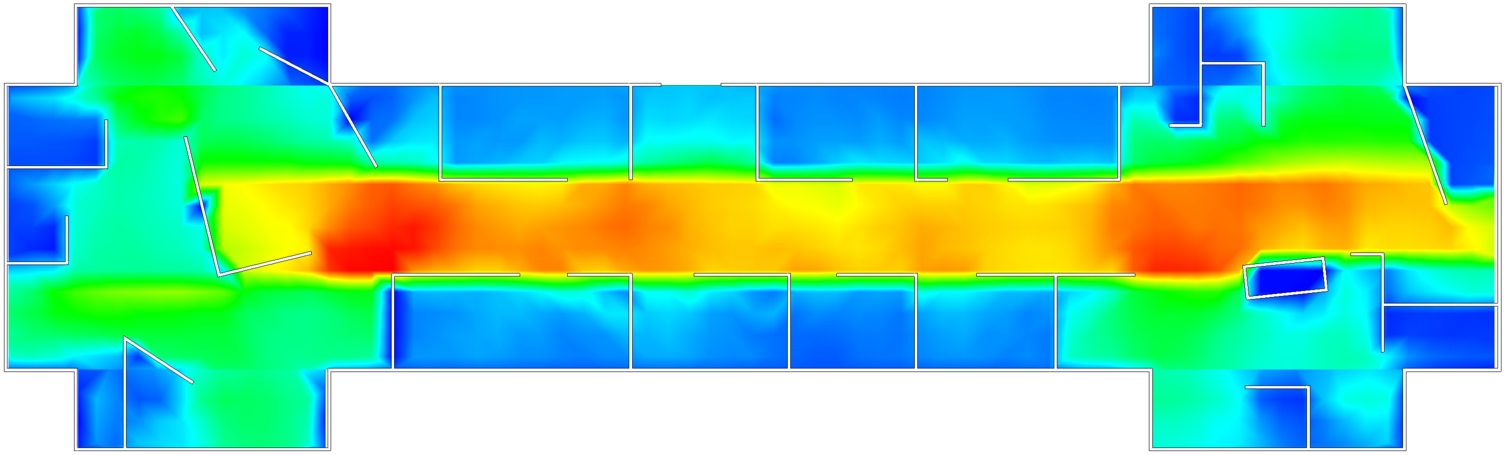} \\
		\objective{\diversityMembers, \vectorObjectiveWeights} = \num{4.2741} \\
		\includegraphics[width=0.9\linewidth]{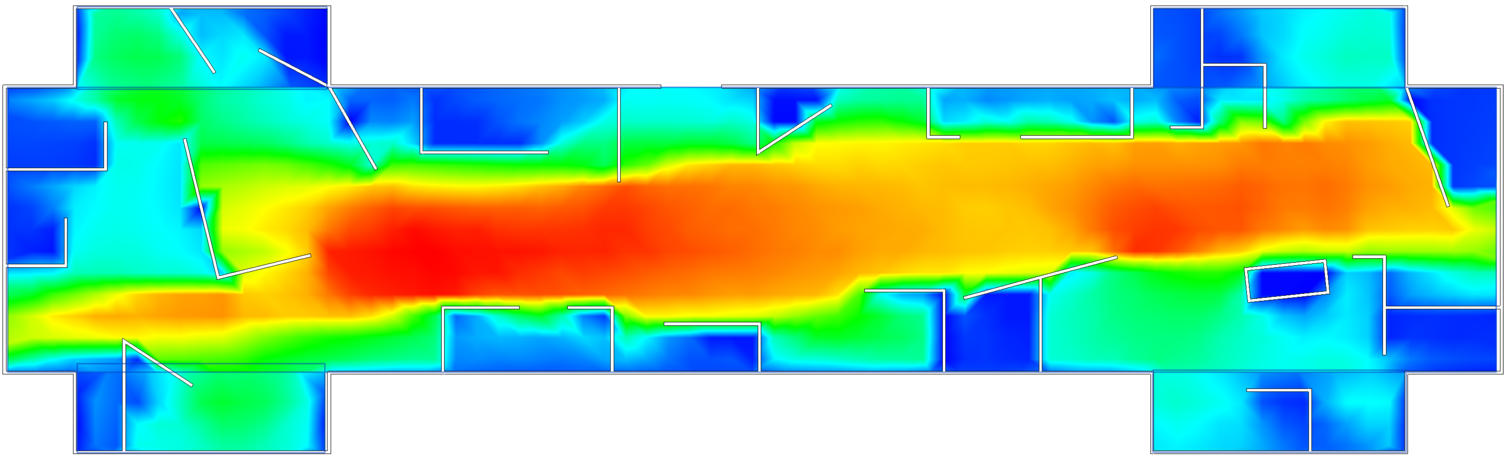} \\
		\objective{\diversityMembers, \vectorObjectiveWeights} = \num{8.5863} \\
		
	\end{tabular}
	\caption{\label{figure:art-gallery-reduced} Optimizing an art gallery. From top to bottom, the default gallery and three consecutive rounds of optimization. Each round is performed with a combination of degree, depth, and entropy. The combined objective $\objective{\diversityMembers, \vectorObjectiveWeights}$ at the end of each round is visualized as a heat map. Red is high value, blue is low.}
\end{figure}

\noindent \textbf{Art Gallery A}. 
~\reffig{figure:art-gallery-reduced} illustrates the iterative design of an art gallery.
Particularly, we are interested in increasing the degree of visibility, a reduction in the depth (which indicates an increase in accessibility), and an increase in entropy or organization of the gallery. The design process is performed over $3$ optimization rounds and a $4$ times improvement in the combined objective measure is discovered.
In the heatmaps, red and blue show high and low values respectively. 

\begin{sidewaysfigure*}
	\centering
	\vspace{60em}
	\setlength{\tabcolsep}{0.1em}
	\begin{tabular}{c | ccccc}
		\begin{tikzpicture}
		\node[anchor=south west,inner sep=0] at (0,0) {\includegraphics[width=0.16\linewidth]{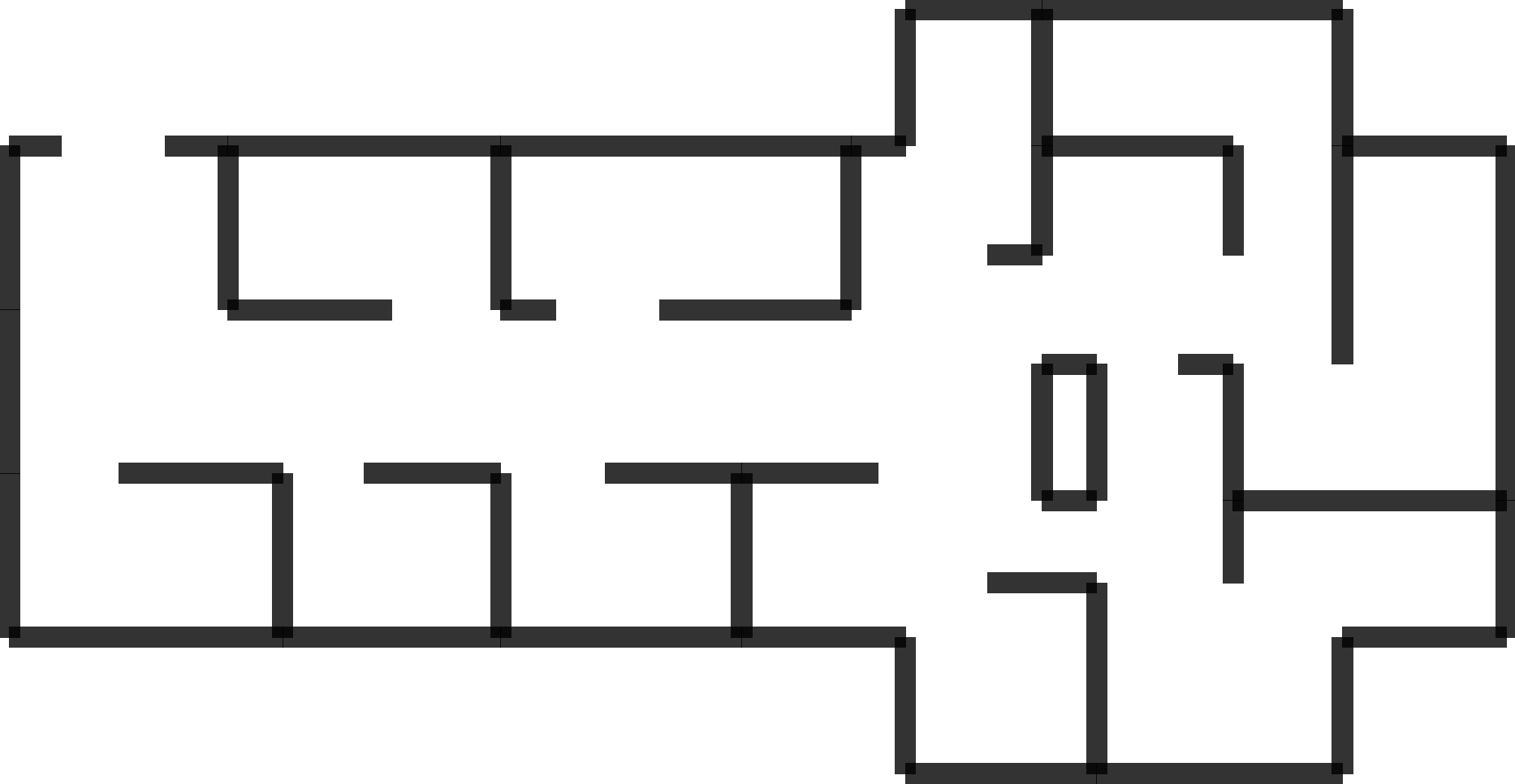}};
		\draw[magenta,thick,rounded corners] (0.5,0.89) -- (0.7,1.08);
		\draw[magenta,thick,rounded corners] (0.5,0.89) -- (0.7,0.69);
		\draw[magenta,thick,rounded corners] (0.61,0.77) arc (-22:22:0.3cm);
		\end{tikzpicture} & 
		\begin{tikzpicture}
		\node[anchor=south west,inner sep=0] at (0,0) {\includegraphics[width=0.16\linewidth]{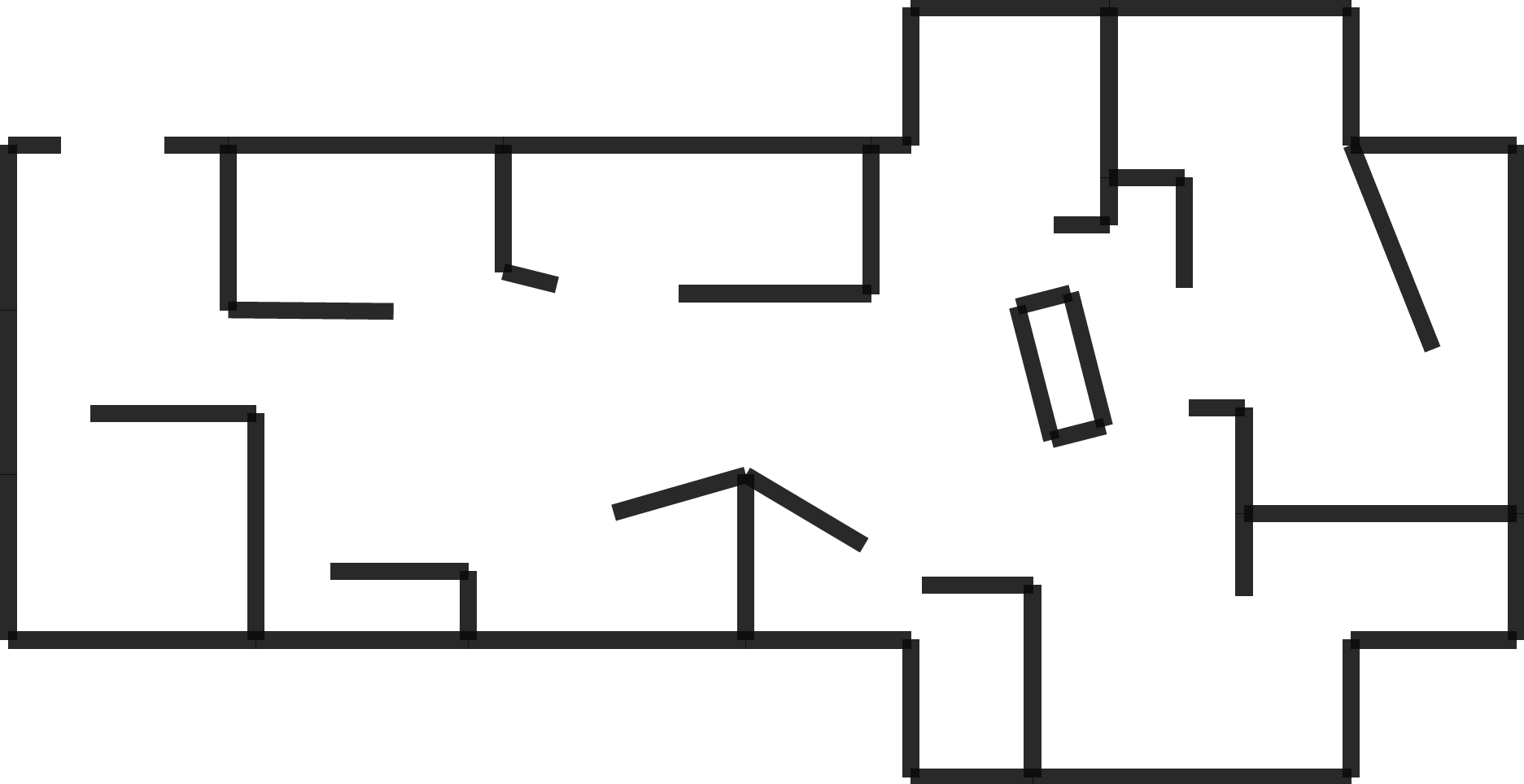}};
		\draw[magenta,thick,rounded corners] (0.5,0.89) -- (0.7,1.08);
		\draw[magenta,thick,rounded corners] (0.5,0.89) -- (0.7,0.69);
		\draw[magenta,thick,rounded corners] (0.61,0.77) arc (-22:22:0.3cm);
		\end{tikzpicture} &
		\begin{tikzpicture}
		\node[anchor=south west,inner sep=0] at (0,0) {\includegraphics[width=0.16\linewidth]{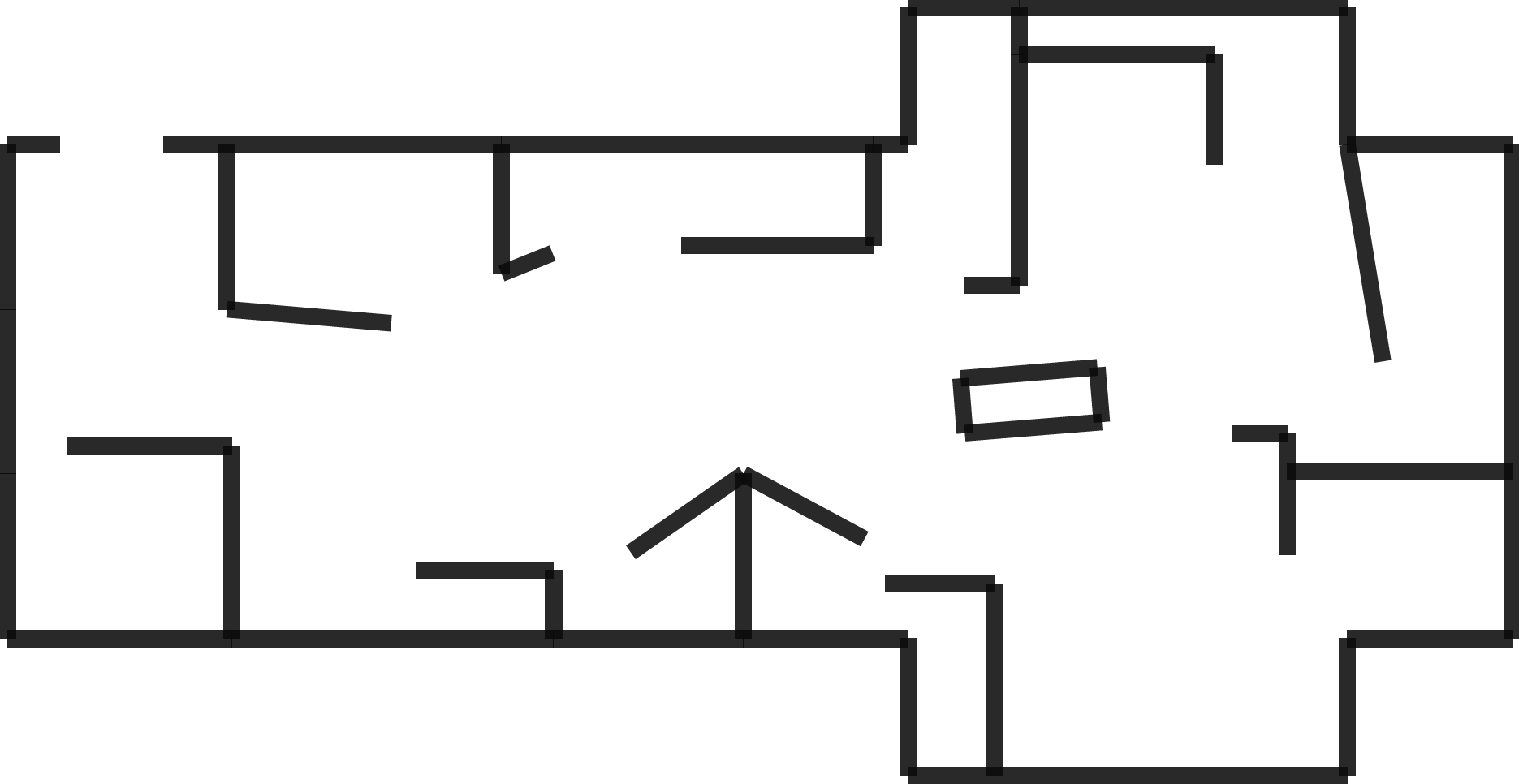}};
		\draw[magenta,thick,rounded corners] (0.5,0.89) -- (0.7,1.08);
		\draw[magenta,thick,rounded corners] (0.5,0.89) -- (0.7,0.69);
		\draw[magenta,thick,rounded corners] (0.61,0.77) arc (-22:22:0.3cm);
		\end{tikzpicture} &
		\begin{tikzpicture}
		\node[anchor=south west,inner sep=0] at (0,0) {\includegraphics[width=0.16\linewidth]{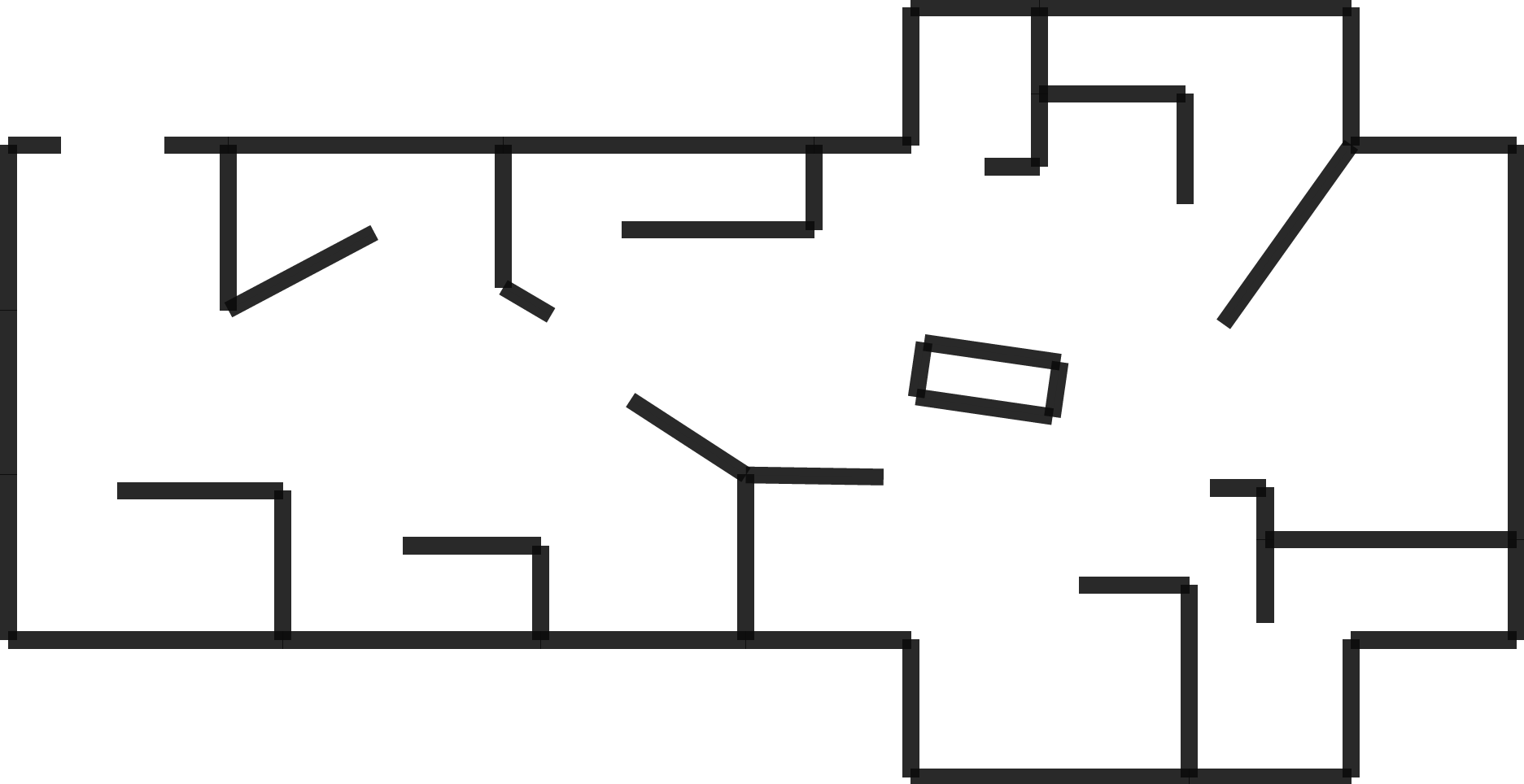}};
		\draw[magenta,thick,rounded corners] (0.5,0.89) -- (0.7,1.08);
		\draw[magenta,thick,rounded corners] (0.5,0.89) -- (0.7,0.69);
		\draw[magenta,thick,rounded corners] (0.61,0.77) arc (-22:22:0.3cm);
		\end{tikzpicture} &
		\begin{tikzpicture}
		\node[anchor=south west,inner sep=0] at (0,0) {\includegraphics[width=0.16\linewidth]{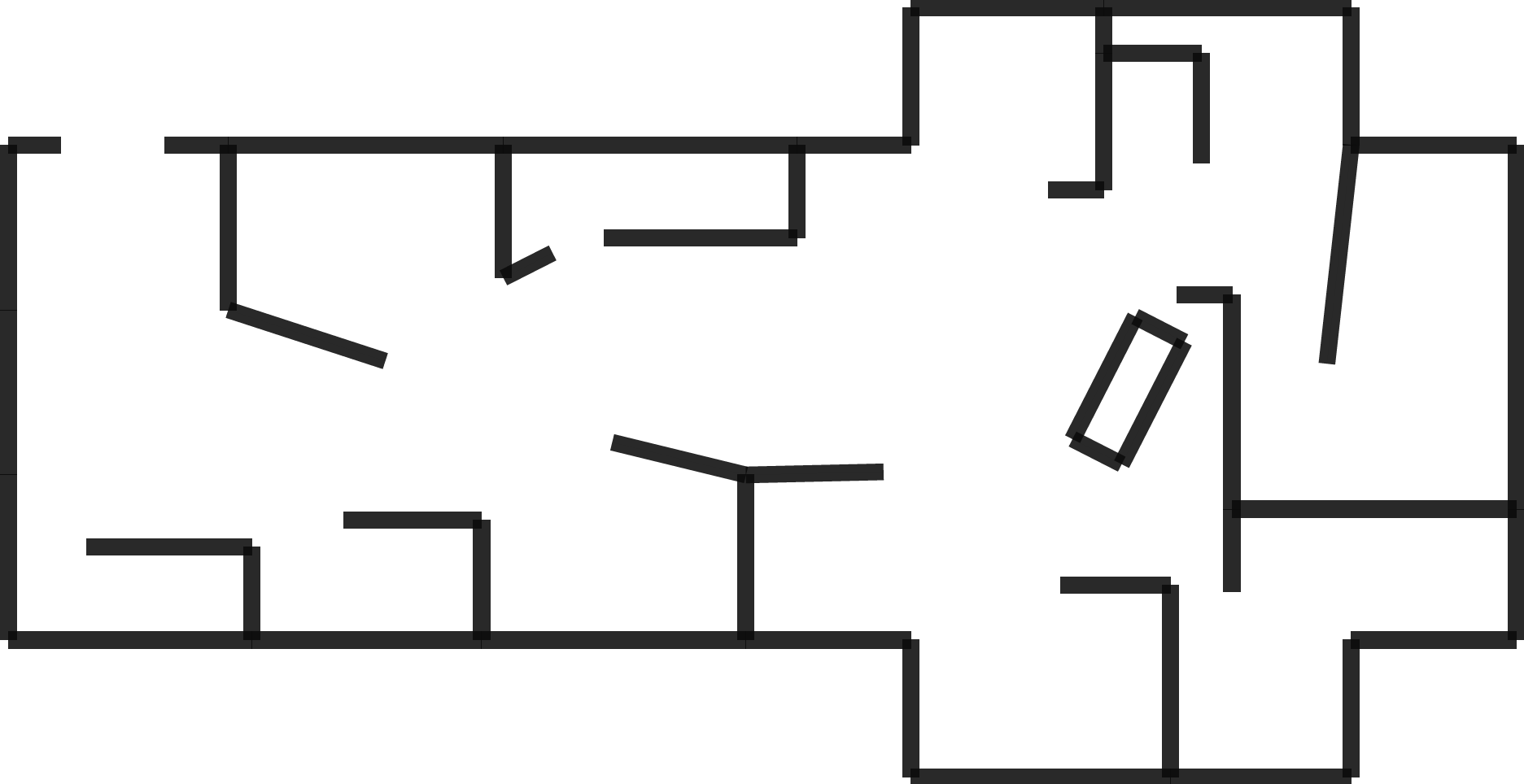}};
		\draw[magenta,thick,rounded corners] (0.5,0.89) -- (0.7,1.08);
		\draw[magenta,thick,rounded corners] (0.5,0.89) -- (0.7,0.69);
		\draw[magenta,thick,rounded corners] (0.61,0.77) arc (-22:22:0.3cm);
		\end{tikzpicture} &
		\begin{tikzpicture}
		\node[anchor=south west,inner sep=0] at (0,0) {\includegraphics[width=0.16\linewidth]{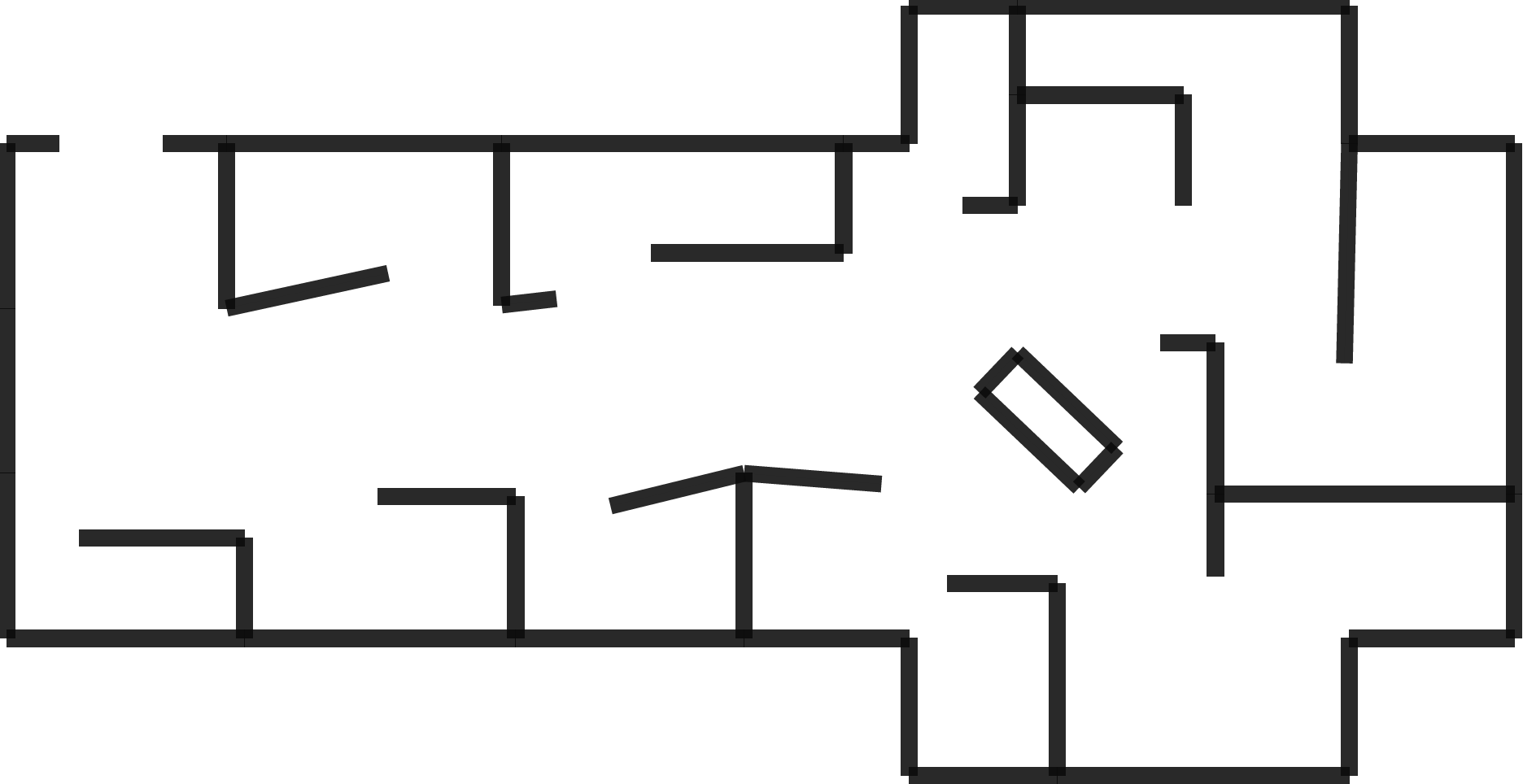}};
		\draw[magenta,thick,rounded corners] (0.5,0.89) -- (0.7,1.08);
		\draw[magenta,thick,rounded corners] (0.5,0.89) -- (0.7,0.69);
		\draw[magenta,thick,rounded corners] (0.61,0.77) arc (-22:22:0.3cm);
		\end{tikzpicture} \\
		\includegraphics[width=0.16\linewidth]{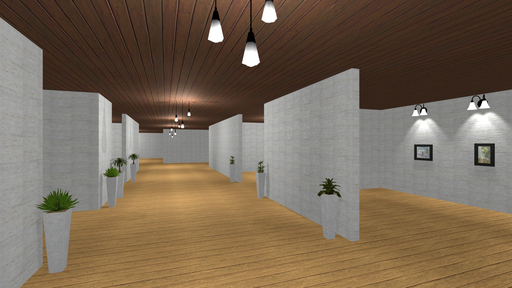} &
		\includegraphics[width=0.16\linewidth]{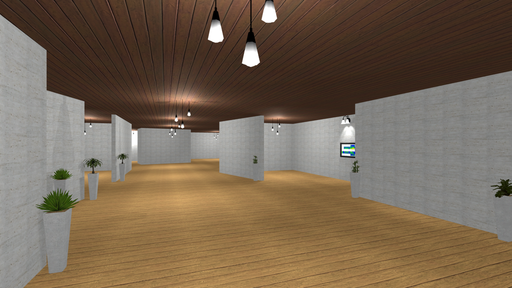} &
		\includegraphics[width=0.16\linewidth]{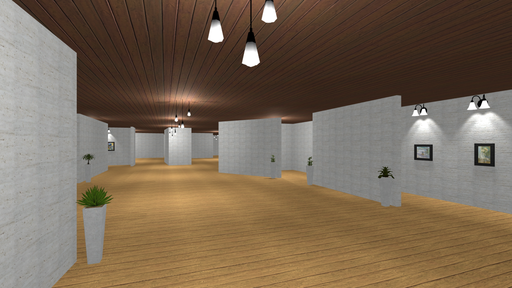} &
		\includegraphics[width=0.16\linewidth]{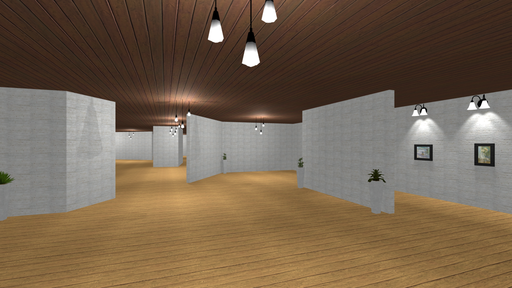} &
		\includegraphics[width=0.16\linewidth]{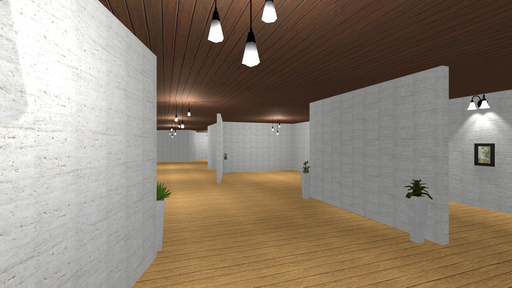} &
		\includegraphics[width=0.16\linewidth]{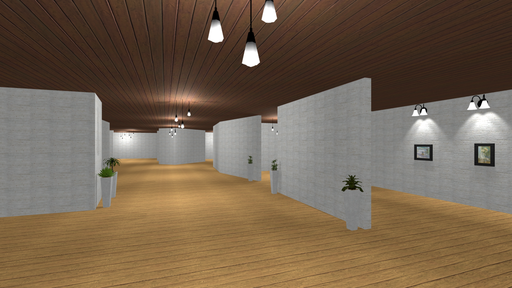} \\
		\includegraphics[width=0.16\linewidth]{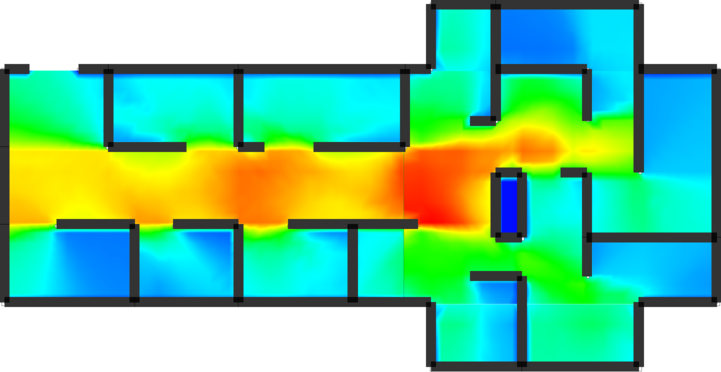} &
		\includegraphics[width=0.16\linewidth]{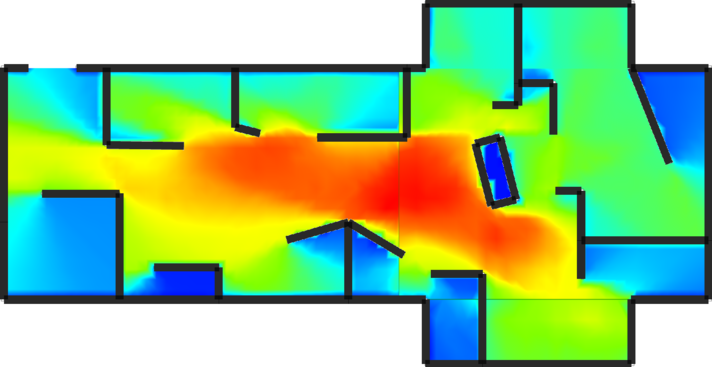} &
		\includegraphics[width=0.16\linewidth]{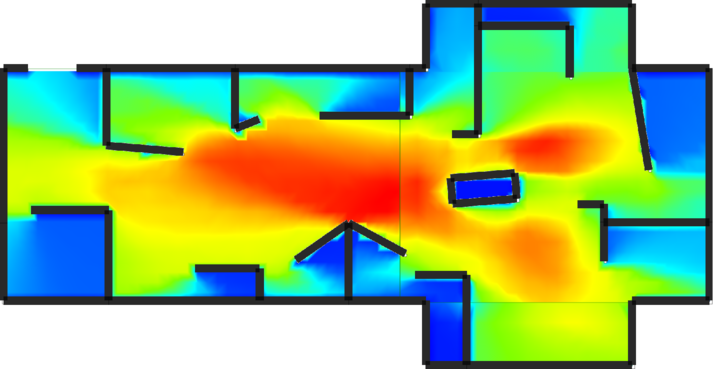} &
		\includegraphics[width=0.16\linewidth]{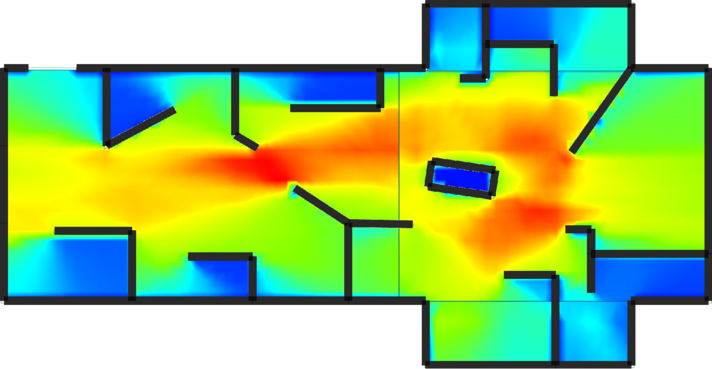} &
		\includegraphics[width=0.16\linewidth]{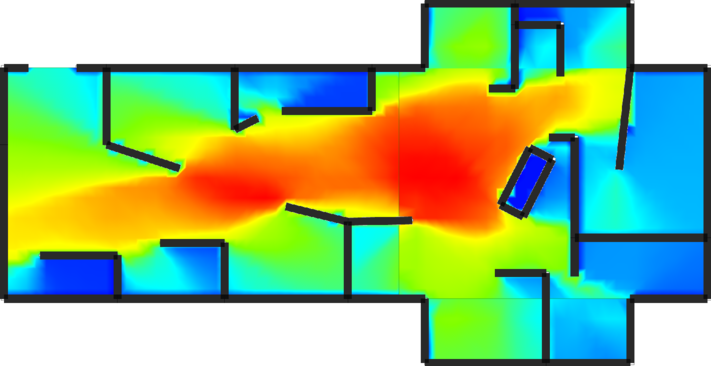} &
		\includegraphics[width=0.16\linewidth]{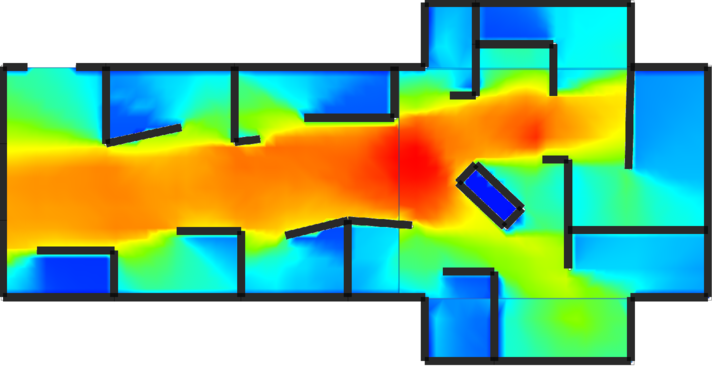} \\
		(a) & (b) & (c) & (d)* & (e)* & (f)*
	\end{tabular}
	\caption{\label{figure:half-art-diversity}\newChanges{This figure has been added to further illustrate the diversity set value.}{Optimizing an art gallery and exemplifying the power of having diverse near optimal designs. The top row of figures shows the blueprint of the wall designs for the art gallery with a particular viewpoint shown in magenta.  The middle row of figures shows the rendered environment from the viewpoint shown in the top row.  The final row of figures shows the combined metric values as a heatmap over the entirety of each design. Column (a) is the original design of the art gallery.  The columns (b) - (f) show the diversity members provided by the \DOME system for a particular parametrization of the environment. (b) is a member that opens up the floor space and the overall visibility down the corridor of the gallery. (c) is a member that balances the corridor visibility of (b) with the visibility of particular exhibits. (d) is a member that balances the visibility from (b) while reducing the number of path decisions further down the corridor and being particularly accessible. (e) is a member that mainly reduces path decisions while increasing gallery sizes. (f) is a member that balances the best of all designs being open, accessible, and easy to navigate.  The (*)s identify the designs  that six expert architects independently designated as preferred.}}
\end{sidewaysfigure*}

\noindent \textbf{Art Gallery B}. 
\newChanges{This section has been added to demonstrate the value the diversity set.}{
~\reffig{figure:half-art-diversity} illustrates the benefits of the diversity member set in the design process of an art gallery. This gallery design was parametrized to allow for interesting reconfigurations of the exhibit rooms which directly modify the open space of the main corridor. The optimization process produced a diversity set that includes both highly angular and interesting designs as well as more balanced designs that carefully reconfigure the view of individual exhibits and the open space in the corridor.}

\begin{figure*}[!htb]
\centering

\begin{tabular}{c c} 

\begin{tikzpicture}[      
     every node/.style={anchor=south west,inner sep=0pt},
     x=1mm, y=1mm,
   ]   
  \node[rotate=90,yscale=1.0] (fig1) at (0,0)
    {\includegraphics[width=0.4\linewidth]{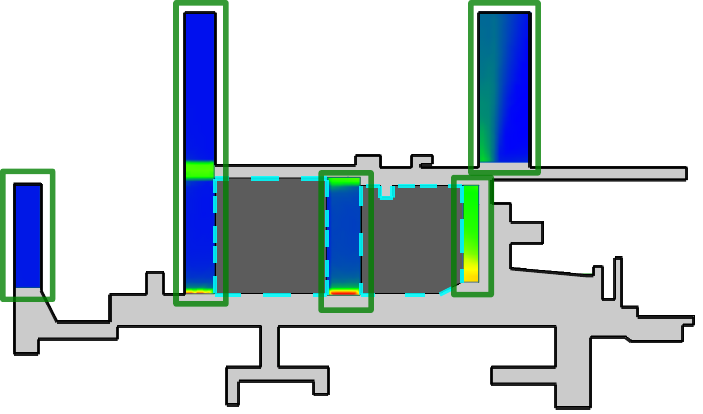}};  
\end{tikzpicture} &
\begin{tikzpicture}[      
     every node/.style={anchor=south west,inner sep=0pt},
     x=1mm, y=1mm,
   ]   
  \node[rotate=90,yscale=1.0] (fig1) at (0,0)
    {\includegraphics[width=0.4\linewidth]{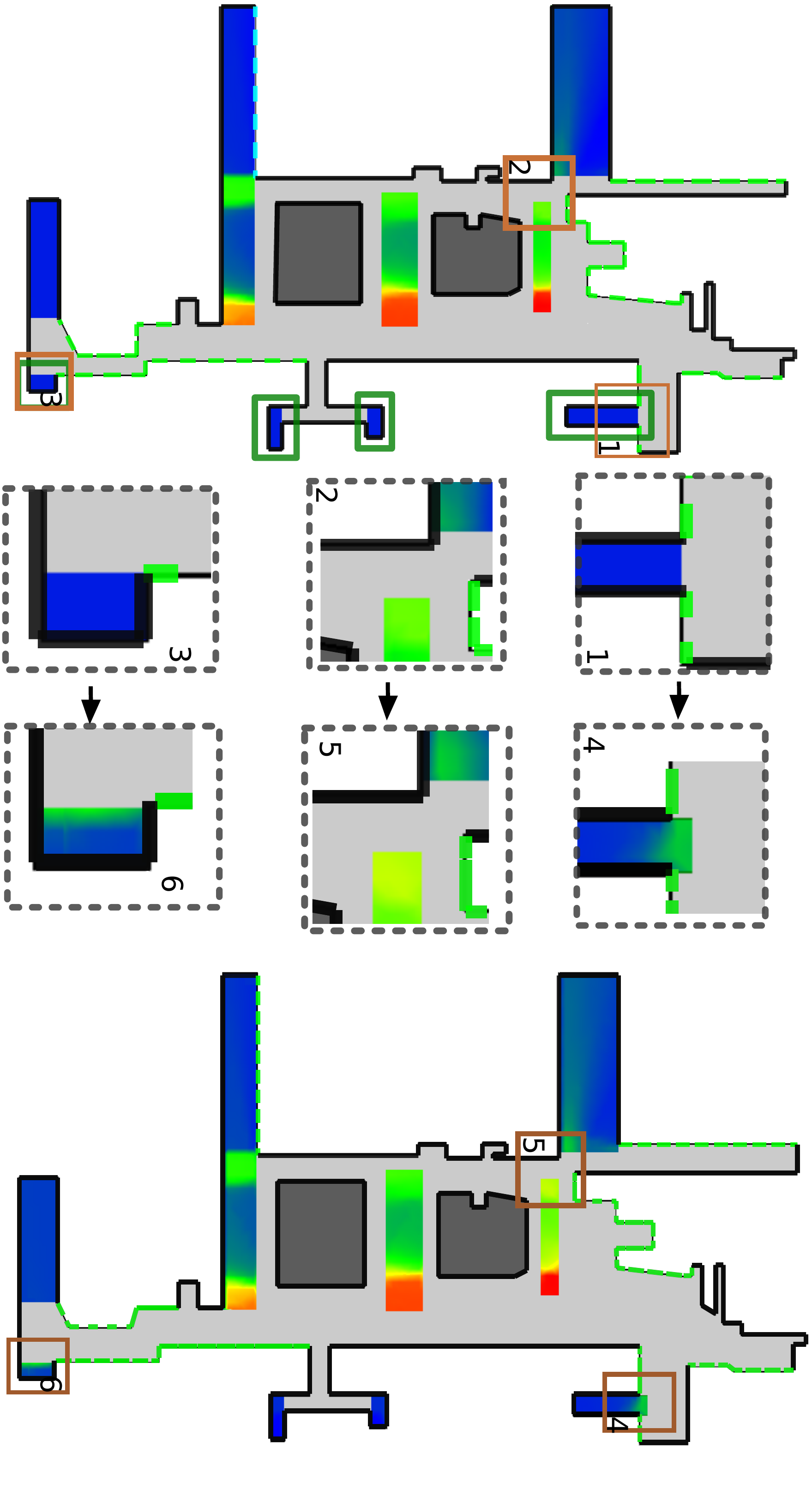}};
    
\end{tikzpicture} \\
 (a) Initial: \num{6.33} & (c) Round 2: $\num{10.49}$ \\
 
\begin{tikzpicture}[      
      every node/.style={anchor=south west,inner sep=0pt},
      x=1mm, y=1mm,
    ]   
  \node[rotate=90,yscale=1.0] (fig1) at (0,0)
     {\includegraphics[width=0.4\linewidth]{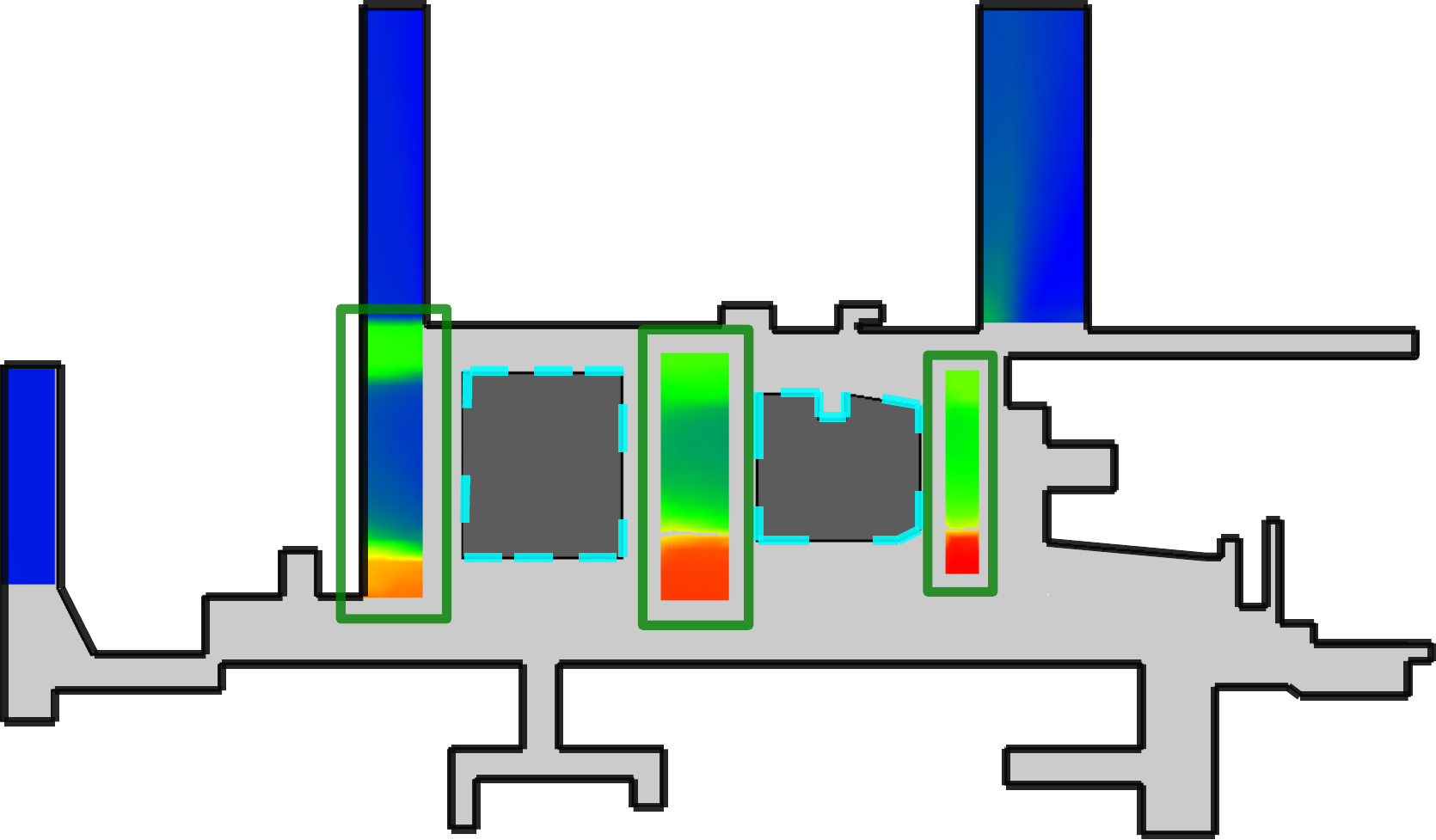}};  
 
\end{tikzpicture} & 

\begin{tikzpicture}[      
       every node/.style={anchor=south west,inner sep=0pt},
       x=1mm, y=1mm,
     ]   
  \node[rotate=90,yscale=1.2] (fig1) at (0,0)
      {\includegraphics[width=0.4\linewidth]{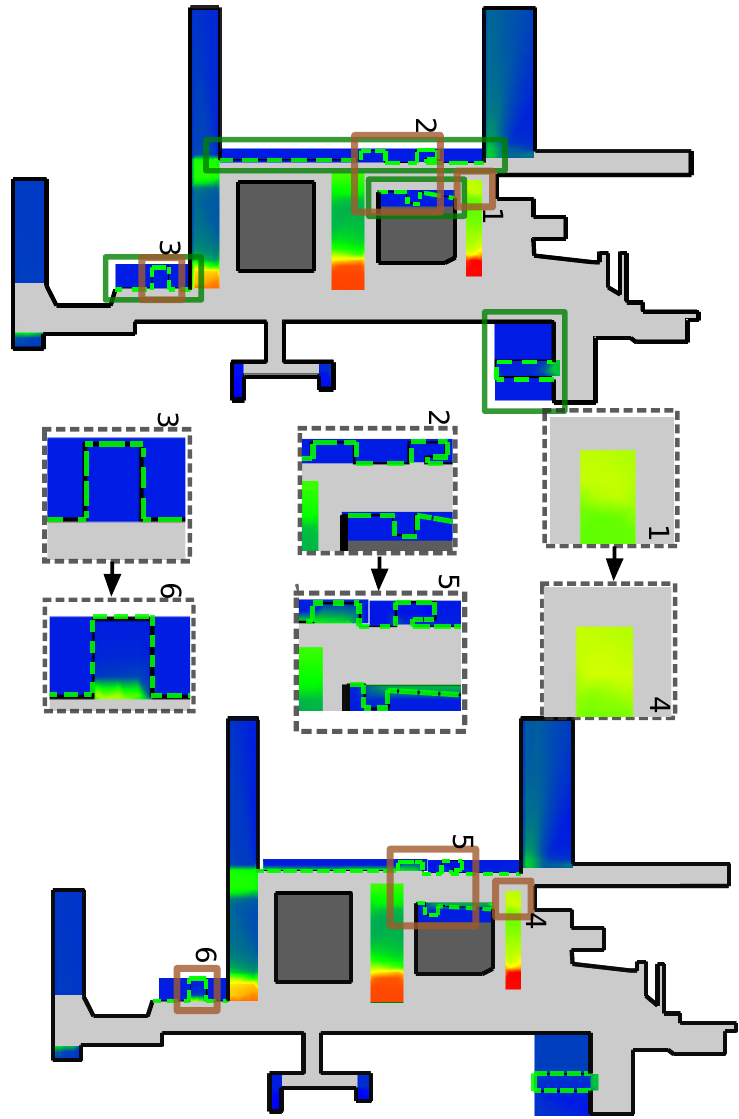}};  
\end{tikzpicture} \\
 (b) Round 1: $\num{7.21}$  & (d) Round 3: $\num{11.68}$  \\


\end{tabular}

\caption{\label{figure:subway-station-example} Optimization of Penn Station, NYC. 
This figure illustrates how the framework can be used on a large complex environment of $\sim 10,000$ vertices.
Additional \roq are incrementally added, to resolve issues in the layout that were identified during the previous design optimization rounds.
The light grey area is the \ror.
The heat map areas are {\roq} with significant changes outlined in brown rectangles.
The dashed cyan lines show the structure of interest that was optimized between each round. 
The green boxes highlight the new areas of interest that were considered during the optimization round. %
Round 1(a-b) regions are chosen to increase the accessibility and visibility of subway platform access. 
Round 2(c) regions are chosen to increase the accessibility and visibility of exits. 
Round 3(d) the placement of washrooms and elevators are improved by making them more viewable and accessible from additional areas in the environment.
}


\end{figure*}

\noindent \textbf{Subway Station}. We use \DOME to optimize a level of the NYC Penn Station. 
The user-in-the-loop approach affords an iterative design process, where a user may initially set up the problem by defining the movable elements, and the \roq and \ror. Upon selecting a suitable revision to the layout from a set of diverse candidates provided by the system, the user may modify the problem formulation.
\reffig{figure:subway-station-example} illustrates results from three iterations. 
By adding additional parameters or changing the regions in an effort the user can resolve issues that may have been identified over the course of previous optimization rounds. In this example, the user iteratively includes new query regions for the stairwell and elevators to account for additional aspects of the layout. What appear as minor alterations to the wall configuration in the subway increase the objective from $6.3$ to $11.68$ leading to a design that significantly improves the pedestrian environment. 

\begin{figure*}[!htb]
	\centering
	\begin{tabular}{c c c}		
		\includegraphics[width=0.30\linewidth]{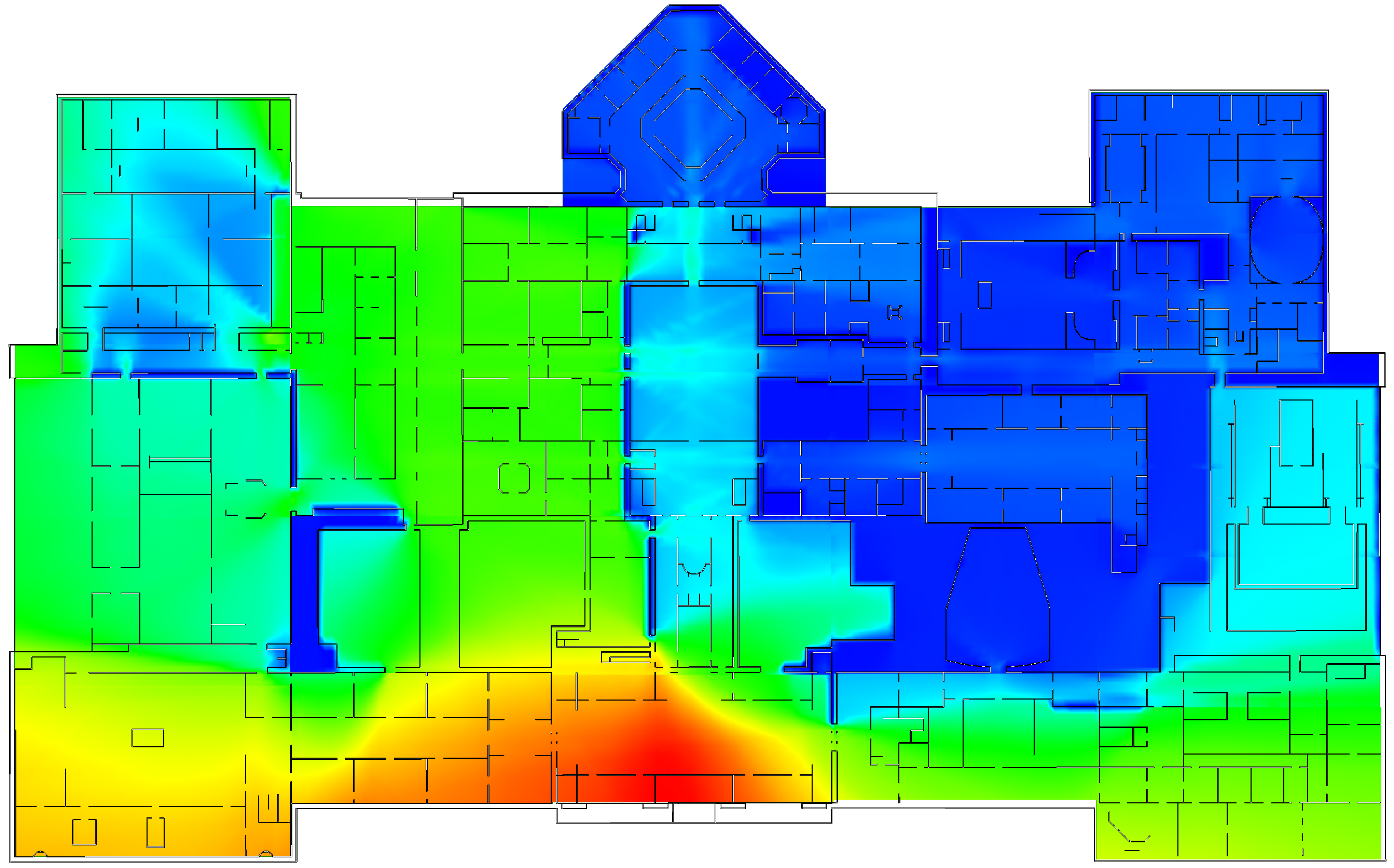} &
		\includegraphics[width=0.30\linewidth]{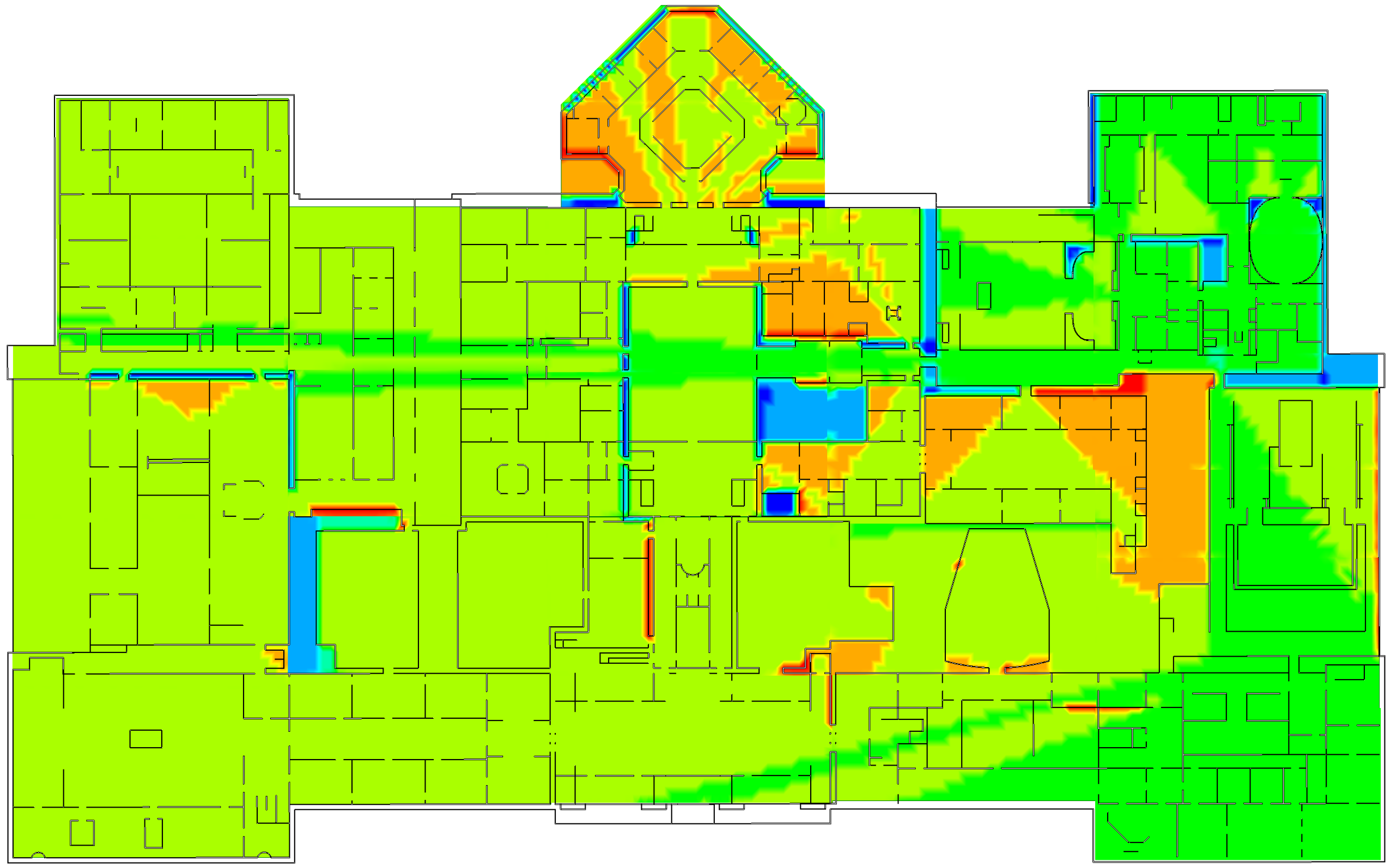} &
		\includegraphics[width=0.30\linewidth]{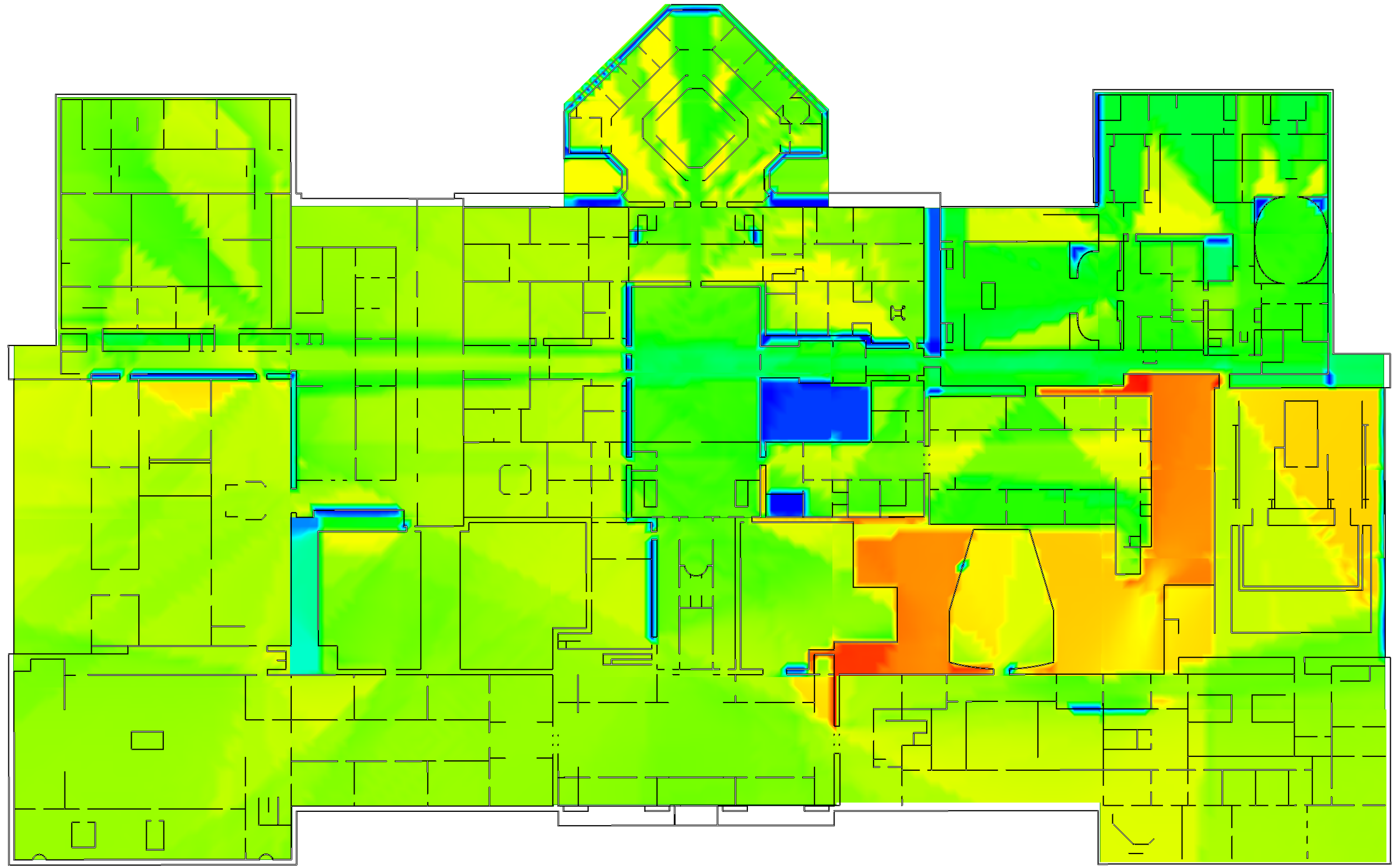} \\
		Degree & Depth & Entropy \\		
		
	\end{tabular}
	\caption{\label{figure:museum-example} 
		Degree, depth, and entropy for the Metropolitan Museum of Art.
	}
\end{figure*}

\begin{figure}[!htb]
	\centering
	\begin{tabular}{c}
		\includegraphics[width=\linewidth]{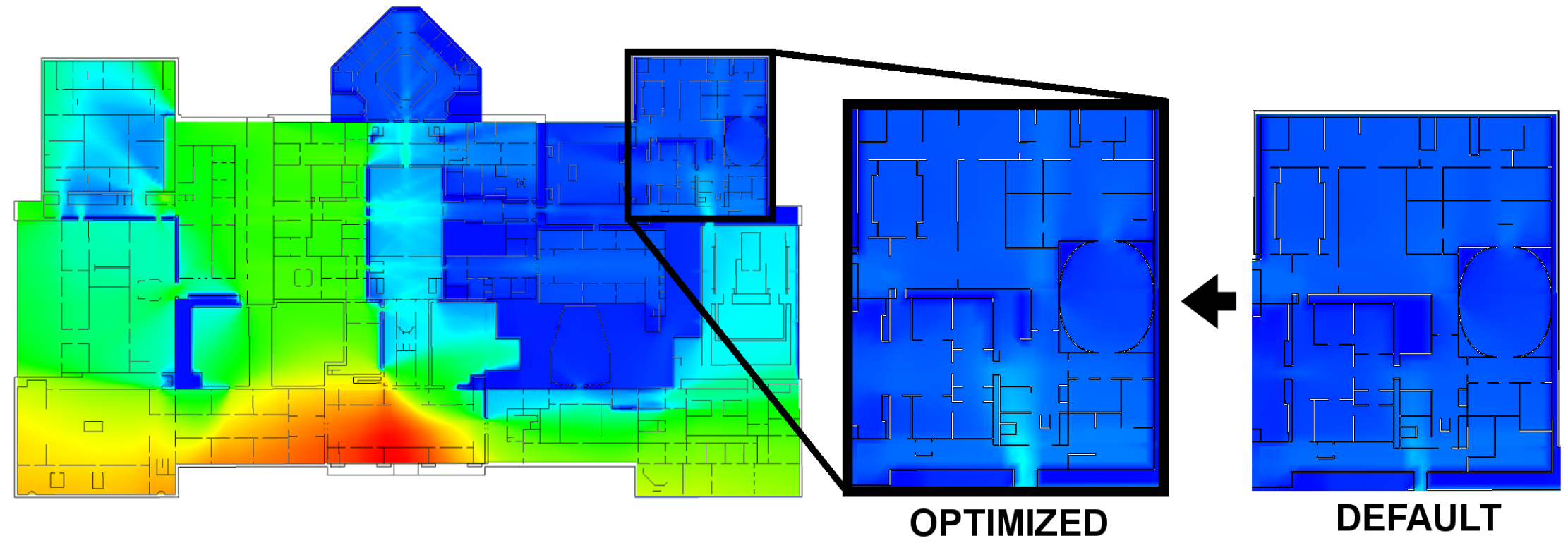}	
		
	\end{tabular}
	\caption{\label{figure:museum-optimized-degree} 
	 Analysis of the metrics reveal low visibility in the top-right section of the museum which we mitigate using \DOME.
	}
\end{figure}

\noindent \textbf{Museum of Metropolitan Art.} In \reffig{figure:museum-example} we visually analyze the layout of the museum by inspecting its degree, depth, and entropy values over the entire layout. In the top-right hand corner of the museum contains an area with very low visibility, specifically of the entrance. Therefore, we optimize the top-right area, shown in \reffig{figure:museum-optimized-degree}, to improve its visibility while maintaining the amount of wall surface area, which is necessary for displaying works of art.

\begin{figure*}
\tikzset{tikzMazeStyle/.style={line cap=rect,double, line width=0.1mm}}
\tikzset{tikzMazeScale/.style={scale=1.8}}
\centering
\begin{tabular}{c | c c | c c c }
	\begin{tikzpicture}[tikzMazeScale]
\begin{scope}[cm={1,0,0,-1,(0.00mm,13.94mm)}]
\begin{scope}[cm={1,0,0,-1,(0.00mm,13.94mm)}]
\definecolor{fc}{RGB}{255,20,147}
\draw [fill=fc,draw=none] (0.16mm,13.94mm) -- ++(2.49mm,0.00mm) -- ++(0.00mm,-1.63mm) -- ++(-2.49mm,0.00mm) -- ++(0.00mm,1.63mm) ;
\definecolor{fc}{RGB}{255,20,147}
\draw [fill=fc,draw=none] (10.55mm,13.90mm) -- ++(2.35mm,0.00mm) -- ++(0.00mm,-1.91mm) -- ++(-2.35mm,0.00mm) -- ++(0.00mm,1.91mm) ;
\definecolor{fc}{RGB}{128,128,128}
\draw [fill=fc,draw=none] (0.12mm,12.59mm) -- ++(12.83mm,0.00mm) -- ++(0.00mm,-12.43mm) -- ++(-12.83mm,0.00mm) -- ++(0.00mm,12.43mm) ;
\end{scope}
\end{scope}

\begin{scope}[cm={1,0,0,-1,(0.00mm,13.94mm)}]
\draw [tikzMazeStyle]  (0.14mm,13.80mm)
-- ++(12.80mm,0.00mm)
(12.94mm,13.80mm)
-- ++(0.00mm,-0.80mm)
(12.94mm,1.00mm)
-- ++(-0.55mm,0.00mm)
(0.14mm,1.00mm)
-- ++(0.00mm,0.80mm)
(0.14mm,7.40mm)
-- ++(0.76mm,0.00mm)
(0.14mm,10.60mm)
-- ++(3.86mm,0.00mm)
(0.14mm,12.20mm)
-- ++(0.76mm,0.00mm)
(0.14mm,9.00mm)
-- ++(2.02mm,0.00mm)
(0.14mm,4.20mm)
-- ++(0.76mm,0.00mm)
(0.14mm,5.80mm)
-- ++(4.02mm,0.00mm)
(0.14mm,2.60mm)
-- ++(3.86mm,0.00mm)
(3.80mm,9.00mm)
-- ++(2.74mm,0.00mm)
(5.80mm,5.80mm)
-- ++(0.74mm,0.00mm)
(5.64mm,2.60mm)
-- ++(0.90mm,0.00mm)
(0.70mm,1.00mm)
-- ++(-0.56mm,0.00mm)
(6.54mm,12.20mm)
-- ++(0.00mm,-0.80mm)
(6.54mm,11.40mm)
-- ++(0.00mm,-0.80mm)
(6.54mm,10.60mm)
-- ++(0.00mm,-0.80mm)
(6.54mm,9.80mm)
-- ++(0.00mm,-0.80mm)
(6.54mm,9.00mm)
-- ++(0.00mm,-0.80mm)
(6.54mm,8.20mm)
-- ++(0.00mm,-0.80mm)
(6.54mm,7.40mm)
-- ++(0.00mm,-0.80mm)
(6.54mm,6.60mm)
-- ++(0.00mm,-0.80mm)
(6.54mm,5.80mm)
-- ++(0.00mm,-0.80mm)
(6.54mm,5.00mm)
-- ++(0.00mm,-0.80mm)
(6.54mm,4.20mm)
-- ++(0.00mm,-0.80mm)
(6.54mm,3.40mm)
-- ++(0.00mm,-0.80mm)
(6.54mm,2.60mm)
-- ++(0.00mm,-0.80mm)
(6.54mm,1.80mm)
-- ++(0.00mm,-0.80mm)
(12.94mm,1.80mm)
-- ++(0.00mm,-0.80mm)
(12.94mm,2.60mm)
-- ++(0.00mm,-0.80mm)
(12.94mm,3.40mm)
-- ++(0.00mm,-0.80mm)
(12.94mm,4.20mm)
-- ++(0.00mm,-0.80mm)
(12.94mm,5.00mm)
-- ++(0.00mm,-0.80mm)
(12.94mm,5.80mm)
-- ++(0.00mm,-0.80mm)
(12.94mm,6.60mm)
-- ++(0.00mm,-0.80mm)
(12.94mm,7.40mm)
-- ++(0.00mm,-0.80mm)
(12.94mm,8.20mm)
-- ++(0.00mm,-0.80mm)
(12.94mm,9.80mm)
-- ++(0.00mm,-0.80mm)
(12.94mm,10.60mm)
-- ++(0.00mm,-0.80mm)
(12.94mm,11.40mm)
-- ++(0.00mm,-0.80mm)
(12.94mm,12.20mm)
-- ++(0.00mm,-0.80mm)
(12.94mm,13.00mm)
-- ++(0.00mm,-0.80mm)
(10.78mm,1.00mm)
-- ++(-8.62mm,0.00mm)
(0.14mm,1.80mm)
-- ++(0.00mm,0.80mm)
(0.14mm,2.60mm)
-- ++(0.00mm,0.80mm)
(0.14mm,3.40mm)
-- ++(0.00mm,0.80mm)
(0.14mm,4.20mm)
-- ++(0.00mm,0.80mm)
(0.14mm,5.00mm)
-- ++(0.00mm,0.80mm)
(0.14mm,5.80mm)
-- ++(0.00mm,0.80mm)
(0.14mm,6.60mm)
-- ++(0.00mm,0.80mm)
(0.14mm,7.40mm)
-- ++(0.00mm,0.80mm)
(0.14mm,8.20mm)
-- ++(0.00mm,0.80mm)
(0.14mm,9.00mm)
-- ++(0.00mm,0.80mm)
(0.14mm,9.80mm)
-- ++(0.00mm,0.80mm)
(0.14mm,10.60mm)
-- ++(0.00mm,0.80mm)
(0.14mm,11.40mm)
-- ++(0.00mm,0.80mm)
(0.14mm,12.20mm)
-- ++(0.00mm,0.80mm)
(0.14mm,13.00mm)
-- ++(0.00mm,0.80mm)
(10.64mm,10.60mm)
-- ++(2.30mm,0.00mm)
(8.68mm,9.00mm)
-- ++(4.26mm,0.00mm)
(10.01mm,5.80mm)
-- ++(2.93mm,0.00mm)
(11.52mm,4.20mm)
-- ++(1.42mm,0.00mm)
(10.17mm,2.60mm)
-- ++(2.77mm,0.00mm)
(12.06mm,7.40mm)
-- ++(0.88mm,0.00mm)
(9.20mm,12.20mm)
-- ++(3.75mm,0.00mm)
(2.54mm,12.20mm)
-- ++(4.00mm,0.00mm)
(5.64mm,10.60mm)
-- ++(0.90mm,0.00mm)
(2.54mm,7.40mm)
-- ++(4.00mm,0.00mm)
(2.54mm,4.20mm)
-- ++(4.00mm,0.00mm)
(12.94mm,9.00mm)
-- ++(0.00mm,-0.80mm)
(6.54mm,12.20mm)
-- ++(0.93mm,0.00mm)
(6.54mm,10.60mm)
-- ++(2.37mm,0.00mm)
(6.54mm,9.00mm)
-- ++(0.41mm,0.00mm)
(6.54mm,7.40mm)
-- ++(3.79mm,0.00mm)
(6.54mm,5.80mm)
-- ++(1.74mm,0.00mm)
(6.54mm,4.20mm)
-- ++(3.26mm,0.00mm)
(6.54mm,2.60mm)
-- ++(1.91mm,0.00mm)
(0.90mm,12.04mm)
-- ++(0.00mm,0.16mm)
(2.54mm,12.04mm)
-- ++(0.00mm,0.16mm)
(4.00mm,10.45mm)
-- ++(0.00mm,0.16mm)
(2.16mm,9.16mm)
-- ++(0.00mm,-0.16mm)
(0.90mm,7.24mm)
-- ++(0.00mm,0.16mm)
(4.16mm,5.64mm)
-- ++(0.00mm,0.16mm)
(0.90mm,4.04mm)
-- ++(0.00mm,0.16mm)
(4.00mm,2.44mm)
-- (4.00mm,2.60mm)
(5.64mm,2.76mm)
-- ++(0.00mm,-0.16mm)
(2.54mm,4.36mm)
-- ++(0.00mm,-0.16mm)
(5.80mm,5.96mm)
-- ++(0.00mm,-0.16mm)
(2.54mm,7.56mm)
-- ++(0.00mm,-0.16mm)
(3.80mm,9.16mm)
-- ++(0.00mm,-0.16mm)
(5.64mm,10.76mm)
-- ++(0.00mm,-0.16mm)
(7.47mm,12.36mm)
-- ++(0.00mm,-0.16mm)
(9.20mm,12.36mm)
-- ++(0.00mm,-0.16mm)
(8.91mm,10.76mm)
-- ++(0.00mm,-0.16mm)
(10.64mm,10.76mm)
-- ++(0.00mm,-0.16mm)
(6.95mm,9.16mm)
-- ++(0.00mm,-0.16mm)
(8.68mm,9.16mm)
-- ++(0.00mm,-0.16mm)
(10.33mm,7.56mm)
-- ++(0.00mm,-0.16mm)
(12.06mm,7.56mm)
-- ++(0.00mm,-0.16mm)
(8.28mm,5.96mm)
-- ++(0.00mm,-0.16mm)
(10.01mm,5.96mm)
-- ++(0.00mm,-0.16mm)
(9.80mm,4.36mm)
-- ++(0.00mm,-0.16mm)
(11.52mm,4.36mm)
-- ++(0.00mm,-0.16mm)
(8.45mm,2.76mm)
-- ++(0.00mm,-0.16mm)
(10.17mm,2.76mm)
-- ++(0.00mm,-0.16mm)
(0.90mm,12.20mm)
-- ++(0.00mm,0.16mm)
(2.54mm,12.20mm)
-- ++(0.00mm,0.16mm)
(5.64mm,10.60mm)
-- ++(0.00mm,-0.16mm)
(4.00mm,10.60mm)
-- ++(0.00mm,0.16mm)
(3.80mm,9.00mm)
-- ++(0.00mm,-0.16mm)
(2.16mm,9.00mm)
-- ++(0.00mm,-0.16mm)
(0.90mm,7.40mm)
-- ++(0.00mm,0.16mm)
(2.54mm,7.40mm)
-- ++(0.00mm,-0.16mm)
(4.16mm,5.80mm)
-- ++(0.00mm,0.16mm)
(5.80mm,5.80mm)
-- ++(0.00mm,-0.16mm)
(2.54mm,4.20mm)
-- ++(0.00mm,-0.16mm)
(0.90mm,4.20mm)
-- ++(0.00mm,0.16mm)
(4.00mm,2.60mm)
-- ++(0.00mm,0.16mm)
(5.64mm,2.60mm)
-- ++(0.00mm,-0.16mm)
(10.17mm,2.60mm)
-- ++(0.00mm,-0.16mm)
(8.45mm,2.60mm)
-- ++(0.00mm,-0.16mm)
(11.52mm,4.20mm)
-- ++(0.00mm,-0.16mm)
(9.80mm,4.20mm)
-- ++(0.00mm,-0.16mm)
(10.01mm,5.80mm)
-- ++(0.00mm,-0.16mm)
(8.28mm,5.80mm)
-- ++(0.00mm,-0.16mm)
(10.33mm,7.40mm)
-- ++(0.00mm,-0.16mm)
(12.06mm,7.40mm)
-- ++(0.00mm,-0.16mm)
(6.95mm,9.00mm)
-- ++(0.00mm,-0.16mm)
(8.68mm,9.00mm)
-- ++(0.00mm,-0.16mm)
(8.91mm,10.60mm)
-- ++(0.00mm,-0.16mm)
(10.64mm,10.60mm)
-- ++(0.00mm,-0.16mm)
(9.20mm,12.20mm)
-- ++(0.00mm,-0.16mm)
(7.47mm,12.20mm)
-- ++(0.00mm,-0.16mm)
;
\end{scope}
\end{tikzpicture} &		
	\begin{tikzpicture}[tikzMazeScale]
\begin{scope}[cm={1,0,0,-1,(0.00mm,13.94mm)}]
\draw [tikzMazeStyle]  (0.47mm,13.82mm)
-- ++(12.80mm,0.00mm)
(13.27mm,13.82mm)
-- ++(0.00mm,-0.80mm)
(13.27mm,1.02mm)
-- ++(-0.56mm,0.00mm)
(0.47mm,1.02mm)
-- ++(0.00mm,0.80mm)
(0.47mm,7.42mm)
-- ++(0.57mm,0.00mm)
(0.47mm,10.62mm)
-- ++(0.56mm,0.00mm)
(0.47mm,12.22mm)
-- ++(0.56mm,0.00mm)
(0.47mm,9.02mm)
-- ++(0.56mm,0.00mm)
(0.47mm,4.22mm)
-- ++(0.56mm,0.00mm)
(0.47mm,5.82mm)
-- ++(0.56mm,0.00mm)
(0.47mm,2.62mm)
-- ++(0.57mm,0.00mm)
(2.66mm,9.02mm)
-- ++(4.21mm,0.00mm)
(2.66mm,5.82mm)
-- ++(4.21mm,0.00mm)
(2.67mm,2.62mm)
-- ++(4.20mm,0.00mm)
(1.03mm,1.02mm)
-- ++(-0.56mm,0.00mm)
(11.35mm,1.02mm)
-- ++(-8.96mm,0.00mm)
(6.87mm,12.22mm)
-- ++(0.00mm,-0.80mm)
(6.87mm,11.42mm)
-- ++(0.00mm,-0.80mm)
(6.87mm,10.62mm)
-- ++(0.00mm,-0.80mm)
(6.87mm,9.82mm)
-- ++(0.00mm,-0.80mm)
(6.87mm,9.02mm)
-- ++(0.00mm,-0.80mm)
(6.87mm,8.22mm)
-- ++(0.00mm,-0.80mm)
(6.87mm,7.42mm)
-- ++(0.00mm,-0.80mm)
(6.87mm,6.62mm)
-- ++(0.00mm,-0.80mm)
(6.87mm,5.82mm)
-- ++(0.00mm,-0.80mm)
(6.87mm,5.02mm)
-- ++(0.00mm,-0.80mm)
(6.87mm,4.22mm)
-- ++(0.00mm,-0.80mm)
(6.87mm,3.42mm)
-- ++(0.00mm,-0.80mm)
(6.87mm,2.62mm)
-- ++(0.00mm,-0.80mm)
(6.87mm,1.82mm)
-- ++(0.00mm,-0.80mm)
(13.27mm,1.82mm)
-- ++(0.00mm,-0.80mm)
(13.27mm,2.62mm)
-- ++(0.00mm,-0.80mm)
(13.27mm,3.42mm)
-- ++(0.00mm,-0.80mm)
(13.27mm,4.22mm)
-- ++(0.00mm,-0.80mm)
(13.27mm,5.02mm)
-- ++(0.00mm,-0.80mm)
(13.27mm,5.82mm)
-- ++(0.00mm,-0.80mm)
(13.27mm,6.62mm)
-- ++(0.00mm,-0.80mm)
(13.27mm,7.42mm)
-- ++(0.00mm,-0.80mm)
(13.27mm,8.22mm)
-- ++(0.00mm,-0.80mm)
(13.27mm,9.82mm)
-- ++(0.00mm,-0.80mm)
(13.27mm,10.62mm)
-- ++(0.00mm,-0.80mm)
(13.27mm,11.42mm)
-- ++(0.00mm,-0.80mm)
(13.27mm,12.22mm)
-- ++(0.00mm,-0.80mm)
(13.27mm,13.02mm)
-- ++(0.00mm,-0.80mm)
(0.47mm,1.82mm)
-- ++(0.00mm,0.80mm)
(0.47mm,2.62mm)
-- ++(0.00mm,0.80mm)
(0.47mm,3.42mm)
-- ++(0.00mm,0.80mm)
(0.47mm,4.22mm)
-- ++(0.00mm,0.80mm)
(0.47mm,5.02mm)
-- ++(0.00mm,0.80mm)
(0.47mm,5.82mm)
-- ++(0.00mm,0.80mm)
(0.47mm,6.62mm)
-- ++(0.00mm,0.80mm)
(0.47mm,7.42mm)
-- ++(0.00mm,0.80mm)
(0.47mm,8.22mm)
-- ++(0.00mm,0.80mm)
(0.47mm,9.02mm)
-- ++(0.00mm,0.80mm)
(0.47mm,9.82mm)
-- ++(0.00mm,0.80mm)
(0.47mm,10.62mm)
-- ++(0.00mm,0.80mm)
(0.47mm,11.42mm)
-- ++(0.00mm,0.80mm)
(0.47mm,12.22mm)
-- ++(0.00mm,0.80mm)
(0.47mm,13.02mm)
-- ++(0.00mm,0.80mm)
(9.13mm,10.62mm)
-- ++(4.13mm,0.00mm)
(9.23mm,9.02mm)
-- ++(4.04mm,0.00mm)
(10.66mm,5.82mm)
-- ++(2.61mm,0.00mm)
(11.16mm,4.22mm)
-- ++(2.11mm,0.00mm)
(11.94mm,2.62mm)
-- ++(1.33mm,0.00mm)
(9.79mm,7.42mm)
-- ++(3.48mm,0.00mm)
(9.12mm,12.22mm)
-- ++(4.14mm,0.00mm)
(2.66mm,12.22mm)
-- ++(4.21mm,0.00mm)
(2.66mm,10.62mm)
-- ++(4.20mm,0.00mm)
(2.67mm,7.42mm)
-- ++(4.20mm,0.00mm)
(2.66mm,4.22mm)
-- ++(4.21mm,0.00mm)
(13.27mm,9.02mm)
-- ++(0.00mm,-0.80mm)
(6.87mm,12.22mm)
-- ++(0.53mm,0.00mm)
(6.87mm,10.62mm)
-- ++(0.54mm,0.00mm)
(6.87mm,9.02mm)
-- ++(0.63mm,0.00mm)
(6.87mm,7.42mm)
-- ++(1.19mm,0.00mm)
(6.87mm,5.82mm)
-- ++(2.06mm,0.00mm)
(6.87mm,4.22mm)
-- ++(2.57mm,0.00mm)
(6.87mm,2.62mm)
-- ++(3.34mm,0.00mm)
(1.02mm,12.06mm)
-- ++(0.00mm,0.16mm)
(2.66mm,12.06mm)
-- ++(0.00mm,0.16mm)
(1.03mm,10.46mm)
-- ++(0.00mm,0.16mm)
(1.02mm,9.18mm)
-- ++(0.00mm,-0.16mm)
(1.03mm,7.26mm)
-- ++(0.00mm,0.16mm)
(1.02mm,5.66mm)
-- ++(0.00mm,0.16mm)
(1.02mm,4.06mm)
-- ++(0.00mm,0.16mm)
(1.03mm,2.46mm)
-- ++(0.00mm,0.16mm)
(2.67mm,2.78mm)
-- ++(0.00mm,-0.16mm)
(2.66mm,4.38mm)
-- ++(0.00mm,-0.16mm)
(2.66mm,5.98mm)
-- ++(0.00mm,-0.16mm)
(2.67mm,7.58mm)
-- ++(0.00mm,-0.16mm)
(2.66mm,9.18mm)
-- ++(0.00mm,-0.16mm)
(2.66mm,10.78mm)
-- ++(0.00mm,-0.16mm)
(7.40mm,12.38mm)
-- ++(0.00mm,-0.16mm)
(9.12mm,12.38mm)
-- ++(0.00mm,-0.16mm)
(7.40mm,10.78mm)
-- ++(0.00mm,-0.16mm)
(9.13mm,10.78mm)
-- ++(0.00mm,-0.16mm)
(7.50mm,9.18mm)
-- ++(0.00mm,-0.16mm)
(9.23mm,9.18mm)
-- ++(0.00mm,-0.16mm)
(8.06mm,7.58mm)
-- ++(0.00mm,-0.16mm)
(9.79mm,7.58mm)
-- ++(0.00mm,-0.16mm)
(8.93mm,5.98mm)
-- ++(0.00mm,-0.16mm)
(10.66mm,5.98mm)
-- ++(0.00mm,-0.16mm)
(9.43mm,4.38mm)
-- ++(0.00mm,-0.16mm)
(11.16mm,4.38mm)
-- ++(0.00mm,-0.16mm)
(10.21mm,2.78mm)
-- ++(0.00mm,-0.16mm)
(11.94mm,2.78mm)
-- ++(0.00mm,-0.16mm)
(1.02mm,12.22mm)
-- ++(0.00mm,0.16mm)
(2.66mm,12.22mm)
-- ++(0.00mm,0.16mm)
(2.66mm,10.62mm)
-- ++(0.00mm,-0.16mm)
(1.03mm,10.62mm)
-- ++(0.00mm,0.16mm)
(2.66mm,9.02mm)
-- ++(0.00mm,-0.16mm)
(1.02mm,9.02mm)
-- ++(0.00mm,-0.16mm)
(1.03mm,7.42mm)
-- ++(0.00mm,0.16mm)
(2.67mm,7.42mm)
-- ++(0.00mm,-0.16mm)
(1.02mm,5.82mm)
-- ++(0.00mm,0.16mm)
(2.66mm,5.82mm)
-- ++(0.00mm,-0.16mm)
(2.66mm,4.22mm)
-- ++(0.00mm,-0.16mm)
(1.02mm,4.22mm)
-- ++(0.00mm,0.16mm)
(1.03mm,2.62mm)
-- ++(0.00mm,0.16mm)
(2.67mm,2.62mm)
-- ++(0.00mm,-0.16mm)
(11.94mm,2.62mm)
-- ++(0.00mm,-0.16mm)
(10.21mm,2.62mm)
-- ++(0.00mm,-0.16mm)
(11.16mm,4.22mm)
-- ++(0.00mm,-0.16mm)
(9.43mm,4.22mm)
-- ++(0.00mm,-0.16mm)
(10.66mm,5.82mm)
-- ++(0.00mm,-0.16mm)
(8.93mm,5.82mm)
-- ++(0.00mm,-0.16mm)
(8.06mm,7.42mm)
-- ++(0.00mm,-0.16mm)
(9.79mm,7.42mm)
-- ++(0.00mm,-0.16mm)
(7.50mm,9.02mm)
-- ++(0.00mm,-0.16mm)
(9.23mm,9.02mm)
-- ++(0.00mm,-0.16mm)
(7.40mm,10.62mm)
-- ++(0.00mm,-0.16mm)
(9.13mm,10.62mm)
-- ++(0.00mm,-0.16mm)
(9.12mm,12.22mm)
-- ++(0.00mm,-0.16mm)
(7.40mm,12.22mm)
-- ++(0.00mm,-0.16mm)
;
\end{scope}
\end{tikzpicture} &
	\begin{tikzpicture}[tikzMazeScale]
\begin{scope}[cm={1,0,0,-1,(0.00mm,13.94mm)}]
\draw [tikzMazeStyle]  (0.47mm,13.82mm)
-- ++(12.80mm,0.00mm)
(13.27mm,13.82mm)
-- ++(0.00mm,-0.80mm)
(13.27mm,1.02mm)
-- ++(-0.56mm,0.00mm)
(0.47mm,1.02mm)
-- ++(0.00mm,0.80mm)
(0.47mm,7.42mm)
-- ++(2.21mm,0.00mm)
(0.47mm,10.62mm)
-- ++(0.57mm,0.00mm)
(0.47mm,12.22mm)
-- ++(4.14mm,0.00mm)
(0.47mm,9.02mm)
-- ++(2.89mm,0.00mm)
(0.47mm,4.22mm)
-- ++(1.19mm,0.00mm)
(0.47mm,5.82mm)
-- ++(1.70mm,0.00mm)
(0.47mm,2.62mm)
-- ++(0.77mm,0.00mm)
(4.99mm,9.02mm)
-- ++(1.88mm,0.00mm)
(3.80mm,5.82mm)
-- ++(3.07mm,0.00mm)
(2.87mm,2.62mm)
-- ++(4.00mm,0.00mm)
(1.03mm,1.02mm)
-- ++(-0.56mm,0.00mm)
(11.35mm,1.02mm)
-- ++(-8.96mm,0.00mm)
(6.87mm,12.22mm)
-- ++(0.00mm,-0.80mm)
(6.87mm,11.42mm)
-- ++(0.00mm,-0.80mm)
(6.87mm,10.62mm)
-- ++(0.00mm,-0.80mm)
(6.87mm,9.82mm)
-- ++(0.00mm,-0.80mm)
(6.87mm,9.02mm)
-- ++(0.00mm,-0.80mm)
(6.87mm,8.22mm)
-- ++(0.00mm,-0.80mm)
(6.87mm,7.42mm)
-- ++(0.00mm,-0.80mm)
(6.87mm,6.62mm)
-- ++(0.00mm,-0.80mm)
(6.87mm,5.82mm)
-- ++(0.00mm,-0.80mm)
(6.87mm,5.02mm)
-- ++(0.00mm,-0.80mm)
(6.87mm,4.22mm)
-- ++(0.00mm,-0.80mm)
(6.87mm,3.42mm)
-- ++(0.00mm,-0.80mm)
(6.87mm,2.62mm)
-- ++(0.00mm,-0.80mm)
(6.87mm,1.82mm)
-- ++(0.00mm,-0.80mm)
(13.27mm,1.82mm)
-- ++(0.00mm,-0.80mm)
(13.27mm,2.62mm)
-- ++(0.00mm,-0.80mm)
(13.27mm,3.42mm)
-- ++(0.00mm,-0.80mm)
(13.27mm,4.22mm)
-- ++(0.00mm,-0.80mm)
(13.27mm,5.02mm)
-- ++(0.00mm,-0.80mm)
(13.27mm,5.82mm)
-- ++(0.00mm,-0.80mm)
(13.27mm,6.62mm)
-- ++(0.00mm,-0.80mm)
(13.27mm,7.42mm)
-- ++(0.00mm,-0.80mm)
(13.27mm,8.22mm)
-- ++(0.00mm,-0.80mm)
(13.27mm,9.82mm)
-- ++(0.00mm,-0.80mm)
(13.27mm,10.62mm)
-- ++(0.00mm,-0.80mm)
(13.27mm,11.42mm)
-- ++(0.00mm,-0.80mm)
(13.27mm,12.22mm)
-- ++(0.00mm,-0.80mm)
(13.27mm,13.02mm)
-- ++(0.00mm,-0.80mm)
(0.47mm,1.82mm)
-- ++(0.00mm,0.80mm)
(0.47mm,2.62mm)
-- ++(0.00mm,0.80mm)
(0.47mm,3.42mm)
-- ++(0.00mm,0.80mm)
(0.47mm,4.22mm)
-- ++(0.00mm,0.80mm)
(0.47mm,5.02mm)
-- ++(0.00mm,0.80mm)
(0.47mm,5.82mm)
-- ++(0.00mm,0.80mm)
(0.47mm,6.62mm)
-- ++(0.00mm,0.80mm)
(0.47mm,7.42mm)
-- ++(0.00mm,0.80mm)
(0.47mm,8.22mm)
-- ++(0.00mm,0.80mm)
(0.47mm,9.02mm)
-- ++(0.00mm,0.80mm)
(0.47mm,9.82mm)
-- ++(0.00mm,0.80mm)
(0.47mm,10.62mm)
-- ++(0.00mm,0.80mm)
(0.47mm,11.42mm)
-- ++(0.00mm,0.80mm)
(0.47mm,12.22mm)
-- ++(0.00mm,0.80mm)
(0.47mm,13.02mm)
-- ++(0.00mm,0.80mm)
(10.23mm,10.62mm)
-- ++(3.03mm,0.00mm)
(10.86mm,9.02mm)
-- ++(2.41mm,0.00mm)
(11.69mm,5.82mm)
-- ++(1.57mm,0.00mm)
(12.21mm,4.22mm)
-- ++(1.05mm,0.00mm)
(12.71mm,2.62mm)
-- ++(0.56mm,0.00mm)
(11.35mm,7.42mm)
-- ++(1.91mm,0.00mm)
(9.76mm,12.22mm)
-- ++(3.51mm,0.00mm)
(6.24mm,12.22mm)
-- ++(0.62mm,0.00mm)
(2.67mm,10.62mm)
-- ++(4.20mm,0.00mm)
(4.31mm,7.42mm)
-- ++(2.55mm,0.00mm)
(3.29mm,4.22mm)
-- ++(3.58mm,0.00mm)
(13.27mm,9.02mm)
-- ++(0.00mm,-0.80mm)
(6.87mm,12.22mm)
-- ++(1.16mm,0.00mm)
(6.87mm,10.62mm)
-- ++(1.64mm,0.00mm)
(6.87mm,9.02mm)
-- ++(2.26mm,0.00mm)
(6.87mm,7.42mm)
-- ++(2.75mm,0.00mm)
(6.87mm,5.82mm)
-- ++(3.10mm,0.00mm)
(6.87mm,4.22mm)
-- ++(3.62mm,0.00mm)
(6.87mm,2.62mm)
-- ++(4.12mm,0.00mm)
(4.61mm,12.06mm)
-- ++(0.00mm,0.16mm)
(6.24mm,12.06mm)
-- ++(0.00mm,0.16mm)
(1.03mm,10.46mm)
-- ++(0.00mm,0.16mm)
(3.35mm,9.18mm)
-- ++(0.00mm,-0.16mm)
(2.68mm,7.26mm)
-- ++(0.00mm,0.16mm)
(2.16mm,5.66mm)
-- ++(0.00mm,0.16mm)
(1.65mm,4.06mm)
-- ++(0.00mm,0.16mm)
(1.23mm,2.46mm)
-- ++(0.00mm,0.16mm)
(2.87mm,2.78mm)
-- ++(0.00mm,-0.16mm)
(3.29mm,4.38mm)
-- ++(0.00mm,-0.16mm)
(3.80mm,5.98mm)
-- ++(0.00mm,-0.16mm)
(4.31mm,7.58mm)
-- ++(0.00mm,-0.16mm)
(4.99mm,9.18mm)
-- ++(0.00mm,-0.16mm)
(2.67mm,10.78mm)
-- ++(0.00mm,-0.16mm)
(8.03mm,12.38mm)
-- ++(0.00mm,-0.16mm)
(9.76mm,12.38mm)
-- ++(0.00mm,-0.16mm)
(8.51mm,10.78mm)
-- ++(0.00mm,-0.16mm)
(10.23mm,10.78mm)
-- ++(0.00mm,-0.16mm)
(9.13mm,9.18mm)
-- ++(0.00mm,-0.16mm)
(10.86mm,9.18mm)
-- ++(0.00mm,-0.16mm)
(9.62mm,7.58mm)
-- ++(0.00mm,-0.16mm)
(11.35mm,7.58mm)
-- ++(0.00mm,-0.16mm)
(9.96mm,5.98mm)
-- ++(0.00mm,-0.16mm)
(11.69mm,5.98mm)
-- ++(0.00mm,-0.16mm)
(10.48mm,4.38mm)
-- ++(0.00mm,-0.16mm)
(12.21mm,4.38mm)
-- ++(0.00mm,-0.16mm)
(10.98mm,2.78mm)
-- ++(0.00mm,-0.16mm)
(12.71mm,2.78mm)
-- ++(0.00mm,-0.16mm)
(4.61mm,12.22mm)
-- ++(0.00mm,0.16mm)
(6.24mm,12.22mm)
-- ++(0.00mm,0.16mm)
(2.67mm,10.62mm)
-- ++(0.00mm,-0.16mm)
(1.03mm,10.62mm)
-- ++(0.00mm,0.16mm)
(4.99mm,9.02mm)
-- ++(0.00mm,-0.16mm)
(3.35mm,9.02mm)
-- ++(0.00mm,-0.16mm)
(2.68mm,7.42mm)
-- ++(0.00mm,0.16mm)
(4.31mm,7.42mm)
-- ++(0.00mm,-0.16mm)
(2.16mm,5.82mm)
-- ++(0.00mm,0.16mm)
(3.80mm,5.82mm)
-- ++(0.00mm,-0.16mm)
(3.29mm,4.22mm)
-- ++(0.00mm,-0.16mm)
(1.65mm,4.22mm)
-- ++(0.00mm,0.16mm)
(1.23mm,2.62mm)
-- ++(0.00mm,0.16mm)
(2.87mm,2.62mm)
-- ++(0.00mm,-0.16mm)
(12.71mm,2.62mm)
-- ++(0.00mm,-0.16mm)
(10.98mm,2.62mm)
-- ++(0.00mm,-0.16mm)
(12.21mm,4.22mm)
-- ++(0.00mm,-0.16mm)
(10.48mm,4.22mm)
-- ++(0.00mm,-0.16mm)
(11.69mm,5.82mm)
-- ++(0.00mm,-0.16mm)
(9.96mm,5.82mm)
-- ++(0.00mm,-0.16mm)
(9.62mm,7.42mm)
-- ++(0.00mm,-0.16mm)
(11.35mm,7.42mm)
-- ++(0.00mm,-0.16mm)
(9.13mm,9.02mm)
-- ++(0.00mm,-0.16mm)
(10.86mm,9.02mm)
-- ++(0.00mm,-0.16mm)
(8.51mm,10.62mm)
-- ++(0.00mm,-0.16mm)
(10.23mm,10.62mm)
-- ++(0.00mm,-0.16mm)
(9.76mm,12.22mm)
-- ++(0.00mm,-0.16mm)
(8.03mm,12.22mm)
-- ++(0.00mm,-0.16mm)
;
\end{scope}
\end{tikzpicture} &
	\begin{tikzpicture}[tikzMazeScale]
\begin{scope}[cm={1,0,0,-1,(0.00mm,13.94mm)}]
\draw [tikzMazeStyle]  (0.47mm,13.82mm)
-- ++(12.80mm,0.00mm)
(13.27mm,13.82mm)
-- ++(0.00mm,-0.80mm)
(13.27mm,1.02mm)
-- ++(-0.56mm,0.00mm)
(0.47mm,1.02mm)
-- ++(0.00mm,0.80mm)
(0.47mm,7.42mm)
-- ++(4.28mm,0.00mm)
(0.47mm,10.62mm)
-- ++(4.30mm,0.00mm)
(0.47mm,12.22mm)
-- ++(0.56mm,0.00mm)
(0.47mm,9.02mm)
-- ++(0.56mm,0.00mm)
(0.47mm,4.22mm)
-- ++(4.23mm,0.00mm)
(0.47mm,5.82mm)
-- ++(1.37mm,0.00mm)
(0.47mm,2.62mm)
-- ++(4.23mm,0.00mm)
(2.66mm,9.02mm)
-- ++(4.21mm,0.00mm)
(3.47mm,5.82mm)
-- ++(3.40mm,0.00mm)
(6.33mm,2.62mm)
-- ++(0.54mm,0.00mm)
(1.03mm,1.02mm)
-- ++(-0.56mm,0.00mm)
(11.35mm,1.02mm)
-- ++(-8.96mm,0.00mm)
(6.87mm,12.22mm)
-- ++(0.00mm,-0.80mm)
(6.87mm,11.42mm)
-- ++(0.00mm,-0.80mm)
(6.87mm,10.62mm)
-- ++(0.00mm,-0.80mm)
(6.87mm,9.82mm)
-- ++(0.00mm,-0.80mm)
(6.87mm,9.02mm)
-- ++(0.00mm,-0.80mm)
(6.87mm,8.22mm)
-- ++(0.00mm,-0.80mm)
(6.87mm,7.42mm)
-- ++(0.00mm,-0.80mm)
(6.87mm,6.62mm)
-- ++(0.00mm,-0.80mm)
(6.87mm,5.82mm)
-- ++(0.00mm,-0.80mm)
(6.87mm,5.02mm)
-- ++(0.00mm,-0.80mm)
(6.87mm,4.22mm)
-- ++(0.00mm,-0.80mm)
(6.87mm,3.42mm)
-- ++(0.00mm,-0.80mm)
(6.87mm,2.62mm)
-- ++(0.00mm,-0.80mm)
(6.87mm,1.82mm)
-- ++(0.00mm,-0.80mm)
(13.27mm,1.82mm)
-- ++(0.00mm,-0.80mm)
(13.27mm,2.62mm)
-- ++(0.00mm,-0.80mm)
(13.27mm,3.42mm)
-- ++(0.00mm,-0.80mm)
(13.27mm,4.22mm)
-- ++(0.00mm,-0.80mm)
(13.27mm,5.02mm)
-- ++(0.00mm,-0.80mm)
(13.27mm,5.82mm)
-- ++(0.00mm,-0.80mm)
(13.27mm,6.62mm)
-- ++(0.00mm,-0.80mm)
(13.27mm,7.42mm)
-- ++(0.00mm,-0.80mm)
(13.27mm,8.22mm)
-- ++(0.00mm,-0.80mm)
(13.27mm,9.82mm)
-- ++(0.00mm,-0.80mm)
(13.27mm,10.62mm)
-- ++(0.00mm,-0.80mm)
(13.27mm,11.42mm)
-- ++(0.00mm,-0.80mm)
(13.27mm,12.22mm)
-- ++(0.00mm,-0.80mm)
(13.27mm,13.02mm)
-- ++(0.00mm,-0.80mm)
(0.47mm,1.82mm)
-- ++(0.00mm,0.80mm)
(0.47mm,2.62mm)
-- ++(0.00mm,0.80mm)
(0.47mm,3.42mm)
-- ++(0.00mm,0.80mm)
(0.47mm,4.22mm)
-- ++(0.00mm,0.80mm)
(0.47mm,5.02mm)
-- ++(0.00mm,0.80mm)
(0.47mm,5.82mm)
-- ++(0.00mm,0.80mm)
(0.47mm,6.62mm)
-- ++(0.00mm,0.80mm)
(0.47mm,7.42mm)
-- ++(0.00mm,0.80mm)
(0.47mm,8.22mm)
-- ++(0.00mm,0.80mm)
(0.47mm,9.02mm)
-- ++(0.00mm,0.80mm)
(0.47mm,9.82mm)
-- ++(0.00mm,0.80mm)
(0.47mm,10.62mm)
-- ++(0.00mm,0.80mm)
(0.47mm,11.42mm)
-- ++(0.00mm,0.80mm)
(0.47mm,12.22mm)
-- ++(0.00mm,0.80mm)
(0.47mm,13.02mm)
-- ++(0.00mm,0.80mm)
(12.80mm,10.62mm)
-- ++(0.46mm,0.00mm)
(9.12mm,9.02mm)
-- ++(4.14mm,0.00mm)
(9.13mm,5.82mm)
-- ++(4.13mm,0.00mm)
(9.13mm,4.22mm)
-- ++(4.14mm,0.00mm)
(9.13mm,2.62mm)
-- ++(4.14mm,0.00mm)
(9.14mm,7.42mm)
-- ++(4.12mm,0.00mm)
(9.12mm,12.22mm)
-- ++(4.14mm,0.00mm)
(2.66mm,12.22mm)
-- ++(4.21mm,0.00mm)
(6.40mm,10.62mm)
-- ++(0.47mm,0.00mm)
(6.38mm,7.42mm)
-- ++(0.49mm,0.00mm)
(6.33mm,4.22mm)
-- ++(0.54mm,0.00mm)
(13.27mm,9.02mm)
-- ++(0.00mm,-0.80mm)
(6.87mm,12.22mm)
-- ++(0.53mm,0.00mm)
(6.87mm,10.62mm)
-- ++(4.21mm,0.00mm)
(6.87mm,9.02mm)
-- ++(0.53mm,0.00mm)
(6.87mm,7.42mm)
-- ++(0.54mm,0.00mm)
(6.87mm,5.82mm)
-- ++(0.54mm,0.00mm)
(6.87mm,4.22mm)
-- ++(0.54mm,0.00mm)
(6.87mm,2.62mm)
-- ++(0.53mm,0.00mm)
(1.02mm,12.06mm)
-- ++(0.00mm,0.16mm)
(2.66mm,12.06mm)
-- ++(0.00mm,0.16mm)
(4.77mm,10.46mm)
-- ++(0.00mm,0.16mm)
(1.02mm,9.18mm)
-- ++(0.00mm,-0.16mm)
(4.74mm,7.26mm)
-- ++(0.00mm,0.16mm)
(1.83mm,5.66mm)
-- ++(0.00mm,0.16mm)
(4.70mm,4.06mm)
-- ++(0.00mm,0.16mm)
(4.70mm,2.46mm)
-- ++(0.00mm,0.16mm)
(6.33mm,2.78mm)
-- ++(0.00mm,-0.16mm)
(6.33mm,4.38mm)
-- ++(0.00mm,-0.16mm)
(3.47mm,5.98mm)
-- ++(0.00mm,-0.16mm)
(6.38mm,7.58mm)
-- ++(0.00mm,-0.16mm)
(2.66mm,9.18mm)
-- ++(0.00mm,-0.16mm)
(6.40mm,10.78mm)
-- ++(0.00mm,-0.16mm)
(7.40mm,12.38mm)
-- ++(0.00mm,-0.16mm)
(9.12mm,12.38mm)
-- ++(0.00mm,-0.16mm)
(11.07mm,10.78mm)
-- ++(0.00mm,-0.16mm)
(12.80mm,10.78mm)
-- ++(0.00mm,-0.16mm)
(7.40mm,9.18mm)
-- ++(0.00mm,-0.16mm)
(9.12mm,9.18mm)
-- ++(0.00mm,-0.16mm)
(7.41mm,7.58mm)
-- ++(0.00mm,-0.16mm)
(9.14mm,7.58mm)
-- ++(0.00mm,-0.16mm)
(7.41mm,5.98mm)
-- ++(0.00mm,-0.16mm)
(9.13mm,5.98mm)
-- ++(0.00mm,-0.16mm)
(7.40mm,4.38mm)
-- ++(0.00mm,-0.16mm)
(9.13mm,4.38mm)
-- ++(0.00mm,-0.16mm)
(7.40mm,2.78mm)
-- ++(0.00mm,-0.16mm)
(9.13mm,2.78mm)
-- ++(0.00mm,-0.16mm)
(1.02mm,12.22mm)
-- ++(0.00mm,0.16mm)
(2.66mm,12.22mm)
-- ++(0.00mm,0.16mm)
(6.40mm,10.62mm)
-- ++(0.00mm,-0.16mm)
(4.77mm,10.62mm)
-- ++(0.00mm,0.16mm)
(2.66mm,9.02mm)
-- ++(0.00mm,-0.16mm)
(1.02mm,9.02mm)
-- ++(0.00mm,-0.16mm)
(4.74mm,7.42mm)
-- ++(0.00mm,0.16mm)
(6.38mm,7.42mm)
-- ++(0.00mm,-0.16mm)
(1.83mm,5.82mm)
-- ++(0.00mm,0.16mm)
(3.47mm,5.82mm)
-- ++(0.00mm,-0.16mm)
(6.33mm,4.22mm)
-- ++(0.00mm,-0.16mm)
(4.70mm,4.22mm)
-- ++(0.00mm,0.16mm)
(4.70mm,2.62mm)
-- ++(0.00mm,0.16mm)
(6.33mm,2.62mm)
-- ++(0.00mm,-0.16mm)
(9.13mm,2.62mm)
-- ++(0.00mm,-0.16mm)
(7.40mm,2.62mm)
-- ++(0.00mm,-0.16mm)
(9.13mm,4.22mm)
-- ++(0.00mm,-0.16mm)
(7.40mm,4.22mm)
-- ++(0.00mm,-0.16mm)
(9.13mm,5.82mm)
-- ++(0.00mm,-0.16mm)
(7.41mm,5.82mm)
-- ++(0.00mm,-0.16mm)
(7.41mm,7.42mm)
-- ++(0.00mm,-0.16mm)
(9.14mm,7.42mm)
-- ++(0.00mm,-0.16mm)
(7.40mm,9.02mm)
-- ++(0.00mm,-0.16mm)
(9.12mm,9.02mm)
-- ++(0.00mm,-0.16mm)
(11.07mm,10.62mm)
-- ++(0.00mm,-0.16mm)
(12.80mm,10.62mm)
-- ++(0.00mm,-0.16mm)
(9.12mm,12.22mm)
-- ++(0.00mm,-0.16mm)
(7.40mm,12.22mm)
-- ++(0.00mm,-0.16mm)
;
\end{scope}
\end{tikzpicture} &
	\begin{tikzpicture}[tikzMazeScale]
\begin{scope}[cm={1,0,0,-1,(0.00mm,13.94mm)}]
\draw [tikzMazeStyle]  (0.47mm,13.82mm)
-- ++(12.80mm,0.00mm)
(13.27mm,13.82mm)
-- ++(0.00mm,-0.80mm)
(13.27mm,1.02mm)
-- ++(-0.56mm,0.00mm)
(0.47mm,1.02mm)
-- ++(0.00mm,0.80mm)
(0.47mm,7.42mm)
-- ++(0.56mm,0.00mm)
(0.47mm,10.62mm)
-- ++(4.30mm,0.00mm)
(0.47mm,12.22mm)
-- ++(0.56mm,0.00mm)
(0.47mm,9.02mm)
-- ++(4.26mm,0.00mm)
(0.47mm,4.22mm)
-- ++(0.56mm,0.00mm)
(0.47mm,5.82mm)
-- ++(0.56mm,0.00mm)
(0.47mm,2.62mm)
-- ++(0.56mm,0.00mm)
(6.36mm,9.02mm)
-- ++(0.51mm,0.00mm)
(2.66mm,5.82mm)
-- ++(4.21mm,0.00mm)
(2.66mm,2.62mm)
-- ++(4.21mm,0.00mm)
(1.03mm,1.02mm)
-- ++(-0.56mm,0.00mm)
(11.35mm,1.02mm)
-- ++(-8.96mm,0.00mm)
(6.87mm,12.22mm)
-- ++(0.00mm,-0.80mm)
(6.87mm,11.42mm)
-- ++(0.00mm,-0.80mm)
(6.87mm,10.62mm)
-- ++(0.00mm,-0.80mm)
(6.87mm,9.82mm)
-- ++(0.00mm,-0.80mm)
(6.87mm,9.02mm)
-- ++(0.00mm,-0.80mm)
(6.87mm,8.22mm)
-- ++(0.00mm,-0.80mm)
(6.87mm,7.42mm)
-- ++(0.00mm,-0.80mm)
(6.87mm,6.62mm)
-- ++(0.00mm,-0.80mm)
(6.87mm,5.82mm)
-- ++(0.00mm,-0.80mm)
(6.87mm,5.02mm)
-- ++(0.00mm,-0.80mm)
(6.87mm,4.22mm)
-- ++(0.00mm,-0.80mm)
(6.87mm,3.42mm)
-- ++(0.00mm,-0.80mm)
(6.87mm,2.62mm)
-- ++(0.00mm,-0.80mm)
(6.87mm,1.82mm)
-- ++(0.00mm,-0.80mm)
(13.27mm,1.82mm)
-- ++(0.00mm,-0.80mm)
(13.27mm,2.62mm)
-- ++(0.00mm,-0.80mm)
(13.27mm,3.42mm)
-- ++(0.00mm,-0.80mm)
(13.27mm,4.22mm)
-- ++(0.00mm,-0.80mm)
(13.27mm,5.02mm)
-- ++(0.00mm,-0.80mm)
(13.27mm,5.82mm)
-- ++(0.00mm,-0.80mm)
(13.27mm,6.62mm)
-- ++(0.00mm,-0.80mm)
(13.27mm,7.42mm)
-- ++(0.00mm,-0.80mm)
(13.27mm,8.22mm)
-- ++(0.00mm,-0.80mm)
(13.27mm,9.82mm)
-- ++(0.00mm,-0.80mm)
(13.27mm,10.62mm)
-- ++(0.00mm,-0.80mm)
(13.27mm,11.42mm)
-- ++(0.00mm,-0.80mm)
(13.27mm,12.22mm)
-- ++(0.00mm,-0.80mm)
(13.27mm,13.02mm)
-- ++(0.00mm,-0.80mm)
(0.47mm,1.82mm)
-- ++(0.00mm,0.80mm)
(0.47mm,2.62mm)
-- ++(0.00mm,0.80mm)
(0.47mm,3.42mm)
-- ++(0.00mm,0.80mm)
(0.47mm,4.22mm)
-- ++(0.00mm,0.80mm)
(0.47mm,5.02mm)
-- ++(0.00mm,0.80mm)
(0.47mm,5.82mm)
-- ++(0.00mm,0.80mm)
(0.47mm,6.62mm)
-- ++(0.00mm,0.80mm)
(0.47mm,7.42mm)
-- ++(0.00mm,0.80mm)
(0.47mm,8.22mm)
-- ++(0.00mm,0.80mm)
(0.47mm,9.02mm)
-- ++(0.00mm,0.80mm)
(0.47mm,9.82mm)
-- ++(0.00mm,0.80mm)
(0.47mm,10.62mm)
-- ++(0.00mm,0.80mm)
(0.47mm,11.42mm)
-- ++(0.00mm,0.80mm)
(0.47mm,12.22mm)
-- ++(0.00mm,0.80mm)
(0.47mm,13.02mm)
-- ++(0.00mm,0.80mm)
(12.80mm,10.62mm)
-- ++(0.46mm,0.00mm)
(9.12mm,9.02mm)
-- ++(4.14mm,0.00mm)
(12.78mm,5.82mm)
-- ++(0.48mm,0.00mm)
(12.78mm,4.22mm)
-- ++(0.48mm,0.00mm)
(9.12mm,2.62mm)
-- ++(4.14mm,0.00mm)
(12.73mm,7.42mm)
-- ++(0.53mm,0.00mm)
(12.80mm,12.22mm)
-- ++(0.46mm,0.00mm)
(2.66mm,12.22mm)
-- ++(4.21mm,0.00mm)
(6.41mm,10.62mm)
-- ++(0.46mm,0.00mm)
(2.66mm,7.42mm)
-- ++(4.21mm,0.00mm)
(2.66mm,4.22mm)
-- ++(4.21mm,0.00mm)
(13.27mm,9.02mm)
-- ++(0.00mm,-0.80mm)
(6.87mm,12.22mm)
-- ++(4.21mm,0.00mm)
(6.87mm,10.62mm)
-- ++(4.21mm,0.00mm)
(6.87mm,9.02mm)
-- ++(0.53mm,0.00mm)
(6.87mm,7.42mm)
-- ++(4.14mm,0.00mm)
(6.87mm,5.82mm)
-- ++(4.19mm,0.00mm)
(6.87mm,4.22mm)
-- ++(4.19mm,0.00mm)
(6.87mm,2.62mm)
-- ++(0.53mm,0.00mm)
(1.02mm,12.06mm)
-- ++(0.00mm,0.16mm)
(2.66mm,12.06mm)
-- ++(0.00mm,0.16mm)
(4.77mm,10.46mm)
-- ++(0.00mm,0.16mm)
(4.72mm,9.18mm)
-- ++(0.00mm,-0.16mm)
(1.02mm,7.26mm)
-- ++(0.00mm,0.16mm)
(1.02mm,5.66mm)
-- ++(0.00mm,0.16mm)
(1.02mm,4.06mm)
-- ++(0.00mm,0.16mm)
(1.02mm,2.46mm)
-- ++(0.00mm,0.16mm)
(2.66mm,2.78mm)
-- ++(0.00mm,-0.16mm)
(2.66mm,4.38mm)
-- ++(0.00mm,-0.16mm)
(2.66mm,5.98mm)
-- ++(0.00mm,-0.16mm)
(2.66mm,7.58mm)
-- ++(0.00mm,-0.16mm)
(6.36mm,9.18mm)
-- ++(0.00mm,-0.16mm)
(6.41mm,10.78mm)
-- ++(0.00mm,-0.16mm)
(11.08mm,12.38mm)
-- ++(0.00mm,-0.16mm)
(12.80mm,12.38mm)
-- ++(0.00mm,-0.16mm)
(11.08mm,10.78mm)
-- ++(0.00mm,-0.16mm)
(12.80mm,10.78mm)
-- ++(0.00mm,-0.16mm)
(7.40mm,9.18mm)
-- ++(0.00mm,-0.16mm)
(9.12mm,9.18mm)
-- ++(0.00mm,-0.16mm)
(11.01mm,7.58mm)
-- ++(0.00mm,-0.16mm)
(12.73mm,7.58mm)
-- ++(0.00mm,-0.16mm)
(11.05mm,5.98mm)
-- ++(0.00mm,-0.16mm)
(12.78mm,5.98mm)
-- ++(0.00mm,-0.16mm)
(11.05mm,4.38mm)
-- ++(0.00mm,-0.16mm)
(12.78mm,4.38mm)
-- ++(0.00mm,-0.16mm)
(7.40mm,2.78mm)
-- ++(0.00mm,-0.16mm)
(9.12mm,2.78mm)
-- ++(0.00mm,-0.16mm)
(1.02mm,12.22mm)
-- ++(0.00mm,0.16mm)
(2.66mm,12.22mm)
-- ++(0.00mm,0.16mm)
(6.41mm,10.62mm)
-- ++(0.00mm,-0.16mm)
(4.77mm,10.62mm)
-- ++(0.00mm,0.16mm)
(6.36mm,9.02mm)
-- ++(0.00mm,-0.16mm)
(4.72mm,9.02mm)
-- ++(0.00mm,-0.16mm)
(1.02mm,7.42mm)
-- ++(0.00mm,0.16mm)
(2.66mm,7.42mm)
-- ++(0.00mm,-0.16mm)
(1.02mm,5.82mm)
-- ++(0.00mm,0.16mm)
(2.66mm,5.82mm)
-- ++(0.00mm,-0.16mm)
(2.66mm,4.22mm)
-- ++(0.00mm,-0.16mm)
(1.02mm,4.22mm)
-- ++(0.00mm,0.16mm)
(1.02mm,2.62mm)
-- ++(0.00mm,0.16mm)
(2.66mm,2.62mm)
-- ++(0.00mm,-0.16mm)
(9.12mm,2.62mm)
-- ++(0.00mm,-0.16mm)
(7.40mm,2.62mm)
-- ++(0.00mm,-0.16mm)
(12.78mm,4.22mm)
-- ++(0.00mm,-0.16mm)
(11.05mm,4.22mm)
-- ++(0.00mm,-0.16mm)
(12.78mm,5.82mm)
-- ++(0.00mm,-0.16mm)
(11.05mm,5.82mm)
-- ++(0.00mm,-0.16mm)
(11.01mm,7.42mm)
-- ++(0.00mm,-0.16mm)
(12.73mm,7.42mm)
-- ++(0.00mm,-0.16mm)
(7.40mm,9.02mm)
-- ++(0.00mm,-0.16mm)
(9.12mm,9.02mm)
-- ++(0.00mm,-0.16mm)
(11.08mm,10.62mm)
-- ++(0.00mm,-0.16mm)
(12.80mm,10.62mm)
-- ++(0.00mm,-0.16mm)
(12.80mm,12.22mm)
-- ++(0.00mm,-0.16mm)
(11.08mm,12.22mm)
-- ++(0.00mm,-0.16mm)
;
\end{scope}
\end{tikzpicture} &
	\begin{tikzpicture}[tikzMazeScale]
\begin{scope}[cm={1,0,0,-1,(0.00mm,13.94mm)}]
\draw [tikzMazeStyle]  (0.47mm,13.82mm)
-- ++(12.80mm,0.00mm)
(13.27mm,13.82mm)
-- ++(0.00mm,-0.80mm)
(13.27mm,1.02mm)
-- ++(-0.56mm,0.00mm)
(0.47mm,1.02mm)
-- ++(0.00mm,0.80mm)
(0.47mm,7.42mm)
-- ++(4.28mm,0.00mm)
(0.47mm,10.62mm)
-- ++(0.56mm,0.00mm)
(0.47mm,12.22mm)
-- ++(4.30mm,0.00mm)
(0.47mm,9.02mm)
-- ++(0.56mm,0.00mm)
(0.47mm,4.22mm)
-- ++(4.23mm,0.00mm)
(0.47mm,5.82mm)
-- ++(0.56mm,0.00mm)
(0.47mm,2.62mm)
-- ++(4.23mm,0.00mm)
(2.66mm,9.02mm)
-- ++(4.21mm,0.00mm)
(2.66mm,5.82mm)
-- ++(4.21mm,0.00mm)
(6.34mm,2.62mm)
-- ++(0.53mm,0.00mm)
(1.03mm,1.02mm)
-- ++(-0.56mm,0.00mm)
(11.35mm,1.02mm)
-- ++(-8.96mm,0.00mm)
(6.87mm,12.22mm)
-- ++(0.00mm,-0.80mm)
(6.87mm,11.42mm)
-- ++(0.00mm,-0.80mm)
(6.87mm,10.62mm)
-- ++(0.00mm,-0.80mm)
(6.87mm,9.82mm)
-- ++(0.00mm,-0.80mm)
(6.87mm,9.02mm)
-- ++(0.00mm,-0.80mm)
(6.87mm,8.22mm)
-- ++(0.00mm,-0.80mm)
(6.87mm,7.42mm)
-- ++(0.00mm,-0.80mm)
(6.87mm,6.62mm)
-- ++(0.00mm,-0.80mm)
(6.87mm,5.82mm)
-- ++(0.00mm,-0.80mm)
(6.87mm,5.02mm)
-- ++(0.00mm,-0.80mm)
(6.87mm,4.22mm)
-- ++(0.00mm,-0.80mm)
(6.87mm,3.42mm)
-- ++(0.00mm,-0.80mm)
(6.87mm,2.62mm)
-- ++(0.00mm,-0.80mm)
(6.87mm,1.82mm)
-- ++(0.00mm,-0.80mm)
(13.27mm,1.82mm)
-- ++(0.00mm,-0.80mm)
(13.27mm,2.62mm)
-- ++(0.00mm,-0.80mm)
(13.27mm,3.42mm)
-- ++(0.00mm,-0.80mm)
(13.27mm,4.22mm)
-- ++(0.00mm,-0.80mm)
(13.27mm,5.02mm)
-- ++(0.00mm,-0.80mm)
(13.27mm,5.82mm)
-- ++(0.00mm,-0.80mm)
(13.27mm,6.62mm)
-- ++(0.00mm,-0.80mm)
(13.27mm,7.42mm)
-- ++(0.00mm,-0.80mm)
(13.27mm,8.22mm)
-- ++(0.00mm,-0.80mm)
(13.27mm,9.82mm)
-- ++(0.00mm,-0.80mm)
(13.27mm,10.62mm)
-- ++(0.00mm,-0.80mm)
(13.27mm,11.42mm)
-- ++(0.00mm,-0.80mm)
(13.27mm,12.22mm)
-- ++(0.00mm,-0.80mm)
(13.27mm,13.02mm)
-- ++(0.00mm,-0.80mm)
(0.47mm,1.82mm)
-- ++(0.00mm,0.80mm)
(0.47mm,2.62mm)
-- ++(0.00mm,0.80mm)
(0.47mm,3.42mm)
-- ++(0.00mm,0.80mm)
(0.47mm,4.22mm)
-- ++(0.00mm,0.80mm)
(0.47mm,5.02mm)
-- ++(0.00mm,0.80mm)
(0.47mm,5.82mm)
-- ++(0.00mm,0.80mm)
(0.47mm,6.62mm)
-- ++(0.00mm,0.80mm)
(0.47mm,7.42mm)
-- ++(0.00mm,0.80mm)
(0.47mm,8.22mm)
-- ++(0.00mm,0.80mm)
(0.47mm,9.02mm)
-- ++(0.00mm,0.80mm)
(0.47mm,9.82mm)
-- ++(0.00mm,0.80mm)
(0.47mm,10.62mm)
-- ++(0.00mm,0.80mm)
(0.47mm,11.42mm)
-- ++(0.00mm,0.80mm)
(0.47mm,12.22mm)
-- ++(0.00mm,0.80mm)
(0.47mm,13.02mm)
-- ++(0.00mm,0.80mm)
(12.80mm,10.62mm)
-- ++(0.46mm,0.00mm)
(12.76mm,9.02mm)
-- ++(0.51mm,0.00mm)
(9.12mm,5.82mm)
-- ++(4.14mm,0.00mm)
(12.78mm,4.22mm)
-- ++(0.48mm,0.00mm)
(9.12mm,2.62mm)
-- ++(4.14mm,0.00mm)
(12.73mm,7.42mm)
-- ++(0.53mm,0.00mm)
(12.80mm,12.22mm)
-- ++(0.46mm,0.00mm)
(6.41mm,12.22mm)
-- ++(0.46mm,0.00mm)
(2.66mm,10.62mm)
-- ++(4.21mm,0.00mm)
(6.38mm,7.42mm)
-- ++(0.48mm,0.00mm)
(6.34mm,4.22mm)
-- ++(0.53mm,0.00mm)
(13.27mm,9.02mm)
-- ++(0.00mm,-0.80mm)
(6.87mm,12.22mm)
-- ++(4.21mm,0.00mm)
(6.87mm,10.62mm)
-- ++(4.21mm,0.00mm)
(6.87mm,9.02mm)
-- ++(4.16mm,0.00mm)
(6.87mm,7.42mm)
-- ++(4.14mm,0.00mm)
(6.87mm,5.82mm)
-- ++(0.53mm,0.00mm)
(6.87mm,4.22mm)
-- ++(4.19mm,0.00mm)
(6.87mm,2.62mm)
-- ++(0.53mm,0.00mm)
(4.77mm,12.06mm)
-- ++(0.00mm,0.16mm)
(6.41mm,12.06mm)
-- ++(0.00mm,0.16mm)
(1.02mm,10.46mm)
-- ++(0.00mm,0.16mm)
(1.02mm,9.18mm)
-- ++(0.00mm,-0.16mm)
(4.75mm,7.26mm)
-- ++(0.00mm,0.16mm)
(1.02mm,5.66mm)
-- ++(0.00mm,0.16mm)
(4.70mm,4.06mm)
-- ++(0.00mm,0.16mm)
(4.70mm,2.46mm)
-- ++(0.00mm,0.16mm)
(6.34mm,2.78mm)
-- ++(0.00mm,-0.16mm)
(6.34mm,4.38mm)
-- ++(0.00mm,-0.16mm)
(2.66mm,5.98mm)
-- ++(0.00mm,-0.16mm)
(6.38mm,7.58mm)
-- ++(0.00mm,-0.16mm)
(2.66mm,9.18mm)
-- ++(0.00mm,-0.16mm)
(2.66mm,10.78mm)
-- ++(0.00mm,-0.16mm)
(11.08mm,12.38mm)
-- ++(0.00mm,-0.16mm)
(12.80mm,12.38mm)
-- ++(0.00mm,-0.16mm)
(11.08mm,10.78mm)
-- ++(0.00mm,-0.16mm)
(12.80mm,10.78mm)
-- ++(0.00mm,-0.16mm)
(11.03mm,9.18mm)
-- ++(0.00mm,-0.16mm)
(12.76mm,9.18mm)
-- ++(0.00mm,-0.16mm)
(11.01mm,7.58mm)
-- ++(0.00mm,-0.16mm)
(12.73mm,7.58mm)
-- ++(0.00mm,-0.16mm)
(7.40mm,5.98mm)
-- ++(0.00mm,-0.16mm)
(9.12mm,5.98mm)
-- ++(0.00mm,-0.16mm)
(11.05mm,4.38mm)
-- ++(0.00mm,-0.16mm)
(12.78mm,4.38mm)
-- ++(0.00mm,-0.16mm)
(7.40mm,2.78mm)
-- ++(0.00mm,-0.16mm)
(9.12mm,2.78mm)
-- ++(0.00mm,-0.16mm)
(4.77mm,12.22mm)
-- ++(0.00mm,0.16mm)
(6.41mm,12.22mm)
-- ++(0.00mm,0.16mm)
(2.66mm,10.62mm)
-- ++(0.00mm,-0.16mm)
(1.02mm,10.62mm)
-- ++(0.00mm,0.16mm)
(2.66mm,9.02mm)
-- ++(0.00mm,-0.16mm)
(1.02mm,9.02mm)
-- ++(0.00mm,-0.16mm)
(4.75mm,7.42mm)
-- ++(0.00mm,0.16mm)
(6.38mm,7.42mm)
-- ++(0.00mm,-0.16mm)
(1.02mm,5.82mm)
-- ++(0.00mm,0.16mm)
(2.66mm,5.82mm)
-- ++(0.00mm,-0.16mm)
(6.34mm,4.22mm)
-- ++(0.00mm,-0.16mm)
(4.70mm,4.22mm)
-- ++(0.00mm,0.16mm)
(4.70mm,2.62mm)
-- ++(0.00mm,0.16mm)
(6.34mm,2.62mm)
-- ++(0.00mm,-0.16mm)
(9.12mm,2.62mm)
-- ++(0.00mm,-0.16mm)
(7.40mm,2.62mm)
-- ++(0.00mm,-0.16mm)
(12.78mm,4.22mm)
-- ++(0.00mm,-0.16mm)
(11.05mm,4.22mm)
-- ++(0.00mm,-0.16mm)
(9.12mm,5.82mm)
-- ++(0.00mm,-0.16mm)
(7.40mm,5.82mm)
-- ++(0.00mm,-0.16mm)
(11.01mm,7.42mm)
-- ++(0.00mm,-0.16mm)
(12.73mm,7.42mm)
-- ++(0.00mm,-0.16mm)
(11.03mm,9.02mm)
-- ++(0.00mm,-0.16mm)
(12.76mm,9.02mm)
-- ++(0.00mm,-0.16mm)
(11.08mm,10.62mm)
-- ++(0.00mm,-0.16mm)
(12.80mm,10.62mm)
-- ++(0.00mm,-0.16mm)
(12.80mm,12.22mm)
-- ++(0.00mm,-0.16mm)
(11.08mm,12.22mm)
-- ++(0.00mm,-0.16mm)
;
\end{scope}
\end{tikzpicture} \\
	\includegraphics[width=0.13\linewidth]{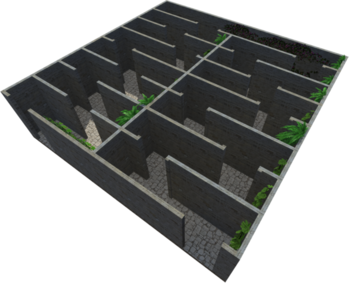} &
	\includegraphics[width=0.13\linewidth]{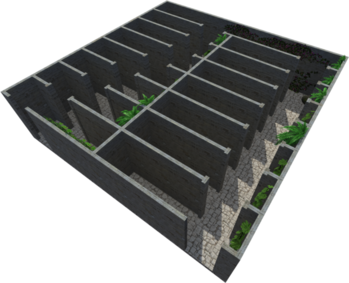} &
	\includegraphics[width=0.13\linewidth]{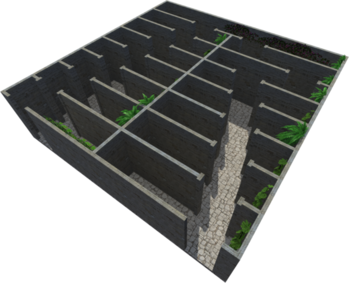} &
	\includegraphics[width=0.13\linewidth]{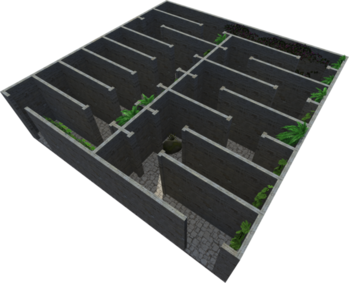} &
	\includegraphics[width=0.13\linewidth]{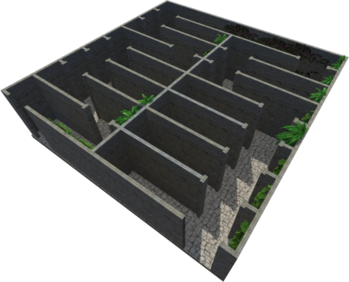} &
	\includegraphics[width=0.13\linewidth]{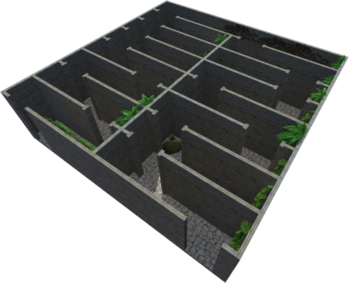} \\
	\includegraphics[width=0.13\linewidth]{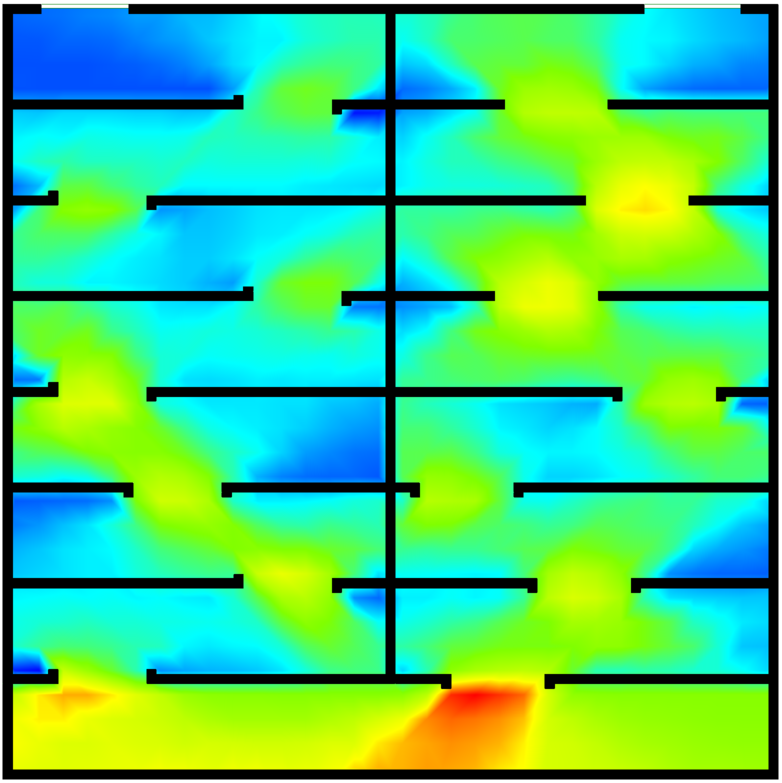} &
	\includegraphics[width=0.13\linewidth]{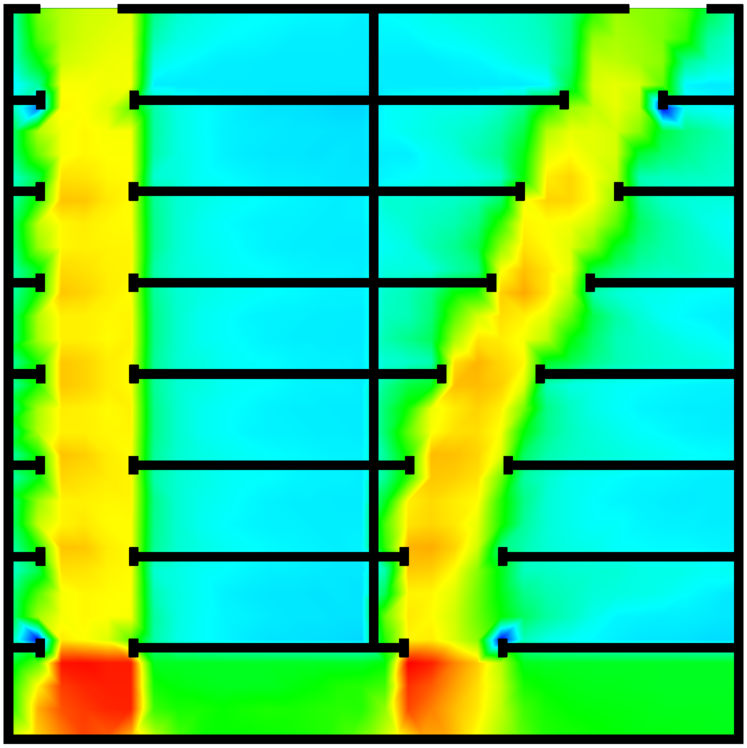} &
	\includegraphics[width=0.13\linewidth]{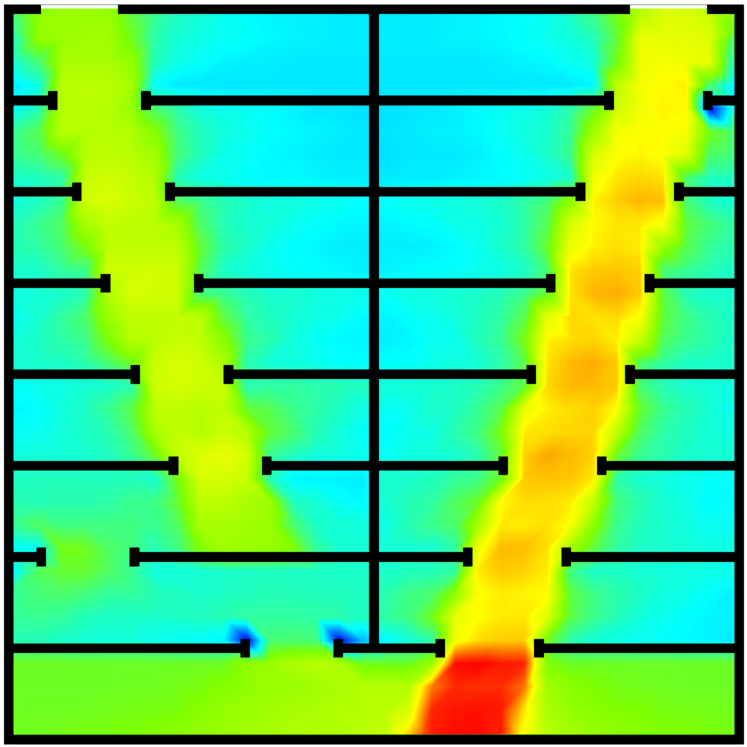} &
	\includegraphics[width=0.13\linewidth]{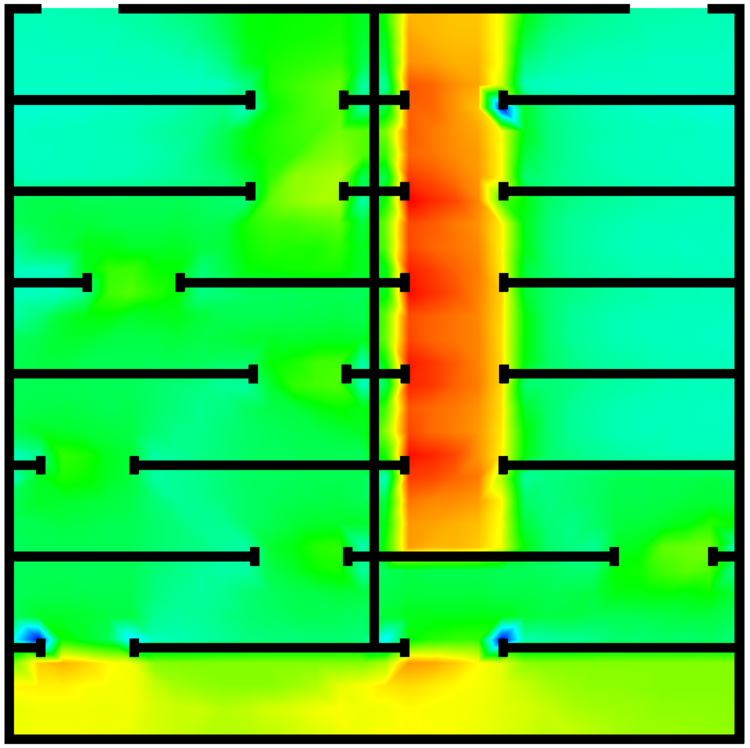} &
	\includegraphics[width=0.13\linewidth]{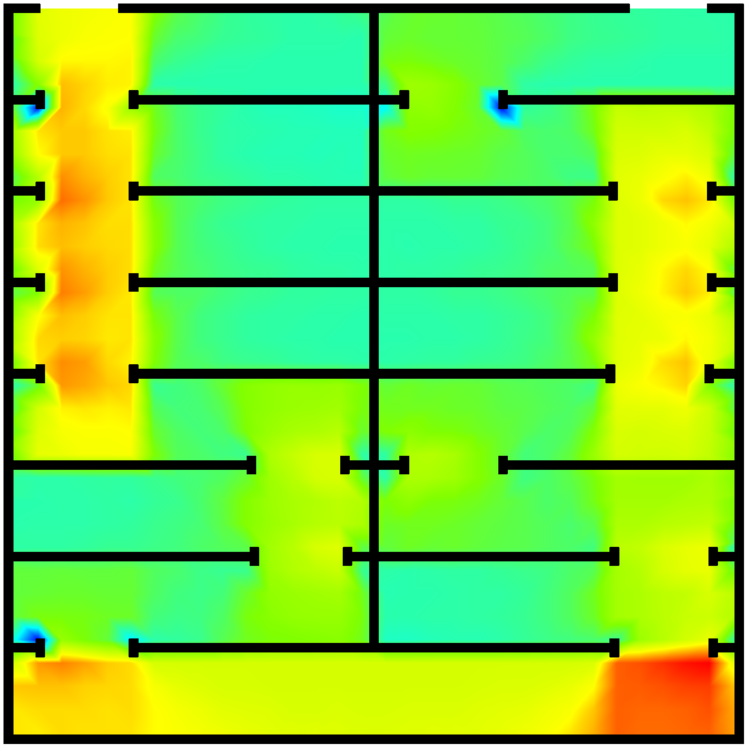} &
	\includegraphics[width=0.13\linewidth]{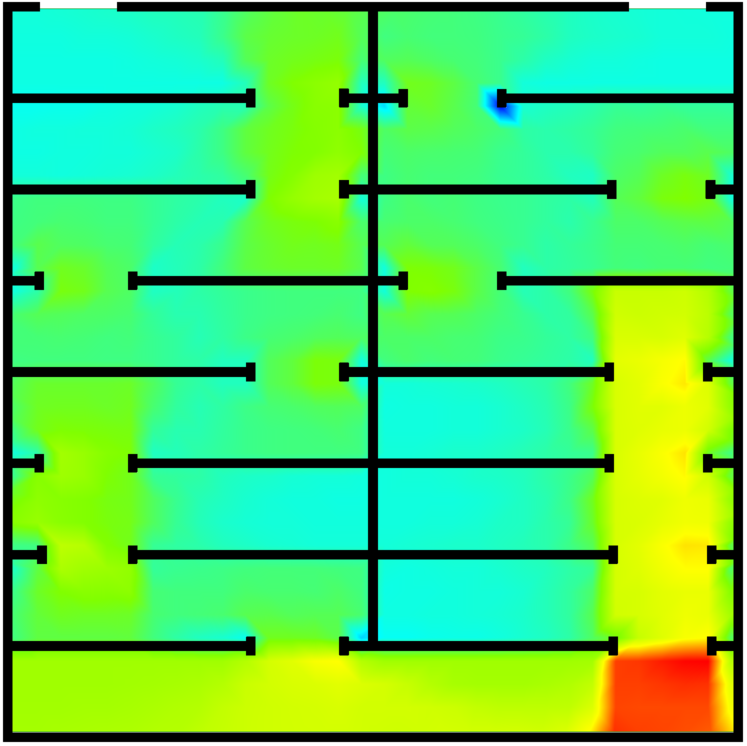} \\
	(a) & (b) & (c) & (d) & (e) & (f) \\

\end{tabular}
\caption{\label{figure:maze-example} 
 \newChanges{This figure has been enhanced with 3D renders and the 2D combined metric heatmaps to better demonstrate certain features of \DOME.}{The top row of figures are blueprints for maze-like designs. The middle row is the rendered environment. The final row of figures shows the combined metric values. (a) Initial maze configuration, with query region (pink) and reference region (gray). (b,c) Optimized maze to reduce environment complexity. (d,e,f), The result from (b) is optimized to increase complexity to produce new mazes. Maze size is approximately 20x20m.}
}
\end{figure*}

\noindent \textbf{Maze.} Interestingly \DOME can  be used to alter the complexity of environments. \reffig{figure:maze-example} illustrates this approach on a maze-like environment. Starting with a standard maze we maximize the visibility, minimize the depth (which maximizes accessibility), and minimize the entropy (which maximizes order). The resulting diverse set of layouts align the doorways to minimize long-windy passageways which have high depth (b,c). The more ordered environments (b) is then fed back into the system, with the objective measure inverted to minimize $-\objective{\cdot}$. The resulting diverse layouts (d,e,f) are of similarly complexity to the original maze, thus providing variations of the original design. Our method is able to automatically remove or introduce complexity in an environment, by altering the objective definition, while producing several diverse designs that meet user-defined criteria.

\section{User Studies}
\label{sec:user-study}
A series of user studies were conducted to assess the usability and design task performance of the \DOME system. Participants were invited to a two part study session. Their first task was to complete an unstructured usability experiment in which they make use of \DOME. The participants were given the opportunity to rate the general usability of \DOME as an assisted user-in-the-loop method. Participants were then randomly assigned to one of three experimental groups, including a control group, to assess the general performance of assisted and unassisted design tools.

For this experiment, $18$ subjects volunteered to participate and gave informed consent. The participants were between the ages of $23$ to $30$ and self identified as $11$ males and $7$ females. All participants were graduate level students in computer science or a closely related field.

\subsection{Usability}
\label{sec:exp-usability}
The goal of this experiment is to evaluate the \DOME method, of automated optimization with high value diverse candidate solutions, as explored from a user perspective. 

\noindent \textbf{Materials and Methods:}
All participants interacted with the method on Desktop PC (Windows 7 64-bit, 8GB RAM, AMD FX(tm)-8320, 8 Computer Cores, 3.5GHz). Using a simple room as a teaching tool, the participants are given short instructions on how to manipulate and set parameter bounds for translation and rotations of environment elements. Participants are then shown how to select candidates from the diversity set. 

The colloquial terms for the set of metrics are explained in general terms, i.e. Degree, Tree Depth, and Entropy are translated to visibility, accessibility, and organization respectively. Since these terms are unfamiliar to novice users outside the field of computational architectural analysis, the task description included simple language with details. The metrics were rephrased as visibility, accessibility, and organization according to previous interpretations~\cite{turner}. Participants were told that: visibility related to how visible any portion of the environment may be to another; accessibility related to how accessible the environment is; and organization related to how confusing the layout of the environment would be. 

After initiation, participants are presented with a complex real world Art Gallery environment in which the ROIs, \roq and \ror, are already defined. The participants are tasked with increasing the visibility, accessibility, and organization of the environment using \DOME for a fixed amount of time ($15$ minutes).  At the end of this task, participants are immediately given a System Usability Scale (SUS) to measure usability of the system~\cite{brooke2013sus,brooke1996sus}. The SUS is a well established and tested method for evaluating the usability of a product.

\noindent \textbf{Results:}
The summary statistics of the SUS scores are reported in Table~\reftab{tab:susmean}. The quartiles for the SUS scores are reported in~\reftab{tab:susq}.

\begin{table}[h]
	\centering
	\begin{tabular}{c | c | c | c}
		Count & Mean & Median & Standard Deviation \\ [0.5ex]
		\toprule
		18 & 70.83 & 73.75 & 14.70 \\
		\bottomrule
	\end{tabular}
	\caption{\label{tab:susmean}Summary statistics for SUS results, where the score range is from 0 to 100.} 
\end{table}

\begin{table}[h]
	\centering
	\begin{tabular}{c | c}
		Quartile & Range \\ [0.5ex]
		\toprule
		$< Q_{1}$ & 22.5 - 68.12 \\
		$[Q_{1}, Q_{2}]$ & 68.12 - 73.75 \\
		$[Q_{2}, Q_{3}]$ & 73.75 - 76.87 \\
		$> Q_{3}$ & 76.87 - 87.5 \\
		$IQR$ & 8.75 \\
		\bottomrule
	\end{tabular}
	\caption{SUS quartile ranges. The ranges for each quartile of the data are reported to show distribution of the results.  The Interquartile Range ($IQR$) is also reported.}
	\label{tab:susq}
\end{table}

\noindent \textbf{Discussion:}	

The SUS score is a composite measure of usability for a system which has been tested on a variety of tasks and proved to be robust and reliable~\cite{sauro2011designing}. As well, a particular advantage of SUS is the ability to provide a reliable measure of usability with as few as $8$ to $12$ participants~\cite{tullis2004comparison}. It has been found that SUS in fact measures two factors: both ``usability" and ``learnability"~\cite{lewis2009factor,borsci2009dimensionality} of a system. SUS scores are scaled to the range of $0$ and $100$, with $68$ being the average score taken over many tasks from different domains with scores above $68$ considered above average and acceptable~\cite{sauro2011designing}. A mapping of scaled SUS scores to common adjectives, based on responses from many participants on several tasks across different domains, provides an intuitive interpretation for each score range~\cite{bangor2009determining}.

The results show that the $18$ participants mean and median scores fall within the adjective range of ``good" and `excellent"~\cite{bangor2009determining}. Furthermore, quartile ranges $> Q_{2}$ show a strong preference for a high SUS score. This can be interpreted as meaning the \DOME system is highly usable and ``learnable" with a degree of confidence.

\subsection{Design Performance}
\label{exp:user-performance}
In this experiment, the effective performance of the method with respect to real world use is evaluated. The hypothesis is that \DOME is better, in terms of objective metric values and efficiency, than manual unassisted design approaches as well as a version of \DOME that does not provide the diversity set. A secondary hypothesis is that, as the complexity of the environment increases, the value of automated optimization and diverse candidate suggestions increases.

\noindent \textbf{Materials and Methods:}
\label{exp:user-performance-mat-meth}
This experiment takes the form of an A-B-C group design wherein the participant pool is divided in to thirds and randomly assigned one of three tools. These groups are provided architectural design tools as follow: the \textbf{A} group is given the unassisted tool (standard Autodesk Revit interface); the \textbf{B} group is given a tool which exposes the optimization portion of \DOME providing only the single most optimal candidate without the diversity set; and the \textbf{C} group is given the full assisted user-in-the-loop \DOME method with diversity. 

Participants are given two different environments. The first is a simple room of an art gallery with three parametrizable walls of the same dimensions. The second is a more complex art gallery with two sides that both have an asymmetric set of parametrizable objects (four square pillars, and three walls) and are connected by a small hallway at the centre.

For each environment, the participant was tasked with improving the metrics, as described in~\refsec{sec:exp-usability}. The participants were given up to $10$ minutes, per environment, to make as many adjustments as they wish.  The participant may finish at any point within the $10$ minutes, concluding their design when satisfied.

\begin{figure}[h]
	\centering
	\includegraphics[trim={4.0cm 8.5cm 4.5cm 8.5cm},clip, width=\linewidth]{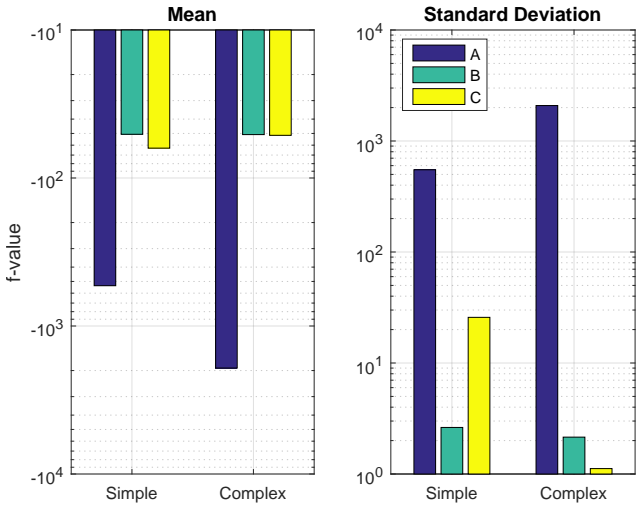}
	\caption{Comparison between participants in the (A) manual group, (B) assisted without diversity, and (c) assisted with diversity (C) by \DOME. \DOME participants designed layouts with significantly higher objective measures on average with have lesser deviation, indicating greater consistency across users. The mean and variance are calculated over f-values as specified in Equation~\ref{eq:scalarObjective}.}
	\label{figure:abcstudy}
\end{figure}

\noindent \textbf{Results:}
The mean and standard deviation of the objective values, Equation~\ref{eq:scalarObjective}, are shown in~\reffig{figure:abcstudy}. The mean number of design iterations made by participants in the \DOME group (B) was: $3.0 \pm 0.63$ for the simple environment; and $3.83 \pm 1.47$ for the complex environment.  The mean number of design iterations made by participants in the \DOME group (C) was: $2.83 \pm 1.72$ for the simple environment; and $2.5 \pm 1.22$ for the complex environment. 

\noindent \textbf{Discussion:}	
The results show that generally, for groups (B) and (C), participants who were given access to optimal results performed better in terms of the objective, than those who made their designs manually~\reffig{figure:abcstudy}. Group (C), who had access to the diversity set, performed on par with those participants in group (B), who were only given the optimal result. In summary, the objective value data shows that using \DOME results in producing environments with much higher objective values.

The number of design iterations performed before participants decided they were satisfied and submitted their work is less when using the full \DOME method. These results indicate that providing solution diversity is helpful, especially in the case of increased environment complexity. Furthermore, as the scenario, and thus task, grows in complexity, diversity becomes more valuable.

It is also noteworthy that the variance in the complex environment is significantly lower when using \DOME with the full diversity set. As well, the group A results show a significantly larger standard deviation compared to B and C. These results suggest that manual optimization can be very inconsistent among different users, while using our system can effectively guide the user and keep the design exploration more focused. Furthermore, this could be a sign that using diversity helps avoid local minima in the design space.

However, it is important to note that solutions returned by group (C), the \DOME users, were still quite diverse, with different users finding new ways to maximize the objective, even for these simple layouts.

\subsection{Expert Validation}
The goal of this experiment was to validate the designs, created by novice participants, from the perspective of architectural and design experts. Experts are asked to provide their perceptual preference of design outputs from novice participant sessions using either the unassisted or assisted tools. Our hypothesis is that there is a preference for designs which are the results of the assisted tool with diverse results as opposed to the standard unassisted tools.

Seven experts in the fields of architecture, interior, and civil design participated in the expert survey. An online questionnaire with a series of binary A/B choices was provided to each participant.

The questionnaire was made up of randomized environment pairs, each with one selection from the manual design set and one from the \DOME tool design set corresponding to participant designs from groups (A) and (C) respectively - described in~\refsec{exp:user-performance}.  Each participant was asked to make a binary choice for each environment pairing based on their expert intuition for which design best fulfilled the metrics for degree of visibility, tree depth, and entropy. The task objectives and metrics described for the \textbf{A}-\textbf{B}-\textbf{C} study were provided to the experts for additional guidance.

\begin{figure}[h]
	\centering
	\includegraphics[clip, width=\linewidth]{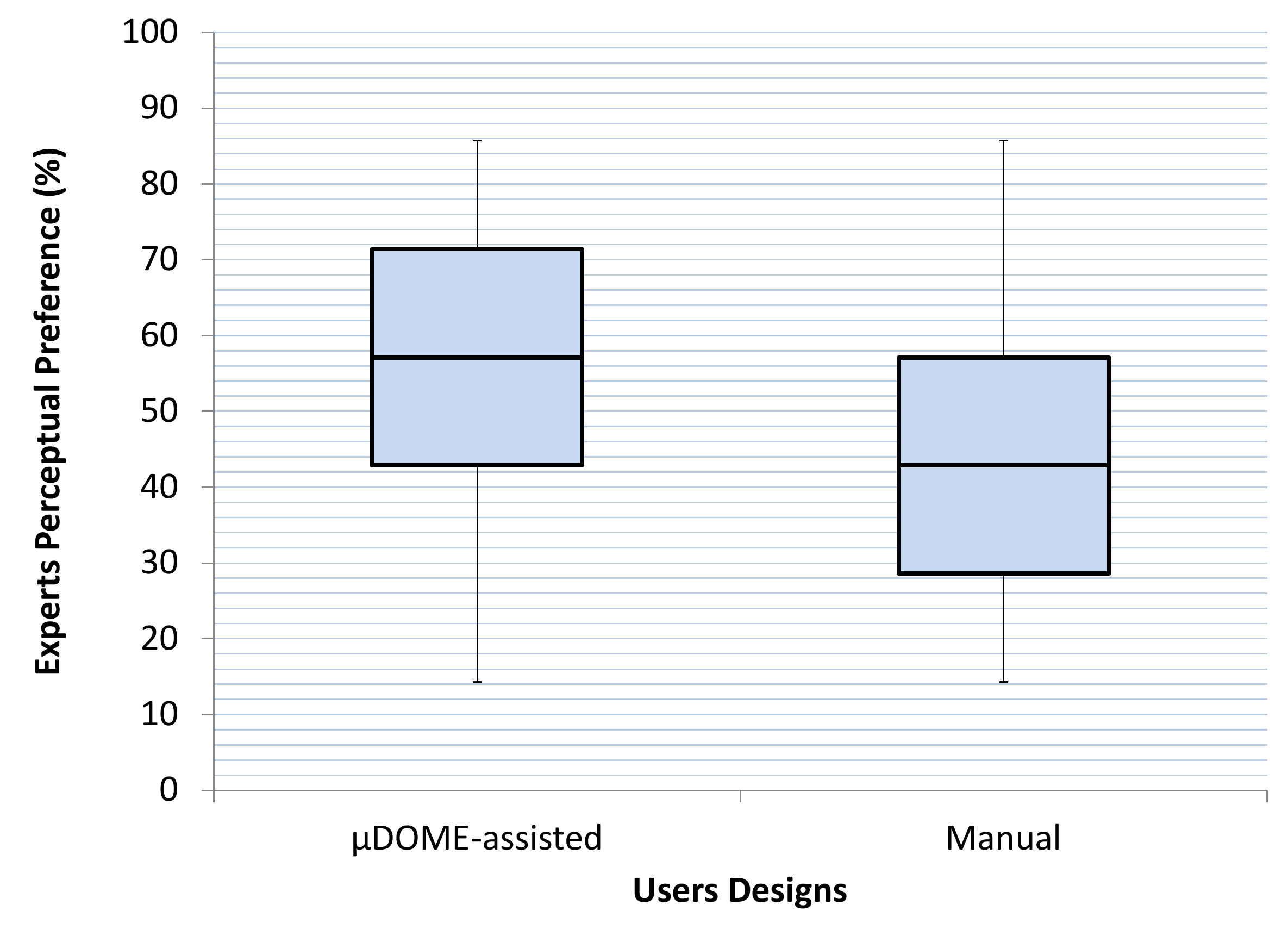}
	\caption{Distribution of experts' perceptual preferences (\%) of the environments designed by novice users.}
	\label{figure:experts-preferences-IQR}
\end{figure}

\noindent \textbf{Results:}
The Interquartile Range (IQR)~\cite{rousseeuw1992statistics} is computed and shown in~\reffig{figure:experts-preferences-IQR}. The horizontal line in the centre of the boxes indicate the Median ($57.1$ and $42.9$) and the boxes cover the Interquartile Range (Q1=$42.9$ to Q3=$71.4$ and Q1=$28.6$ to Q3=$57.1$) for the users designs from \DOME and Manual tools respectively. 

\noindent \textbf{Discussion:}	
The results show that there is high preference for \DOME designs with diverse results as opposed to the standard unassisted tools. This also reveals that \DOME guides participants to preferable design patterns, even if the designers are novices or from a non-related field.

\newChanges{This content has been added to show that \DOME is benificial to experts as well.}{A survey was given to a diverse set of experts in architectural design (N=$6$) to select a perfered design from~\reffig{figure:half-art-diversity}. There were no outlier selections, all three chosen candidates received the same number of votes, as well no experts chose the original gallery design given the context. This indicates that in the multi-objective building design space experts have there own preferences and the diversity optimization can facilitate these many preferences. The experts also agreed that an iterative non-prescriptive (ie. no single solution) approach is necessary and beneficial.}

\section{Conclusion}
\label{sec:conclusion}
We have presented \DOME, a user-in-the-loop system for computer-aided environment design that analysis and optimizes environments with respect to human behaviour. 
The user study indicates that not just design optimization but design diversity can be beneficial to the user. Results revealed that providing multiple diverse designs to the user, especially in the case of more complex environments, allows the user to find better solutions in less iterations.
By providing the user with several candidate designs rather than just analytical data or a single optimal solution, the user remains a crucial part of the process at all stages of the design. In a sense, the system enhances the creative process rather than control it.



\noindent \textbf{Limitations and Future Work.} 

Like most multi-objective optimization frameworks, our approach includes a variety of weights that the user can set to tweak the results. Although one can rely on default values, it might be beneficial for the user to adjust them. We plan to study the effects of these parameters on the resulting configurations with a large scale experiment, and attempt to identify specific relationships which might serve as guidelines.

The system is interactive for moderate scale designs. 
We have identified possibilities for improving performance, such as employing approximate and incremental algorithms for computing the spatial analysis metrics, which we plan to investigate in the future. We also want to investigate the use of more dynamic metrics, for example, crowd flow.

It is worth noting that these measures could be estimated or supplanted by crowd simulations~\cite{berseth2015environment,midlayout}. However, these methods are impractical for repeated use in an interactive applications such as ours.  Furthermore, they tend to be sensitive to the particular crowd simulator and the simulators internal parameters. 
Learning the relationship between an environment parametrization and a realistic crowd from examples is an interesting future project.

\bibliographystyle{ACM-Reference-Format}
\bibliography{paper}


\begin{thebibliography}{00}


\ifx \showCODEN    \undefined \def \showCODEN     #1{\unskip}     \fi
\ifx \showDOI      \undefined \def \showDOI       #1{{\tt DOI:}\penalty0{#1}\ }
  \fi
\ifx \showISBNx    \undefined \def \showISBNx     #1{\unskip}     \fi
\ifx \showISBNxiii \undefined \def \showISBNxiii  #1{\unskip}     \fi
\ifx \showISSN     \undefined \def \showISSN      #1{\unskip}     \fi
\ifx \showLCCN     \undefined \def \showLCCN      #1{\unskip}     \fi
\ifx \shownote     \undefined \def \shownote      #1{#1}          \fi
\ifx \showarticletitle \undefined \def \showarticletitle #1{#1}   \fi
\ifx \showURL      \undefined \def \showURL       #1{#1}          \fi
\providecommand\bibfield[2]{#2}
\providecommand\bibinfo[2]{#2}
\providecommand\natexlab[1]{#1}
\providecommand\showeprint[2][]{arXiv:#2}

\bibitem[\protect\citeauthoryear{Agrawal, Shen, and van~de Panne}{Agrawal
  et~al\mbox{.}}{2014}]%
        {6781622}
\bibfield{author}{\bibinfo{person}{Shailen Agrawal}, \bibinfo{person}{Shuo
  Shen}, {and} \bibinfo{person}{Michiel van~de Panne}.}
  \bibinfo{year}{2014}\natexlab{}.
\newblock \showarticletitle{Diverse Motions and Character Shapes for Simulated
  Skills}.
\newblock \bibinfo{journal}{{\em IEEE Transactions on Visualization and
  Computer Graphics\/}} \bibinfo{volume}{20}, \bibinfo{number}{10}
  (\bibinfo{date}{Oct} \bibinfo{year}{2014}), \bibinfo{pages}{1345--1355}.
\newblock
\showISSN{1077-2626}
\showDOI{%
\url{https://doi.org/10.1109/TVCG.2014.2314658}}


\bibitem[\protect\citeauthoryear{Akase and Okada}{Akase and Okada}{2014}]%
        {Akase:2014:WMR:2636240.2636849}
\bibfield{author}{\bibinfo{person}{Ryuya Akase} {and}
  \bibinfo{person}{Yoshihiro Okada}.} \bibinfo{year}{2014}\natexlab{}.
\newblock \showarticletitle{Web-based Multiuser 3D Room Layout System Using
  Interactive Evolutionary Computation with Conjoint Analysis}. In
  \bibinfo{booktitle}{{\em Proceedings of the 7th International Symposium on
  Visual Information Communication and Interaction}} {\em
  (\bibinfo{series}{VINCI '14})}. \bibinfo{publisher}{ACM},
  \bibinfo{address}{New York, NY, USA}, Article \bibinfo{articleno}{178},
  \bibinfo{numpages}{10}~pages.
\newblock
\showISBNx{978-1-4503-2765-7}
\showDOI{%
\url{https://doi.org/10.1145/2636240.2636849}}


\bibitem[\protect\citeauthoryear{AlHalawani and Mitra}{AlHalawani and
  Mitra}{2015}]%
        {DBLP:conf/isvc/AlHalawaniM15}
\bibfield{author}{\bibinfo{person}{Sawsan AlHalawani} {and}
  \bibinfo{person}{Niloy~J. Mitra}.} \bibinfo{year}{2015}\natexlab{}.
\newblock \showarticletitle{Congestion-Aware Warehouse Flow Analysis and
  Optimization}. In \bibinfo{booktitle}{{\em Advances in Visual Computing -
  11th International Symposium, {ISVC} 2015, December 14-16, 2015, Proceedings,
  Part {II}}}. \bibinfo{pages}{702--711}.
\newblock
\showDOI{%
\url{https://doi.org/10.1007/978-3-319-27863-6_66}}


\bibitem[\protect\citeauthoryear{AlHalawani, Yang, Wonka, and Mitra}{AlHalawani
  et~al\mbox{.}}{2014}]%
        {AlHalawani:2014:MLW:2771589.2771605}
\bibfield{author}{\bibinfo{person}{Sawsan AlHalawani},
  \bibinfo{person}{Yong-Liang Yang}, \bibinfo{person}{Peter Wonka}, {and}
  \bibinfo{person}{Niloy~J. Mitra}.} \bibinfo{year}{2014}\natexlab{}.
\newblock \showarticletitle{What Makes London Work Like London?}
\newblock \bibinfo{journal}{{\em Comput. Graph. Forum\/}} \bibinfo{volume}{33},
  \bibinfo{number}{5} (\bibinfo{date}{Aug.} \bibinfo{year}{2014}),
  \bibinfo{pages}{157--165}.
\newblock
\showISSN{0167-7055}
\showDOI{%
\url{https://doi.org/10.1111/cgf.12441}}


\bibitem[\protect\citeauthoryear{Arvin and House}{Arvin and House}{2002}]%
        {arvin2002modeling}
\bibfield{author}{\bibinfo{person}{Scott~A. Arvin} {and}
  \bibinfo{person}{Donald~H. House}.} \bibinfo{year}{2002}\natexlab{}.
\newblock \showarticletitle{Modeling architectural design objectives in
  physically based space planning}.
\newblock \bibinfo{journal}{{\em Automation in Construction\/}}
  \bibinfo{volume}{11}, \bibinfo{number}{2} (\bibinfo{year}{2002}),
  \bibinfo{pages}{213--225}.
\newblock


\bibitem[\protect\citeauthoryear{Bafna}{Bafna}{2003}]%
        {bafna2003}
\bibfield{author}{\bibinfo{person}{Sonit Bafna}.}
  \bibinfo{year}{2003}\natexlab{}.
\newblock \showarticletitle{{Space Syntax: A Brief Introduction to Its Logic
  and Analytical Techniques}}.
\newblock \bibinfo{journal}{{\em Environment and Behavior\/}}
  \bibinfo{volume}{35}, \bibinfo{number}{1} (\bibinfo{year}{2003}),
  \bibinfo{pages}{17--29}.
\newblock
\showDOI{%
\url{https://doi.org/10.1177/0013916502238863}}


\bibitem[\protect\citeauthoryear{Bangor, Kortum, and Miller}{Bangor
  et~al\mbox{.}}{2009}]%
        {bangor2009determining}
\bibfield{author}{\bibinfo{person}{Aaron Bangor}, \bibinfo{person}{Philip
  Kortum}, {and} \bibinfo{person}{James Miller}.}
  \bibinfo{year}{2009}\natexlab{}.
\newblock \showarticletitle{Determining what individual SUS scores mean: Adding
  an adjective rating scale}.
\newblock \bibinfo{journal}{{\em Journal of usability studies\/}}
  \bibinfo{volume}{4}, \bibinfo{number}{3} (\bibinfo{year}{2009}),
  \bibinfo{pages}{114--123}.
\newblock


\bibitem[\protect\citeauthoryear{Bassuet, Rife, and Dellatorre}{Bassuet
  et~al\mbox{.}}{2014}]%
        {bassuet2014computational}
\bibfield{author}{\bibinfo{person}{Alban Bassuet}, \bibinfo{person}{Dave Rife},
  {and} \bibinfo{person}{Luca Dellatorre}.} \bibinfo{year}{2014}\natexlab{}.
\newblock \showarticletitle{Computational and Optimization Design in Geometric
  Acoustics}.
\newblock \bibinfo{journal}{{\em Building Acoustics\/}} \bibinfo{volume}{21},
  \bibinfo{number}{1} (\bibinfo{year}{2014}), \bibinfo{pages}{75--86}.
\newblock


\bibitem[\protect\citeauthoryear{Berseth, Usman, Haworth, Kapadia, and
  Faloutsos}{Berseth et~al\mbox{.}}{2015}]%
        {berseth2015environment}
\bibfield{author}{\bibinfo{person}{Glen Berseth}, \bibinfo{person}{Muhammad
  Usman}, \bibinfo{person}{Brandon Haworth}, \bibinfo{person}{Mubbasir
  Kapadia}, {and} \bibinfo{person}{Petros Faloutsos}.}
  \bibinfo{year}{2015}\natexlab{}.
\newblock \showarticletitle{Environment optimization for crowd evacuation}.
\newblock \bibinfo{journal}{{\em Computer Animation and Virtual Worlds\/}}
  \bibinfo{volume}{26}, \bibinfo{number}{3--4} (\bibinfo{year}{2015}),
  \bibinfo{pages}{377--386}.
\newblock


\bibitem[\protect\citeauthoryear{Block, Knippers, Mitra, and Wang}{Block
  et~al\mbox{.}}{2014}]%
        {block2014advances}
\bibfield{author}{\bibinfo{person}{Philippe Block}, \bibinfo{person}{Jan
  Knippers}, \bibinfo{person}{Niloy~J. Mitra}, {and} \bibinfo{person}{Wenping
  Wang}.} \bibinfo{year}{2014}\natexlab{}.
\newblock \bibinfo{title}{Advances in Architectural Geometry 2014}.
\newblock   (\bibinfo{year}{2014}).
\newblock


\bibitem[\protect\citeauthoryear{Borsci, Federici, and Lauriola}{Borsci
  et~al\mbox{.}}{2009}]%
        {borsci2009dimensionality}
\bibfield{author}{\bibinfo{person}{Simone Borsci}, \bibinfo{person}{Stefano
  Federici}, {and} \bibinfo{person}{Marco Lauriola}.}
  \bibinfo{year}{2009}\natexlab{}.
\newblock \showarticletitle{On the dimensionality of the System Usability
  Scale: a test of alternative measurement models}.
\newblock \bibinfo{journal}{{\em Cognitive processing\/}} \bibinfo{volume}{10},
  \bibinfo{number}{3} (\bibinfo{year}{2009}), \bibinfo{pages}{193--197}.
\newblock


\bibitem[\protect\citeauthoryear{Brooke}{Brooke}{2013}]%
        {brooke2013sus}
\bibfield{author}{\bibinfo{person}{John Brooke}.}
  \bibinfo{year}{2013}\natexlab{}.
\newblock \showarticletitle{SUS: a retrospective}.
\newblock \bibinfo{journal}{{\em Journal of usability studies\/}}
  \bibinfo{volume}{8}, \bibinfo{number}{2} (\bibinfo{year}{2013}),
  \bibinfo{pages}{29--40}.
\newblock


\bibitem[\protect\citeauthoryear{Brooke et~al\mbox{.}}{Brooke
  et~al\mbox{.}}{1996}]%
        {brooke1996sus}
\bibfield{author}{\bibinfo{person}{John Brooke} {and}
  \bibinfo{person}{others}.} \bibinfo{year}{1996}\natexlab{}.
\newblock \showarticletitle{SUS-A quick and dirty usability scale}.
\newblock \bibinfo{journal}{{\em Usability evaluation in industry\/}}
  \bibinfo{volume}{189}, \bibinfo{number}{194} (\bibinfo{year}{1996}),
  \bibinfo{pages}{4--7}.
\newblock


\bibitem[\protect\citeauthoryear{Caldas and Norford}{Caldas and
  Norford}{2002}]%
        {caldas2002design}
\bibfield{author}{\bibinfo{person}{Luisa~Gama Caldas} {and}
  \bibinfo{person}{Leslie~K. Norford}.} \bibinfo{year}{2002}\natexlab{}.
\newblock \showarticletitle{A design optimization tool based on a genetic
  algorithm}.
\newblock \bibinfo{journal}{{\em Automation in construction\/}}
  \bibinfo{volume}{11}, \bibinfo{number}{2} (\bibinfo{year}{2002}),
  \bibinfo{pages}{173--184}.
\newblock


\bibitem[\protect\citeauthoryear{Coman and Mu{\~n}oz-Avila}{Coman and
  Mu{\~n}oz-Avila}{2011}]%
        {coman2011generating}
\bibfield{author}{\bibinfo{person}{Alexandra Coman} {and}
  \bibinfo{person}{H{\'e}ctor Mu{\~n}oz-Avila}.}
  \bibinfo{year}{2011}\natexlab{}.
\newblock \showarticletitle{Generating Diverse Plans Using Quantitative and
  Qualitative Plan Distance Metrics.}. In \bibinfo{booktitle}{{\em AAAI}}.
  Citeseer, \bibinfo{pages}{946--951}.
\newblock


\bibitem[\protect\citeauthoryear{Dara-Abrams}{Dara-Abrams}{2006}]%
        {dara2006architecture}
\bibfield{author}{\bibinfo{person}{Drew Dara-Abrams}.}
  \bibinfo{year}{2006}\natexlab{}.
\newblock \showarticletitle{Architecture of mind and world: How urban form
  influences spatial cognition}. In \bibinfo{booktitle}{{\em Proceedings of the
  Space Syntax and Spatial Cognition of the Workshop at Spatial Cognition,
  Bremen, Germany}}, Vol.~\bibinfo{volume}{24}.
\newblock


\bibitem[\protect\citeauthoryear{Davies, Mora, and Peebles}{Davies
  et~al\mbox{.}}{2006}]%
        {davies2006isovists}
\bibfield{author}{\bibinfo{person}{Clare Davies}, \bibinfo{person}{Rodrigo
  Mora}, {and} \bibinfo{person}{David Peebles}.}
  \bibinfo{year}{2006}\natexlab{}.
\newblock \showarticletitle{Isovists for Orientation: can space syntax help us
  predict directional confusion?}. In \bibinfo{booktitle}{{\em Space syntax and
  spatial cognition: Proceedings of the workshop held in Bremen}},
  Vol.~\bibinfo{volume}{2}. \bibinfo{pages}{81--92}.
\newblock


\bibitem[\protect\citeauthoryear{Desyllas and Duxbury}{Desyllas and
  Duxbury}{2001}]%
        {desyllas}
\bibfield{author}{\bibinfo{person}{Jake Desyllas} {and}
  \bibinfo{person}{Elspeth Duxbury}.} \bibinfo{year}{2001}\natexlab{}.
\newblock \showarticletitle{Axial Maps and Visibility Graph Analysis: A
  comparison of their methodology and use in models of urban pedestrian
  movement}. In \bibinfo{booktitle}{{\em 3rd International Space Syntax
  Symposium}}. \bibinfo{pages}{27.1--27.13}.
\newblock


\bibitem[\protect\citeauthoryear{Ding, Gregov, Grodzevich, Halevy, Kavazovic,
  Romanko, Seeman, Shioda, and Youbissi}{Ding et~al\mbox{.}}{2006}]%
        {ding2006discussions}
\bibfield{author}{\bibinfo{person}{Yichuan Ding}, \bibinfo{person}{Sandra
  Gregov}, \bibinfo{person}{Oleg Grodzevich}, \bibinfo{person}{Itamar Halevy},
  \bibinfo{person}{Zanin Kavazovic}, \bibinfo{person}{Oleksandr Romanko},
  \bibinfo{person}{Tamar Seeman}, \bibinfo{person}{Romy Shioda}, {and}
  \bibinfo{person}{Fabien Youbissi}.} \bibinfo{year}{2006}\natexlab{}.
\newblock \showarticletitle{Discussions on normalization and other topics in
  multiobjective optimization}. In \bibinfo{booktitle}{{\em Fields-MITACS,
  Fields Industrial Problem Solving Workshop}}.
\newblock


\bibitem[\protect\citeauthoryear{Emo, Hoelscher, Wiener, and Dalton}{Emo
  et~al\mbox{.}}{2012}]%
        {emo2012wayfinding}
\bibfield{author}{\bibinfo{person}{Beatrix Emo}, \bibinfo{person}{Christoph
  Hoelscher}, \bibinfo{person}{Jan Wiener}, {and} \bibinfo{person}{Ruth
  Dalton}.} \bibinfo{year}{2012}\natexlab{}.
\newblock \showarticletitle{Wayfinding and spatial configuration: evidence from
  street corners}.
\newblock  (\bibinfo{year}{2012}).
\newblock


\bibitem[\protect\citeauthoryear{Felkner, Chatzi, and Kotnik}{Felkner
  et~al\mbox{.}}{2013}]%
        {felkner2013interactive}
\bibfield{author}{\bibinfo{person}{Juliana Felkner}, \bibinfo{person}{Eleni
  Chatzi}, {and} \bibinfo{person}{Toni Kotnik}.}
  \bibinfo{year}{2013}\natexlab{}.
\newblock \showarticletitle{Interactive particle swarm optimization for the
  architectural design of truss structures}. In \bibinfo{booktitle}{{\em
  Computational Intelligence for Engineering Solutions (CIES), 2013 IEEE
  Symposium on}}. IEEE, \bibinfo{pages}{15--22}.
\newblock


\bibitem[\protect\citeauthoryear{Feng, Yu, Yeung, Yin, and Zhou}{Feng
  et~al\mbox{.}}{2016}]%
        {midlayout}
\bibfield{author}{\bibinfo{person}{Tian Feng}, \bibinfo{person}{Lap-Fai Yu},
  \bibinfo{person}{Sai-Kit Yeung}, \bibinfo{person}{KangKang Yin}, {and}
  \bibinfo{person}{Kun Zhou}.} \bibinfo{year}{2016}\natexlab{}.
\newblock \showarticletitle{Crowd-driven Mid-scale Layout Design}.
\newblock \bibinfo{journal}{{\em ACM Trans. Graph.\/}} \bibinfo{volume}{35},
  \bibinfo{number}{4}, Article \bibinfo{articleno}{132} (\bibinfo{date}{July}
  \bibinfo{year}{2016}), \bibinfo{numpages}{14}~pages.
\newblock
\showISSN{0730-0301}
\showDOI{%
\url{https://doi.org/10.1145/2897824.2925894}}


\bibitem[\protect\citeauthoryear{Fisher, Savva, Li, Hanrahan, and
  Niessner}{Fisher et~al\mbox{.}}{2015}]%
        {Fisher:2015:ASS:2816795.2818057}
\bibfield{author}{\bibinfo{person}{Matthew Fisher}, \bibinfo{person}{Manolis
  Savva}, \bibinfo{person}{Yangyan Li}, \bibinfo{person}{Pat Hanrahan}, {and}
  \bibinfo{person}{Matthias Niessner}.} \bibinfo{year}{2015}\natexlab{}.
\newblock \showarticletitle{Activity-centric Scene Synthesis for Functional 3D
  Scene Modeling}.
\newblock \bibinfo{journal}{{\em ACM Trans. Graph.\/}} \bibinfo{volume}{34},
  \bibinfo{number}{6}, Article \bibinfo{articleno}{179} (\bibinfo{date}{Oct.}
  \bibinfo{year}{2015}), \bibinfo{numpages}{13}~pages.
\newblock
\showISSN{0730-0301}
\showDOI{%
\url{https://doi.org/10.1145/2816795.2818057}}


\bibitem[\protect\citeauthoryear{Fruin}{Fruin}{1971}]%
        {fruin1971pedestrian}
\bibfield{author}{\bibinfo{person}{John~J Fruin}.}
  \bibinfo{year}{1971}\natexlab{}.
\newblock \bibinfo{booktitle}{{\em Pedestrian planning and design}}.
\newblock \bibinfo{type}{{T}echnical {R}eport}.
\newblock


\bibitem[\protect\citeauthoryear{Galle}{Galle}{1981}]%
        {Galle:1981:AEG:358800.358804}
\bibfield{author}{\bibinfo{person}{Per Galle}.}
  \bibinfo{year}{1981}\natexlab{}.
\newblock \showarticletitle{An Algorithm for Exhaustive Generation of Building
  Floor Plans}.
\newblock \bibinfo{journal}{{\em Commun. ACM\/}} \bibinfo{volume}{24},
  \bibinfo{number}{12} (\bibinfo{date}{Dec.} \bibinfo{year}{1981}),
  \bibinfo{pages}{813--825}.
\newblock
\showISSN{0001-0782}
\showDOI{%
\url{https://doi.org/10.1145/358800.358804}}


\bibitem[\protect\citeauthoryear{Hansen and Ostermeier}{Hansen and
  Ostermeier}{1996}]%
        {542381}
\bibfield{author}{\bibinfo{person}{Nikolaus Hansen} {and}
  \bibinfo{person}{Andreas Ostermeier}.} \bibinfo{year}{1996}\natexlab{}.
\newblock \showarticletitle{Adapting arbitrary normal mutation distributions in
  evolution strategies: the covariance matrix adaptation}. In
  \bibinfo{booktitle}{{\em IEEE International Conference on Evolutionary
  Computation}}. \bibinfo{pages}{312--317}.
\newblock


\bibitem[\protect\citeauthoryear{Harada, Witkin, and Baraff}{Harada
  et~al\mbox{.}}{1995}]%
        {Harada:1995:IPM:218380.218443}
\bibfield{author}{\bibinfo{person}{Mikako Harada}, \bibinfo{person}{Andrew
  Witkin}, {and} \bibinfo{person}{David Baraff}.}
  \bibinfo{year}{1995}\natexlab{}.
\newblock \showarticletitle{Interactive Physically-based Manipulation of
  Discrete/Continuous Models}. In \bibinfo{booktitle}{{\em Proceedings of the
  22Nd Annual Conference on Computer Graphics and Interactive Techniques}} {\em
  (\bibinfo{series}{SIGGRAPH '95})}. \bibinfo{publisher}{ACM},
  \bibinfo{address}{New York, NY, USA}, \bibinfo{pages}{199--208}.
\newblock
\showISBNx{0-89791-701-4}
\showDOI{%
\url{https://doi.org/10.1145/218380.218443}}


\bibitem[\protect\citeauthoryear{Hebrard, Hnich, O'Sullivan, and Walsh}{Hebrard
  et~al\mbox{.}}{2005}]%
        {hebrard2005finding}
\bibfield{author}{\bibinfo{person}{Emmanuel Hebrard}, \bibinfo{person}{Brahim
  Hnich}, \bibinfo{person}{Barry O'Sullivan}, {and} \bibinfo{person}{Toby
  Walsh}.} \bibinfo{year}{2005}\natexlab{}.
\newblock \showarticletitle{Finding diverse and similar solutions in constraint
  programming}. In \bibinfo{booktitle}{{\em AAAI}}, Vol.~\bibinfo{volume}{5}.
  \bibinfo{pages}{372--377}.
\newblock


\bibitem[\protect\citeauthoryear{Hillier and Hanson}{Hillier and
  Hanson}{1984}]%
        {hillier}
\bibfield{author}{\bibinfo{person}{Bill Hillier} {and}
  \bibinfo{person}{Julienne Hanson}.} \bibinfo{year}{1984}\natexlab{}.
\newblock \showarticletitle{The social logic of space, 1984}.
\newblock \bibinfo{journal}{{\em Cambridge: Press syndicate of the University
  of Cambridge\/}} (\bibinfo{year}{1984}).
\newblock


\bibitem[\protect\citeauthoryear{Hillier, Hanson, and Peponis}{Hillier
  et~al\mbox{.}}{1987}]%
        {hillier1987syntactic}
\bibfield{author}{\bibinfo{person}{WRG Hillier}, \bibinfo{person}{Julienne
  Hanson}, {and} \bibinfo{person}{John Peponis}.}
  \bibinfo{year}{1987}\natexlab{}.
\newblock \showarticletitle{Syntactic analysis of settlements}.
\newblock \bibinfo{journal}{{\em Architecture et comportement/Architecture and
  Behaviour\/}} \bibinfo{volume}{3}, \bibinfo{number}{3}
  (\bibinfo{year}{1987}), \bibinfo{pages}{217--231}.
\newblock


\bibitem[\protect\citeauthoryear{H{\"o}lscher, Meilinger, Vrachliotis,
  Br{\"o}samle, and Knauff}{H{\"o}lscher et~al\mbox{.}}{2004}]%
        {holscher}
\bibfield{author}{\bibinfo{person}{Christoph H{\"o}lscher},
  \bibinfo{person}{Tobias Meilinger}, \bibinfo{person}{Georg Vrachliotis},
  \bibinfo{person}{Martin Br{\"o}samle}, {and} \bibinfo{person}{Markus
  Knauff}.} \bibinfo{year}{2004}\natexlab{}.
\newblock \showarticletitle{Finding the way inside: Linking architectural
  design analysis and cognitive processes}.
\newblock In \bibinfo{booktitle}{{\em Spatial Cognition IV. Reasoning, Action,
  Interaction}}. \bibinfo{publisher}{Springer}, \bibinfo{pages}{1--23}.
\newblock


\bibitem[\protect\citeauthoryear{Jiang, Claramunt, and Klarqvist}{Jiang
  et~al\mbox{.}}{2000}]%
        {jiang}
\bibfield{author}{\bibinfo{person}{Bin Jiang}, \bibinfo{person}{Christophe
  Claramunt}, {and} \bibinfo{person}{Bj{\"o}rn Klarqvist}.}
  \bibinfo{year}{2000}\natexlab{}.
\newblock \showarticletitle{Integration of space syntax into GIS for modelling
  urban spaces}.
\newblock \bibinfo{journal}{{\em International Journal of Applied Earth
  Observation and Geoinformation\/}} \bibinfo{volume}{2}, \bibinfo{number}{3}
  (\bibinfo{year}{2000}), \bibinfo{pages}{161--171}.
\newblock


\bibitem[\protect\citeauthoryear{Kapadia, Pelechano, Allbeck, and
  Badler}{Kapadia et~al\mbox{.}}{2015}]%
        {doi:10.2200/S00673ED1V01Y201509CGR020}
\bibfield{author}{\bibinfo{person}{Mubbasir Kapadia}, \bibinfo{person}{Nuria
  Pelechano}, \bibinfo{person}{Jan Allbeck}, {and} \bibinfo{person}{Norm
  Badler}.} \bibinfo{year}{2015}\natexlab{}.
\newblock \showarticletitle{Virtual Crowds: Steps Toward Behavioral Realism}.
\newblock \bibinfo{journal}{{\em Synthesis Lectures on Visual Computing\/}}
  \bibinfo{volume}{7}, \bibinfo{number}{4} (\bibinfo{year}{2015}),
  \bibinfo{pages}{1--270}.
\newblock
\showDOI{%
\url{https://doi.org/10.2200/S00673ED1V01Y201509CGR020}}
\showeprint{http://dx.doi.org/10.2200/S00673ED1V01Y201509CGR020}


\bibitem[\protect\citeauthoryear{Krause, Arbon\`{e}s, and Igel}{Krause
  et~al\mbox{.}}{2016}]%
        {NIPS2016_6457}
\bibfield{author}{\bibinfo{person}{Oswin Krause},
  \bibinfo{person}{D\'{\i}dac~Rodr\'{\i}guez Arbon\`{e}s}, {and}
  \bibinfo{person}{Christian Igel}.} \bibinfo{year}{2016}\natexlab{}.
\newblock \showarticletitle{CMA-ES with Optimal Covariance Update and Storage
  Complexity}.
\newblock In \bibinfo{booktitle}{{\em Advances in Neural Information Processing
  Systems 29}}, \bibfield{editor}{\bibinfo{person}{D.~D. Lee},
  \bibinfo{person}{M.~Sugiyama}, \bibinfo{person}{U.~V. Luxburg},
  \bibinfo{person}{I.~Guyon}, {and} \bibinfo{person}{R.~Garnett}} (Eds.).
  \bibinfo{publisher}{Curran Associates, Inc.}, \bibinfo{pages}{370--378}.
\newblock
\showURL{%
\url{http://papers.nips.cc/paper/6457-cma-es-with-optimal-covariance-update-and-storage-complexity.pdf}}


\bibitem[\protect\citeauthoryear{Lewis and Sauro}{Lewis and Sauro}{2009}]%
        {lewis2009factor}
\bibfield{author}{\bibinfo{person}{James~R Lewis} {and} \bibinfo{person}{Jeff
  Sauro}.} \bibinfo{year}{2009}\natexlab{}.
\newblock \showarticletitle{The factor structure of the system usability
  scale}.
\newblock In \bibinfo{booktitle}{{\em Human centered design}}.
  \bibinfo{publisher}{Springer}, \bibinfo{pages}{94--103}.
\newblock


\bibitem[\protect\citeauthoryear{Liu, Yang, AlHalawani, and Mitra}{Liu
  et~al\mbox{.}}{2013}]%
        {DBLP:journals/vc/LiuYAM13}
\bibfield{author}{\bibinfo{person}{Han Liu}, \bibinfo{person}{Yong{-}Liang
  Yang}, \bibinfo{person}{Sawsan AlHalawani}, {and} \bibinfo{person}{Niloy~J.
  Mitra}.} \bibinfo{year}{2013}\natexlab{}.
\newblock \showarticletitle{Constraint-aware interior layout exploration for
  pre-cast concrete-based buildings}.
\newblock \bibinfo{journal}{{\em The Visual Computer\/}} \bibinfo{volume}{29},
  \bibinfo{number}{6-8} (\bibinfo{year}{2013}), \bibinfo{pages}{663--673}.
\newblock
\showDOI{%
\url{https://doi.org/10.1007/s00371-013-0825-1}}


\bibitem[\protect\citeauthoryear{Ma, Vining, Lefebvre, and Sheffer}{Ma
  et~al\mbox{.}}{2014}]%
        {CGF:CGF12314}
\bibfield{author}{\bibinfo{person}{Chongyang Ma}, \bibinfo{person}{Nicholas
  Vining}, \bibinfo{person}{Sylvain Lefebvre}, {and} \bibinfo{person}{Alla
  Sheffer}.} \bibinfo{year}{2014}\natexlab{}.
\newblock \showarticletitle{Game level layout from design specification}.
\newblock \bibinfo{journal}{{\em Computer Graphics Forum\/}}
  \bibinfo{volume}{33}, \bibinfo{number}{2} (\bibinfo{year}{2014}),
  \bibinfo{pages}{95--104}.
\newblock
\showISSN{1467-8659}
\showDOI{%
\url{https://doi.org/10.1111/cgf.12314}}


\bibitem[\protect\citeauthoryear{Marks, Andalman, Beardsley, Freeman, Gibson,
  Hodgins, Kang, Mirtich, Pfister, Ruml, Ryall, Seims, and Shieber}{Marks
  et~al\mbox{.}}{1997}]%
        {Marks:1997:DGG:258734.258887}
\bibfield{author}{\bibinfo{person}{Joe Marks}, \bibinfo{person}{Brad Andalman},
  \bibinfo{person}{Paul~A. Beardsley}, \bibinfo{person}{William Freeman},
  \bibinfo{person}{Sarah Gibson}, \bibinfo{person}{Jessica Hodgins},
  \bibinfo{person}{Thomas Kang}, \bibinfo{person}{Brian Mirtich},
  \bibinfo{person}{Hanspeter Pfister}, \bibinfo{person}{Wheeler Ruml},
  \bibinfo{person}{Kathy Ryall}, \bibinfo{person}{Joshua Seims}, {and}
  \bibinfo{person}{Stuart Shieber}.} \bibinfo{year}{1997}\natexlab{}.
\newblock \showarticletitle{Design Galleries: A General Approach to Setting
  Parameters for Computer Graphics and Animation}. In \bibinfo{booktitle}{{\em
  Proceedings of ACM SIGGRAPH}}. \bibinfo{pages}{389--400}.
\newblock
\showISBNx{0-89791-896-7}
\showDOI{%
\url{https://doi.org/10.1145/258734.258887}}


\bibitem[\protect\citeauthoryear{Marler and Arora}{Marler and Arora}{2004}]%
        {Marler2004}
\bibfield{author}{\bibinfo{person}{R.~Timothy Marler} {and}
  \bibinfo{person}{Jasbir~S. Arora}.} \bibinfo{year}{2004}\natexlab{}.
\newblock \showarticletitle{Survey of multi-objective optimization methods for
  engineering}.
\newblock \bibinfo{journal}{{\em Structural and Multidisciplinary
  Optimization\/}} \bibinfo{volume}{26}, \bibinfo{number}{6}
  (\bibinfo{year}{2004}), \bibinfo{pages}{369--395}.
\newblock
\showISSN{1615-1488}
\showDOI{%
\url{https://doi.org/10.1007/s00158-003-0368-6}}


\bibitem[\protect\citeauthoryear{Meilinger, Franz, and B{\"u}lthoff}{Meilinger
  et~al\mbox{.}}{2012}]%
        {meilinger2012isovists}
\bibfield{author}{\bibinfo{person}{Tobias Meilinger}, \bibinfo{person}{Gerald
  Franz}, {and} \bibinfo{person}{Heinrich~H B{\"u}lthoff}.}
  \bibinfo{year}{2012}\natexlab{}.
\newblock \showarticletitle{From isovists via mental representations to
  behaviour: first steps toward closing the causal chain}.
\newblock \bibinfo{journal}{{\em Environment and Planning B: Planning and
  Design\/}} \bibinfo{volume}{39}, \bibinfo{number}{1} (\bibinfo{year}{2012}),
  \bibinfo{pages}{48--62}.
\newblock


\bibitem[\protect\citeauthoryear{Merrell, Schkufza, and Koltun}{Merrell
  et~al\mbox{.}}{2010}]%
        {Merrell:2010:CRB:1882261.1866203}
\bibfield{author}{\bibinfo{person}{Paul Merrell}, \bibinfo{person}{Eric
  Schkufza}, {and} \bibinfo{person}{Vladlen Koltun}.}
  \bibinfo{year}{2010}\natexlab{}.
\newblock \showarticletitle{Computer-generated Residential Building Layouts}.
\newblock \bibinfo{journal}{{\em ACM Trans. Graph.\/}} \bibinfo{volume}{29},
  \bibinfo{number}{6}, Article \bibinfo{articleno}{181} (\bibinfo{date}{Dec.}
  \bibinfo{year}{2010}), \bibinfo{numpages}{12}~pages.
\newblock
\showISSN{0730-0301}
\showDOI{%
\url{https://doi.org/10.1145/1882261.1866203}}


\bibitem[\protect\citeauthoryear{Merrell, Schkufza, Li, Agrawala, and
  Koltun}{Merrell et~al\mbox{.}}{2011}]%
        {Merrell:2011:IFL:2010324.1964982}
\bibfield{author}{\bibinfo{person}{Paul Merrell}, \bibinfo{person}{Eric
  Schkufza}, \bibinfo{person}{Zeyang Li}, \bibinfo{person}{Maneesh Agrawala},
  {and} \bibinfo{person}{Vladlen Koltun}.} \bibinfo{year}{2011}\natexlab{}.
\newblock \showarticletitle{Interactive Furniture Layout Using Interior Design
  Guidelines}.
\newblock \bibinfo{journal}{{\em ACM Trans. Graph.\/}} \bibinfo{volume}{30},
  \bibinfo{number}{4}, Article \bibinfo{articleno}{87} (\bibinfo{date}{July}
  \bibinfo{year}{2011}), \bibinfo{numpages}{10}~pages.
\newblock
\showISSN{0730-0301}
\showDOI{%
\url{https://doi.org/10.1145/2010324.1964982}}


\bibitem[\protect\citeauthoryear{Michalek and Papalambros}{Michalek and
  Papalambros}{2002}]%
        {michalek2002interactive}
\bibfield{author}{\bibinfo{person}{Jeremy Michalek} {and}
  \bibinfo{person}{Panos Papalambros}.} \bibinfo{year}{2002}\natexlab{}.
\newblock \showarticletitle{{Interactive design optimization of architectural
  layouts}}.
\newblock \bibinfo{journal}{{\em Engineering Optimization\/}}
  \bibinfo{volume}{34}, \bibinfo{number}{5} (\bibinfo{year}{2002}),
  \bibinfo{pages}{485--501}.
\newblock
\showDOI{%
\url{https://doi.org/10.1080/03052150214021}}


\bibitem[\protect\citeauthoryear{Milgo, Ronoh, Waiganjo, and Manderick}{Milgo
  et~al\mbox{.}}{2017}]%
        {Milgo:2017:ACB:3067695.3075611}
\bibfield{author}{\bibinfo{person}{Edna Milgo}, \bibinfo{person}{Nixon Ronoh},
  \bibinfo{person}{Peter Waiganjo}, {and} \bibinfo{person}{Bernard Manderick}.}
  \bibinfo{year}{2017}\natexlab{}.
\newblock \showarticletitle{Adaptiveness of CMA Based Samplers}. In
  \bibinfo{booktitle}{{\em Proceedings of the Genetic and Evolutionary
  Computation Conference Companion}} {\em (\bibinfo{series}{GECCO '17})}.
  \bibinfo{publisher}{ACM}, \bibinfo{address}{New York, NY, USA},
  \bibinfo{pages}{179--180}.
\newblock
\showISBNx{978-1-4503-4939-0}
\showDOI{%
\url{https://doi.org/10.1145/3067695.3075611}}


\bibitem[\protect\citeauthoryear{Müller and Sbalzarini}{Müller and
  Sbalzarini}{2010}]%
        {5586491}
\bibfield{author}{\bibinfo{person}{C.~L. Müller} {and} \bibinfo{person}{I.~F.
  Sbalzarini}.} \bibinfo{year}{2010}\natexlab{}.
\newblock \showarticletitle{Gaussian Adaptation as a unifying framework for
  continuous black-box optimization and adaptive Monte Carlo sampling}. In
  \bibinfo{booktitle}{{\em IEEE Congress on Evolutionary Computation}}.
  \bibinfo{pages}{1--8}.
\newblock
\showISSN{1089-778X}
\showDOI{%
\url{https://doi.org/10.1109/CEC.2010.5586491}}


\bibitem[\protect\citeauthoryear{Neetil and de~Mendez}{Neetil and
  de~Mendez}{2012}]%
        {Neetil:2012:SGS:2230458}
\bibfield{author}{\bibinfo{person}{Jaroslav Neetil} {and}
  \bibinfo{person}{Patrice~Ossona de Mendez}.} \bibinfo{year}{2012}\natexlab{}.
\newblock \bibinfo{booktitle}{{\em Sparsity: Graphs, Structures, and
  Algorithms}}.
\newblock \bibinfo{publisher}{Springer Publishing Company, Incorporated}.
\newblock
\showISBNx{3642278744, 9783642278747}


\bibitem[\protect\citeauthoryear{Nguyen, Reiter, and Rigo}{Nguyen
  et~al\mbox{.}}{2014}]%
        {NGUYEN20141043}
\bibfield{author}{\bibinfo{person}{Anh-Tuan Nguyen}, \bibinfo{person}{Sigrid
  Reiter}, {and} \bibinfo{person}{Philippe Rigo}.}
  \bibinfo{year}{2014}\natexlab{}.
\newblock \showarticletitle{A review on simulation-based optimization methods
  applied to building performance analysis}.
\newblock \bibinfo{journal}{{\em Applied Energy\/}} \bibinfo{volume}{113},
  \bibinfo{number}{Supplement C} (\bibinfo{year}{2014}), \bibinfo{pages}{1043
  -- 1058}.
\newblock
\showISSN{0306-2619}
\showDOI{%
\url{https://doi.org/10.1016/j.apenergy.2013.08.061}}


\bibitem[\protect\citeauthoryear{Peng, Yang, Bao, Fink, Yan, Wonka, and
  Mitra}{Peng et~al\mbox{.}}{2016}]%
        {Peng:2016:CND:2897824.2925935}
\bibfield{author}{\bibinfo{person}{Chi-Han Peng}, \bibinfo{person}{Yong-Liang
  Yang}, \bibinfo{person}{Fan Bao}, \bibinfo{person}{Daniel Fink},
  \bibinfo{person}{Dong-Ming Yan}, \bibinfo{person}{Peter Wonka}, {and}
  \bibinfo{person}{Niloy~J. Mitra}.} \bibinfo{year}{2016}\natexlab{}.
\newblock \showarticletitle{Computational Network Design from Functional
  Specifications}.
\newblock \bibinfo{journal}{{\em ACM Trans. Graph.\/}} \bibinfo{volume}{35},
  \bibinfo{number}{4}, Article \bibinfo{articleno}{131} (\bibinfo{date}{July}
  \bibinfo{year}{2016}), \bibinfo{numpages}{12}~pages.
\newblock
\showISSN{0730-0301}
\showDOI{%
\url{https://doi.org/10.1145/2897824.2925935}}


\bibitem[\protect\citeauthoryear{Peponis, Zimring, and Choi}{Peponis
  et~al\mbox{.}}{1990}]%
        {peponis}
\bibfield{author}{\bibinfo{person}{John Peponis}, \bibinfo{person}{Craig
  Zimring}, {and} \bibinfo{person}{Yoon~Kyung Choi}.}
  \bibinfo{year}{1990}\natexlab{}.
\newblock \showarticletitle{Finding the building in wayfinding}.
\newblock \bibinfo{journal}{{\em Environment and behavior\/}}
  \bibinfo{volume}{22}, \bibinfo{number}{5} (\bibinfo{year}{1990}),
  \bibinfo{pages}{555--590}.
\newblock


\bibitem[\protect\citeauthoryear{Pottmann, Eigensatz, Vaxman, and
  Wallner}{Pottmann et~al\mbox{.}}{2014}]%
        {pottmann2014architectural}
\bibfield{author}{\bibinfo{person}{Helmut Pottmann}, \bibinfo{person}{Michael
  Eigensatz}, \bibinfo{person}{Amir Vaxman}, {and} \bibinfo{person}{Johannes
  Wallner}.} \bibinfo{year}{2014}\natexlab{}.
\newblock \showarticletitle{Architectural geometry}.
\newblock \bibinfo{journal}{{\em Computers \& Graphics\/}}
  (\bibinfo{year}{2014}).
\newblock


\bibitem[\protect\citeauthoryear{Rousseeuw and Croux}{Rousseeuw and
  Croux}{1992}]%
        {rousseeuw1992statistics}
\bibfield{author}{\bibinfo{person}{Peter~J. Rousseeuw} {and}
  \bibinfo{person}{Christophe Croux}.} \bibinfo{year}{1992}\natexlab{}.
\newblock \showarticletitle{Explicit scale estimators with high breakdown
  point}.
\newblock \bibinfo{journal}{{\em L1-Statistical analysis and related
  methods\/}}  \bibinfo{volume}{1} (\bibinfo{year}{1992}),
  \bibinfo{pages}{77--92}.
\newblock


\bibitem[\protect\citeauthoryear{Sauro and Lewis}{Sauro and Lewis}{2011}]%
        {sauro2011designing}
\bibfield{author}{\bibinfo{person}{Jeff Sauro} {and} \bibinfo{person}{James~R
  Lewis}.} \bibinfo{year}{2011}\natexlab{}.
\newblock \showarticletitle{When designing usability questionnaires, does it
  hurt to be positive?}. In \bibinfo{booktitle}{{\em Proceedings of the SIGCHI
  Conference on Human Factors in Computing Systems}}. ACM,
  \bibinfo{pages}{2215--2224}.
\newblock


\bibitem[\protect\citeauthoryear{Shi and Yang}{Shi and Yang}{2013}]%
        {shi2013performance}
\bibfield{author}{\bibinfo{person}{Xing Shi} {and} \bibinfo{person}{Wenjie
  Yang}.} \bibinfo{year}{2013}\natexlab{}.
\newblock \showarticletitle{Performance-driven architectural design and
  optimization technique from a perspective of architects}.
\newblock \bibinfo{journal}{{\em Automation in Construction\/}}
  \bibinfo{volume}{32} (\bibinfo{year}{2013}), \bibinfo{pages}{125--135}.
\newblock


\bibitem[\protect\citeauthoryear{Srivastava, Nguyen, Gerevini, Kambhampati, Do,
  and Serina}{Srivastava et~al\mbox{.}}{2007}]%
        {srivastava2007domain}
\bibfield{author}{\bibinfo{person}{Biplav Srivastava},
  \bibinfo{person}{Tuan~Anh Nguyen}, \bibinfo{person}{Alfonso Gerevini},
  \bibinfo{person}{Subbarao Kambhampati}, \bibinfo{person}{Minh~Binh Do}, {and}
  \bibinfo{person}{Ivan Serina}.} \bibinfo{year}{2007}\natexlab{}.
\newblock \showarticletitle{Domain Independent Approaches for Finding Diverse
  Plans.}. In \bibinfo{booktitle}{{\em IJCAI}}. \bibinfo{pages}{2016--2022}.
\newblock


\bibitem[\protect\citeauthoryear{Tullis and Stetson}{Tullis and
  Stetson}{2004}]%
        {tullis2004comparison}
\bibfield{author}{\bibinfo{person}{Thomas~S. Tullis} {and}
  \bibinfo{person}{Jacqueline~N. Stetson}.} \bibinfo{year}{2004}\natexlab{}.
\newblock \showarticletitle{A comparison of questionnaires for assessing
  website usability}. In \bibinfo{booktitle}{{\em Usability professional
  association conference}}. \bibinfo{pages}{1--12}.
\newblock


\bibitem[\protect\citeauthoryear{Turner}{Turner}{2001}]%
        {turner}
\bibfield{author}{\bibinfo{person}{Alasdair Turner}.}
  \bibinfo{year}{2001}\natexlab{}.
\newblock \showarticletitle{A program to perform visibility graph analysis}. In
  \bibinfo{booktitle}{{\em Proceedings of the 3rd Space Syntax Symposium,
  Atlanta, University of Michigan}}. \bibinfo{pages}{31--1}.
\newblock


\bibitem[\protect\citeauthoryear{Turner and Penn}{Turner and Penn}{1999}]%
        {turner1999making}
\bibfield{author}{\bibinfo{person}{Alasdair Turner} {and} \bibinfo{person}{Alan
  Penn}.} \bibinfo{year}{1999}\natexlab{}.
\newblock \showarticletitle{Making isovists syntactic: isovist integration
  analysis}. In \bibinfo{booktitle}{{\em 2nd International Symposium on Space
  Syntax, Brasilia}}. Citeseer.
\newblock


\bibitem[\protect\citeauthoryear{Turrin, von Buelow, and Stouffs}{Turrin
  et~al\mbox{.}}{2011}]%
        {turrin2011design}
\bibfield{author}{\bibinfo{person}{Michela Turrin}, \bibinfo{person}{Peter von
  Buelow}, {and} \bibinfo{person}{Rudi Stouffs}.}
  \bibinfo{year}{2011}\natexlab{}.
\newblock \showarticletitle{Design explorations of performance driven geometry
  in architectural design using parametric modeling and genetic algorithms}.
\newblock \bibinfo{journal}{{\em Advanced Engineering Informatics\/}}
  \bibinfo{volume}{25}, \bibinfo{number}{4} (\bibinfo{year}{2011}),
  \bibinfo{pages}{656--675}.
\newblock


\bibitem[\protect\citeauthoryear{Twigg and James}{Twigg and James}{2007}]%
        {Twigg:2007:MBC:1275808.1276395}
\bibfield{author}{\bibinfo{person}{Christopher~D. Twigg} {and}
  \bibinfo{person}{Doug~L. James}.} \bibinfo{year}{2007}\natexlab{}.
\newblock \showarticletitle{Many-worlds Browsing for Control of Multibody
  Dynamics}. In \bibinfo{booktitle}{{\em Proceedings of ACM SIGGRAPH}}.
  \bibinfo{publisher}{ACM}, \bibinfo{address}{New York, NY, USA}, Article
  \bibinfo{articleno}{14}.
\newblock
\showDOI{%
\url{https://doi.org/10.1145/1275808.1276395}}


\bibitem[\protect\citeauthoryear{Ursem}{Ursem}{2002}]%
        {Ursem2002}
\bibfield{author}{\bibinfo{person}{Rasmus~K. Ursem}.}
  \bibinfo{year}{2002}\natexlab{}.
\newblock \bibinfo{booktitle}{{\em Parallel Problem Solving from Nature ---
  PPSN VII: 7th International Conference Granada, Spain, September 7--11, 2002
  Proceedings}}.
\newblock \bibinfo{publisher}{Springer Berlin Heidelberg},
  \bibinfo{address}{Berlin, Heidelberg}, Chapter Diversity-Guided Evolutionary
  Algorithms, \bibinfo{pages}{462--471}.
\newblock
\showISBNx{978-3-540-45712-1}
\showDOI{%
\url{https://doi.org/10.1007/3-540-45712-7_45}}


\bibitem[\protect\citeauthoryear{Usman, Haworth, Berseth, Kapadia, and
  Faloutsos}{Usman et~al\mbox{.}}{2017}]%
        {Usman:2017:PES:3136457.3136458}
\bibfield{author}{\bibinfo{person}{Muhammad Usman}, \bibinfo{person}{Brandon
  Haworth}, \bibinfo{person}{Glen Berseth}, \bibinfo{person}{Mubbasir Kapadia},
  {and} \bibinfo{person}{Petros Faloutsos}.} \bibinfo{year}{2017}\natexlab{}.
\newblock \showarticletitle{Perceptual Evaluation of Space in Virtual
  Environments}. In \bibinfo{booktitle}{{\em Proceedings of the Tenth
  International Conference on Motion in Games}} {\em (\bibinfo{series}{MIG
  '17})}. \bibinfo{publisher}{ACM}, \bibinfo{address}{New York, NY, USA},
  Article \bibinfo{articleno}{16}, \bibinfo{numpages}{10}~pages.
\newblock
\showISBNx{978-1-4503-5541-4}
\showDOI{%
\url{https://doi.org/10.1145/3136457.3136458}}


\bibitem[\protect\citeauthoryear{Wagner, Beume, and Naujoks}{Wagner
  et~al\mbox{.}}{2007}]%
        {wagner2007pareto}
\bibfield{author}{\bibinfo{person}{Tobias Wagner}, \bibinfo{person}{Nicola
  Beume}, {and} \bibinfo{person}{Boris Naujoks}.}
  \bibinfo{year}{2007}\natexlab{}.
\newblock \showarticletitle{Pareto-, aggregation-, and indicator-based methods
  in many-objective optimization}. In \bibinfo{booktitle}{{\em Evolutionary
  multi-criterion optimization}}. Springer, \bibinfo{pages}{742--756}.
\newblock


\bibitem[\protect\citeauthoryear{Yi and Yi}{Yi and Yi}{2014}]%
        {yi2014performance}
\bibfield{author}{\bibinfo{person}{Hwang Yi} {and} \bibinfo{person}{Yun~Kyu
  Yi}.} \bibinfo{year}{2014}\natexlab{}.
\newblock \showarticletitle{Performance Based Architectural Design
  Optimization: Automated 3D Space Layout Using Simulated Annealing}. In
  \bibinfo{booktitle}{{\em ASHRAE/IBPSA-USA Building Simulation Conference}}.
\newblock


\bibitem[\protect\citeauthoryear{Yu, Yeung, Tang, Terzopoulos, Chan, and
  Osher}{Yu et~al\mbox{.}}{2011}]%
        {craigyu2011furniture}
\bibfield{author}{\bibinfo{person}{Lap-Fai Yu}, \bibinfo{person}{Sai~Kit
  Yeung}, \bibinfo{person}{Chi-Keung Tang}, \bibinfo{person}{Demetri
  Terzopoulos}, \bibinfo{person}{Tony~F. Chan}, {and} \bibinfo{person}{Stanley
  Osher}.} \bibinfo{year}{2011}\natexlab{}.
\newblock \showarticletitle{Make it home: automatic optimization of furniture
  arrangement}.
\newblock \bibinfo{journal}{{\em ACM Transactions on Graphics\/}}
  \bibinfo{volume}{30}, \bibinfo{number}{4} (\bibinfo{year}{2011}),
  \bibinfo{pages}{86}.
\newblock


\end{thebibliography}

\addcontentsline{toc}{chapter}{Appendix}
\section{Appendix}

\subsection{Metrics}

This section describes additional details related to the methods used in this work.

\subsection{CMA vs Simulated Annealing + MCMC} 
		
\newChanges{}{
	The choice of optimization algorithm to use for this type of design problem is an important consideration. 
	A recent review of building architecture related optimization frameworks highlights the numerous optimization techniques used in the area, and reasons why some are better than others for particular design problems~\cite{NGUYEN20141043}.
	Here we list the most relevant reasons for using CMA.
	Simulated annealing (SA) may need careful design of special parameter selection methods, like the ones used in ~\cite{midlayout}.
	SA is a poor choice given our desire for imposing design constraints.
	SA can handle noisy objective functions but only under certain conditions that can not be guaranteed for most building metrics. 
	Also, genetic algorithms (GAs), like CMA, are often parallelizable, making the method more efficient.
	CMA should be better at escaping local-minima.
	Last, GAs have also been shown to show better early convergence, leading to quickly finding good local-minima that are often good enough for these types of design problems.
}
	\newChanges{}{CMA is a form of MCMC where the chain is the series of generated covariance distributions~\cite{NIPS2016_6457}.
	You can even formulate MCMC to use a variant of CMA for sampling to improve convergence~\cite{5586491}
	These samplers outperform many variants of MCMC~\cite{Milgo:2017:ACB:3067695.3075611}
	}

\subsection{Multi-Objective Optimization Methods}

\subsubsection{Scalarized}
Computes a weighted combination of the objectives, weighting all of the objective terms with respect to some relative weighting. This method is challenging to use for two reasons.
One, determining the weights to use for a combination of objectives can be a daunting task.
Also, the objectives themselves may not be linear, with some growing faster than others usually precluding the possibility of finding a single set of weights that works well when the environment changes.
Second, If a relative weighting is used it helps to normalize the metrics in some way.
The maximum value for the Degree metric can be found by removing all of the items from the simulation and calculating the degree. There is no simple calculation to find the diversity bound, however, an upper bound can be found via optimization. The diversity metric is very cheap to compute (relative to Degree, etc), an optimization for only diversity can be performed first, to find the upper bound on diversity. Both degree and diversity are non-linear functions, this is okay and could give desirable results when performing a scalarized optimization, but it would still be challenging to find objective weights~\cite{ding2006discussions}.

\subsubsection{Pareto Front Optimization}
This method essentially finds a set of points (non-dominated points) that are optimal trade-offs between a set of objectives. The issue with using a Pareto Optimal Front method is that the computation of diversity between the members is non-trivial. Diversity is a measure of the distance between points in the Pareto front. It is not clear how to accomplish this without introducing a large number of parameters. Possibly, two different objectives could be chosen to optimize with respect to, but those objectives are only proxies for diversity and could be very similar producing results with minimal dissimilarity.

\subsubsection{Hierarchical Optimization}

\begin{figure}[!htb]
\includegraphics[width=0.95\linewidth]{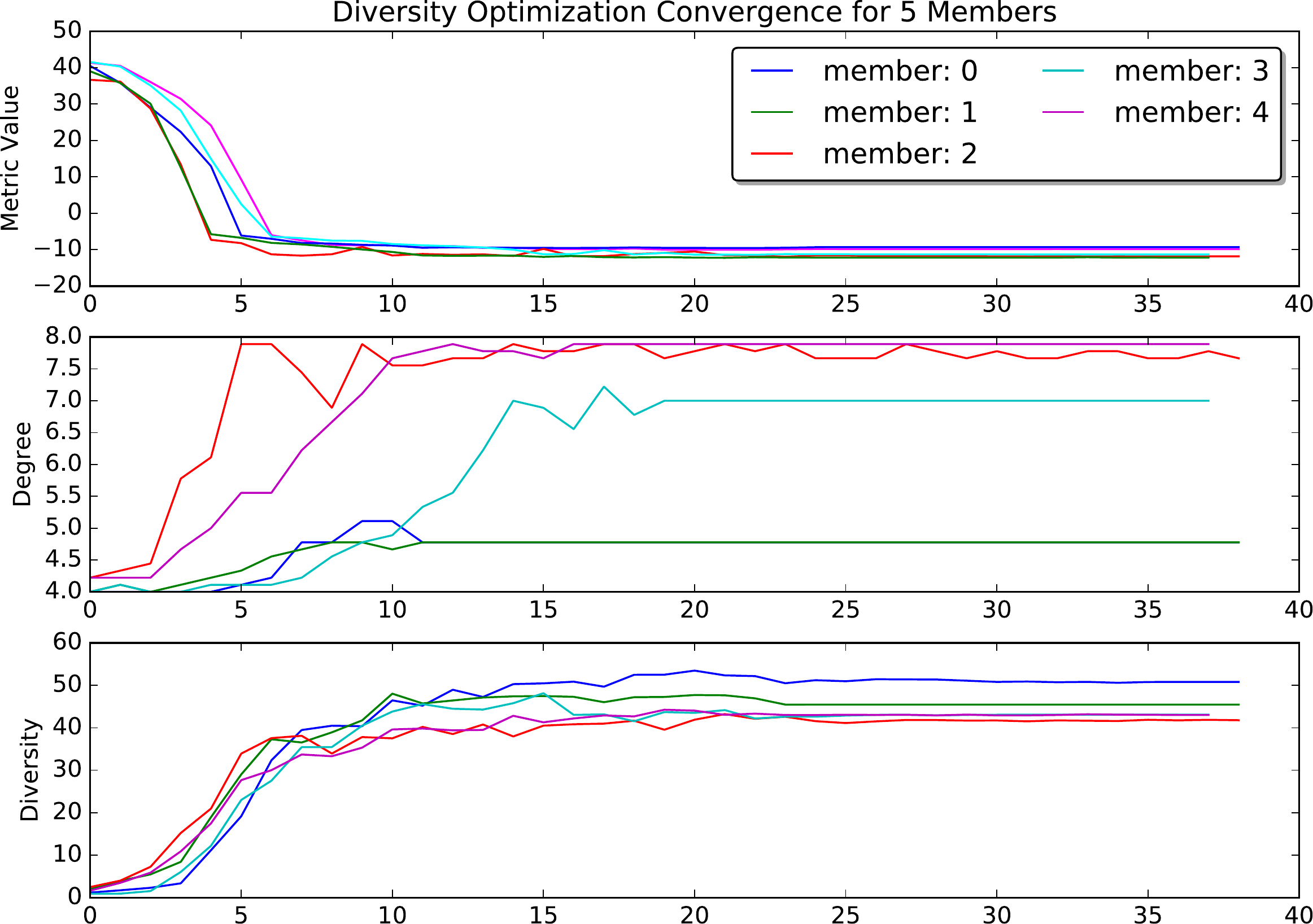}
\caption{Diversity optimization convergence}
\label{fig:optimization-convergence}
\end{figure}

With hierarchical optimization an ordering and objective specific thresholds are used, instead of only relative weights. The objectives are optimized in the order given. Each objective is optimized to find its optimum and from this a constraint is added to the optimization for the next objective. This constraint adds a penalty whenever the  previous objective(s) value goes below the threshold value(s). This gives more control over the trade-offs between objectives. This method works well and converges quickly, as can be seen in \reffig{fig:optimization-convergence}.
In this experiment we optimized \textit{art-gallery B} in~\reffig{figure:half-art-diversity} with a diversity set of size $5$.
This optimization completed in a few seconds and converged before the optimization was terminated. 

\end{document}